\documentstyle[amssym]{mn}
\font\japit = cmti10 at 10truept
\input{epsfig.sty}

\title
     [Real Space Power of PSCz]
{\vglue-3.0truecm
\centerline{\japit Submitted to Monthly Notices}
\vglue 2.5truecm
\noindent
The Real Space Power Spectrum of the PSCz Survey from 0.01 to 300 $\bmath h$\,Mpc$\bmath{^{-1}}$
\author[A. J. S. Hamilton and M. Tegmark]
     {A. J. S. Hamilton$^1$ and Max Tegmark$^2$ \\
	$^1$JILA and Dept.\ Astrophysical \& Planetary Sciences,
	Box 440, U. Colorado, Boulder CO 80309, USA; \\
	\ Andrew.Hamilton@colorado.edu; http:$/\!/$casa.colorado.edu/$\sim$ajsh/ \\
	$^2$Dept. of Physics, Univ. of Pennsylvania, Philadelphia, PA 19104, USA;
	max@physics.upenn.edu; http:$/\!/$www.hep.upenn.edu/$\sim$max/}
}

\newcommand{\rmn}{\rm}
\newcommand{\bmi}{\bmath}

\newcommand{\mx}{\bf}		% Matrix font

\newcommand{\upi}{\pi}
\newcommand{\upartial}{\partial}
\newcommand{\dd}{{\rmn d}}	% MNRAS
\newcommand{\e}{{\rmn e}}	% MNRAS
\newcommand{\im}{{\rmn i}}	% MNRAS

\newcommand{\ddt}{\dd^2 \hspace{-.5pt}}
\newcommand{\ddd}{\dd^3 \hspace{-.5pt}}

\newcommand{\simpropto}{\!\!\begin{array}{c} {\propto} \\
                  [-1.7ex] \sim \end{array}\!\!}

\newcommand{\FFTLog}{FFTL{\sc og}}
\newcommand{\deltaD}{\delta_{3D}}

\newcommand{\kpc}{{\rmn kpc}}
\newcommand{\Mpc}{{\rmn Mpc}}

\newcommand{\el}{\ell}

\newcommand{\k}{{\bmi k}}
\newcommand{\r}{{\bmi r}}

\newcommand{\z}{{\bmi z}}

\newcommand{\mxA}{{\mx A}}
\newcommand{\mxC}{{\mx C}}
\newcommand{\mxH}{{\mx H}}
\newcommand{\mxunit}{{\mx 1}}

\newcommand{\aap}[2]{A\&A, #1, #2}

\newcommand{\aj}[2]{AJ, #1, #2}
\newcommand{\anyas}[2]{Ann.\ NY Acad.\ Sci., #1, #2}
\newcommand{\apj}[2]{ApJ, #1, #2}
\newcommand{\apjs}[2]{ApJS, #1, #2}
\newcommand{\araa}[2]{ARA\&A, #1, #2}
\newcommand{\ass}[2]{Ap\&SS, #1, #2}
\newcommand{\mn}[2]{MNRAS, #1, #2}

\newcommand{\sffig}{
    \begin{figure}
    \begin{center}
    \leavevmode
    \epsfxsize=2.9in	% normal
    \epsfbox{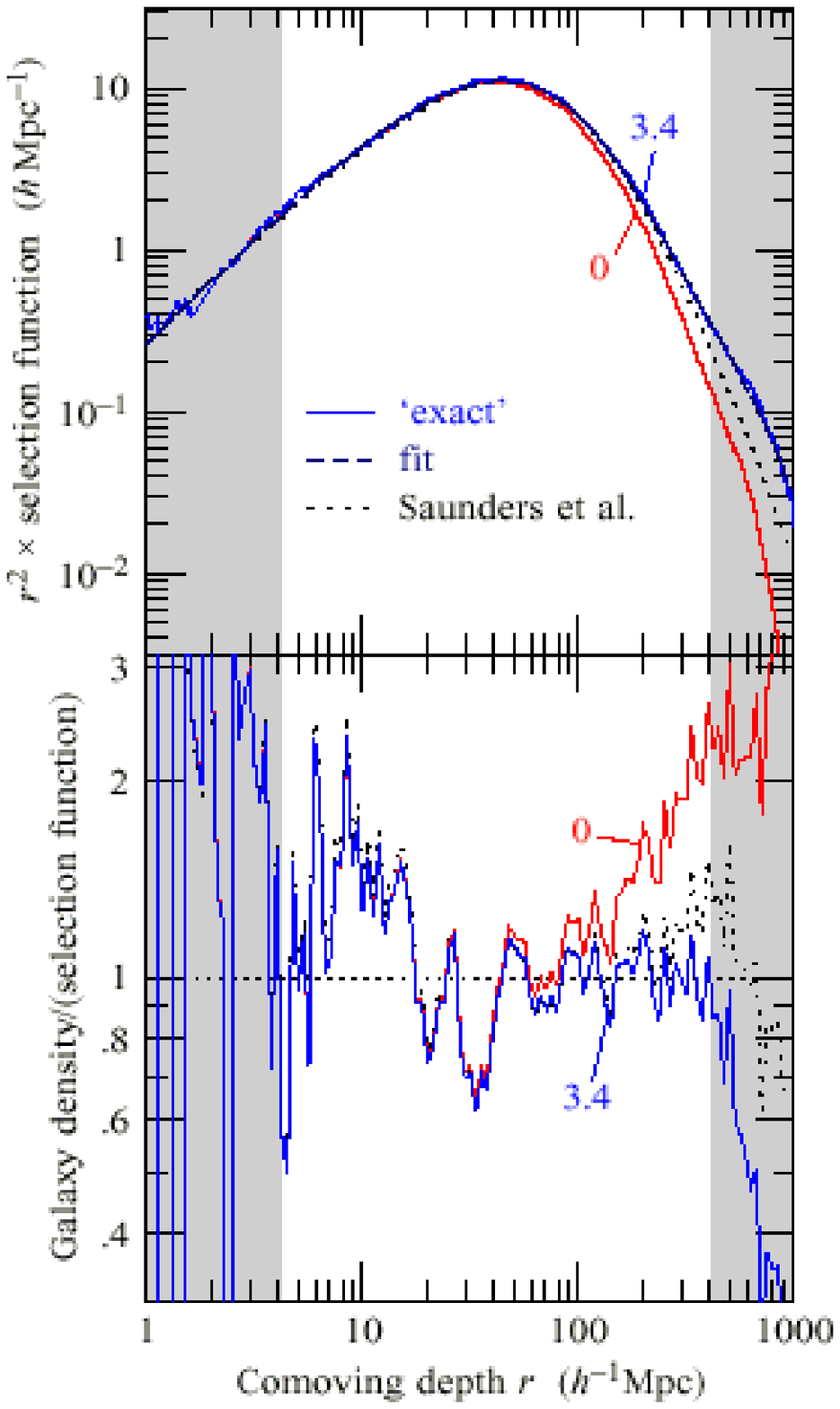}
    \end{center}
    \caption[1]{\small
(Upper panel) Selection function of the PSCz 0.6~Jy survey
as a function of comoving depth $r$.
The selection function is multiplied by $r^2$
in order to reduce its range and hence to bring out more detail.
The thin solid line is the `exact' selection function,
from Lynden-Bell's (1971) $C^-$ method;
the upper solid line assumes that galaxies evolve with
luminosity $\propto (1+z)^{3.4}$,
while the lower solid line assumes no luminosity evolution.
The `exact' selection function is actually a step function
with a step at the limiting distance of each galaxy,
but the steps are so fine (there are $12\,867$ of them)
that the lines look almost continuous.
The dashed line is the smooth analytic fit
to the selection function with evolution,
equation~(\protect\ref{sffit}),
adopted by HTP and here.
The fit lies almost on top of the `exact' solution.
For comparison, the dotted line is the fit suggested by Saunders et al.\ (2000).
(Lower panel)
Ratio of the observed galaxy number density to the fitted
selection function at radial depth $r$ in the PSCz survey,
averaged in depth bins $0.025$~dex wide
(this plot appears also in HTP).
The lower line assumes that galaxies evolve with
luminosity $\propto (1+z)^{3.4}$,
while the upper line assumes no luminosity evolution.
The dotted line corresponds to the fit suggested by Saunders et al.\ (2000).
The unshaded region from radial depth
$10^{0.625} \, h^{-1} \Mpc \approx 4.2 \, h^{-1} \Mpc$
to
$10^{2.625} \, h^{-1} \Mpc \approx 420 \, h^{-1} \Mpc$
is the region retained for analysis in this paper.
    \label{sf}
    }
    \end{figure}
}

\newcommand{\xikcontsfig}{
    \begin{figure*}
    \begin{minipage}{175mm}
    \begin{center}
    \leavevmode
    \epsfxsize=4in	% normal
    \epsfbox{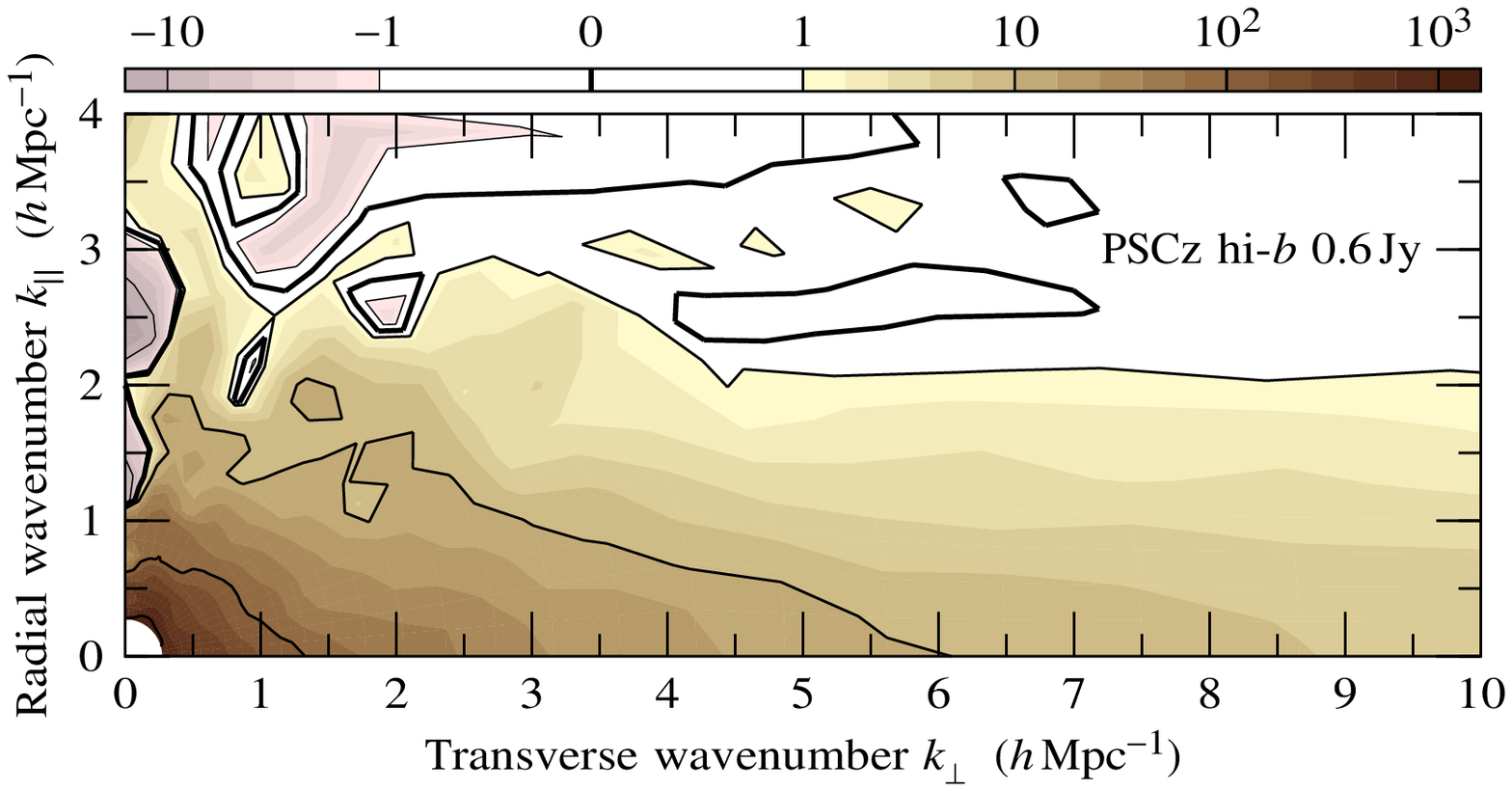}
    \end{center}
    \caption[1]{\small
Contour plot of the redshift space power spectrum
$P^s(k_\perp,k_\parallel)$
of the PSCz 0.6~Jy survey at nonlinear scales.
Power along the transverse (horizontal) axis is unaffected by redshift
distortions, and is therefore equal to the real space power spectrum.
Velocity dispersion suppresses power away from the transverse axis.
The plotted redshift power is constructed from the harmonics of
redshift power, truncated at the $k$-dependent maximum harmonic given by
equations~(\ref{ladopt}) and (\ref{lmax}).
The combination of FKP weightings (\S\S\ref{FKP}, \ref{compress})
is such as to optimize the measurement of power along the transverse axis.
Thin, medium, and thick contours represent
negative, positive, and zero values respectively.
    \label{xikconts}
    }
    \end{minipage}
    \end{figure*}
}

\newcommand{\xikcontsafig}{
    \begin{figure*}
    \begin{minipage}{175mm}
    \begin{center}
    \leavevmode
    \epsfxsize=4in	% normal
    \epsfbox{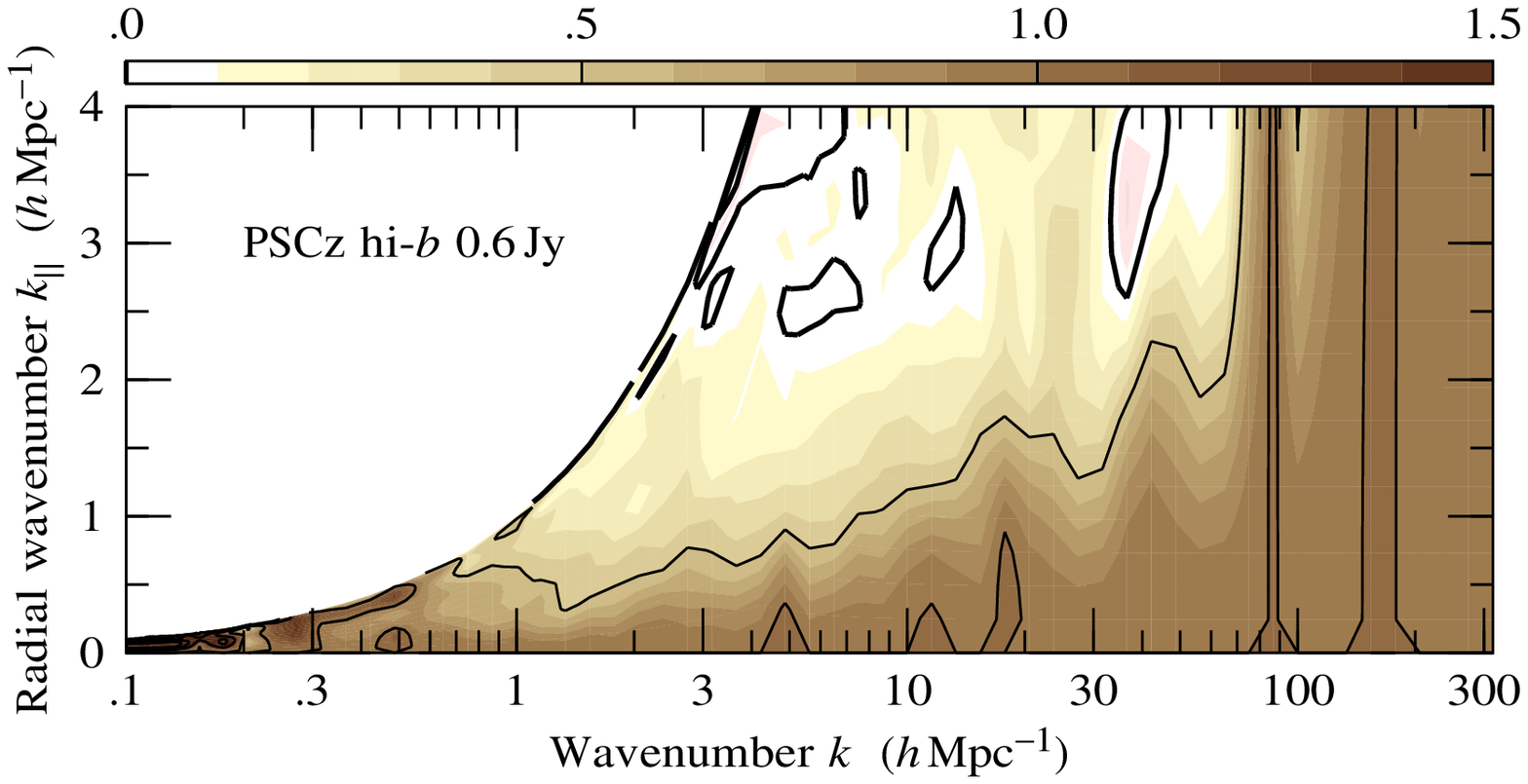}
    \end{center}
    \caption[1]{\small
Contour plot of the ratio
$f(\k) \equiv P^s(\k)/P(k)$
of the redshift to real space power spectrum.
By construction, the ratio $f(\k)$ equals one along the horizontal axis,
where $k_\parallel = 0$.
The width of the ridge along the horizontal axis
is roughly equal to the inverse of the pairwise galaxy velocity dispersion,
$\Delta k_\parallel \sim 1/\sigma$.
If velocity dispersion were independent of scale,
then the contours in this diagram would be horizontal.
The pairwise velocity dispersion reaches a maximum at
$k \approx 1.3 \, h \, \Mpc^{-1}$,
where the contours crowd the horizontal axis most closely.
Medium and thick contours represent
positive and zero values respectively.
The white space to the top left of the diagram
appears because the line-of-sight wavenumber $k_\parallel$ (the vertical axis)
must be less than or equal to the total wavenumber
$k = (k_\perp^2 + k_\parallel^2)^{1/2}$ (the horizontal axis);
the boundary is shaped exponentially because
the plot is linear in $k_\parallel$ but logarithmic in $k$.
    \label{xikcontsa}
    }
    \end{minipage}
    \end{figure*}
}

\newcommand{\wfig}{
    \begin{figure}
    \begin{center}
    \leavevmode
    \epsfxsize=2.6in	% normal
    \epsfbox{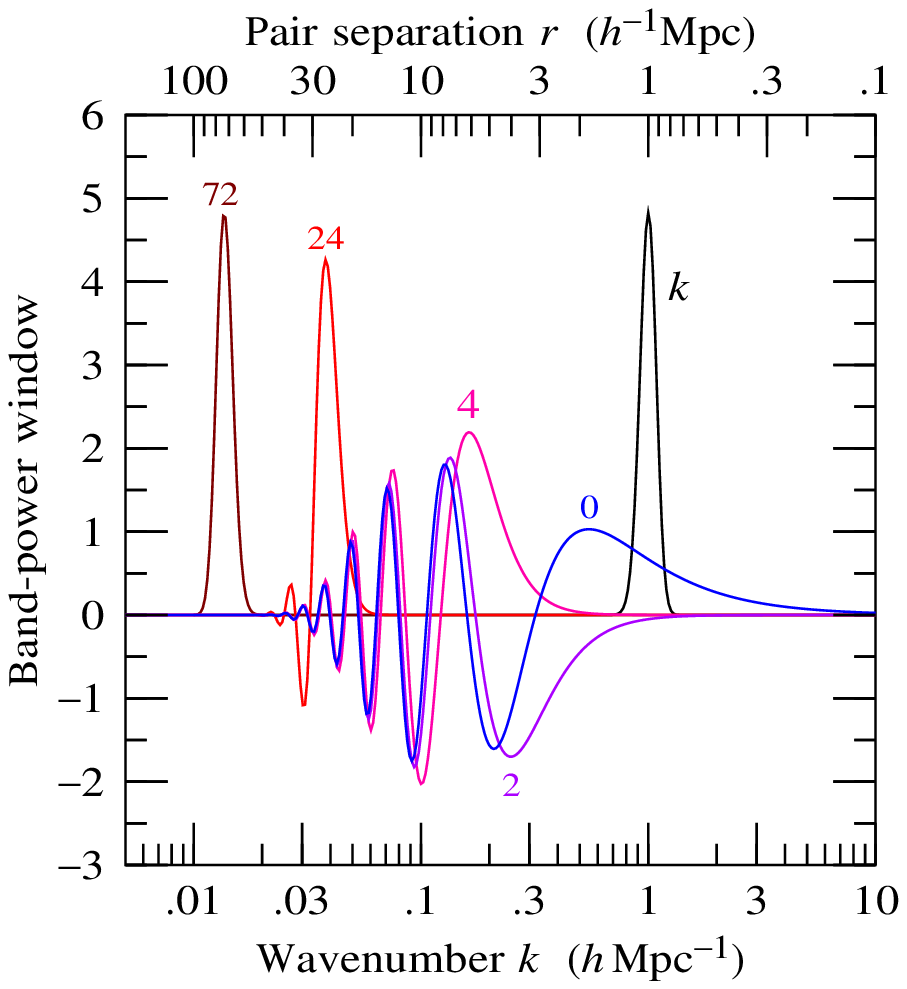}
    \end{center}
    \caption[1]{\small
Band-power windows for $\tilde k = 1 \, h \, \Mpc^{-1}$.
The window marked $k$ is the scaled band-power window
$W(\tilde k, k) \, k^{3/2}/(2\upi^2)$
with $n = 72$,
equation~(\ref{W}),
plotted as a function of the wavenumber $k$ labelled on the lower axis.
The window is scaled with $k^{-3/2} 4 \upi k^3/(2\upi)^3 = k^{3/2}/(2\upi^2)$
to reveal more clearly the effective shape of the window
when a power spectrum $\propto k^{-3/2}$
(as approximately the case in PSCz) is folded through it.
The plotted scaled window has the property that it yields $1$
when integrated over either $\int \dd\ln k$ or $\int k^{3/2} \, \dd\ln k$.
The remaining windows,
each marked with the associated harmonic number $\el$,
are the corresponding windows
$W_\el(\tilde k, r) \, r^{3/2} (2/\upi)^{1/2}$
in real space,
equation~(\ref{Wlag}),
plotted as a function of the separation $r$ labelled on the upper axis.
Again, each window is scaled with
$(2\upi r)^{-3/2} 4 \upi r^3 = r^{3/2} (2/\upi)^{1/2}$
to reveal more clearly the effective shape of the window
when a correlation function $(2\upi r)^{-3/2}$,
corresponding to a power spectrum $k^{-3/2}$,
is folded through it.
The plotted scaled windows have the property that they yield $1$
when integrated over $\int \dd\ln r$, for all $\el$.
Changing the characteristic wavenumber $\tilde k$ of the band-power
shifts all windows sideways on this plot, without changing their shape.
    \label{w}
    }
    \end{figure}
}

\newcommand{\mapfig}{
    \begin{figure*}
    \begin{minipage}{175mm}
    \begin{center}
    \leavevmode
    \epsfig{file=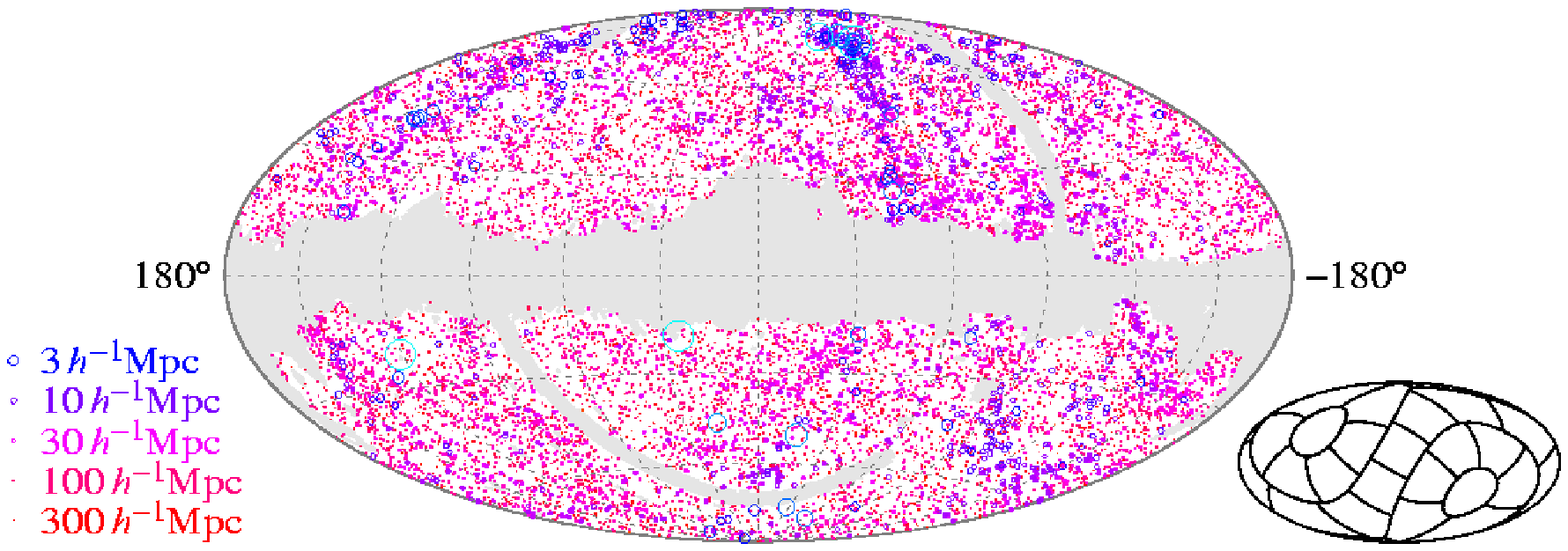, width=175mm, angle=0}
    \end{center}
    \caption[1]{\small
The $12\,871$ galaxies of the PSCz 0.6~Jy survey with
the high-latitude angular mask
({\tt hibpsczmask.dat} in the PSCz package).
The map is a Hammer-Aitoff projection,
in Galactic coordinates, with the Galactic centre at the centre.
Larger points signify closer galaxies
[area $\propto$ 1/(redshift distance)],
as exampled.
The inset shows the 22 angular regions used in the error analysis;
the boundaries of the angular regions are lines of constant
ecliptic longitude and latitude.
    \label{map}
    }
    \end{minipage}
    \end{figure*}
}

\newcommand{\xikfig}{
    \begin{figure*}
    \begin{minipage}{175mm}
    \begin{center}
    \leavevmode
    \epsfxsize=5in	% normal
    \epsfbox{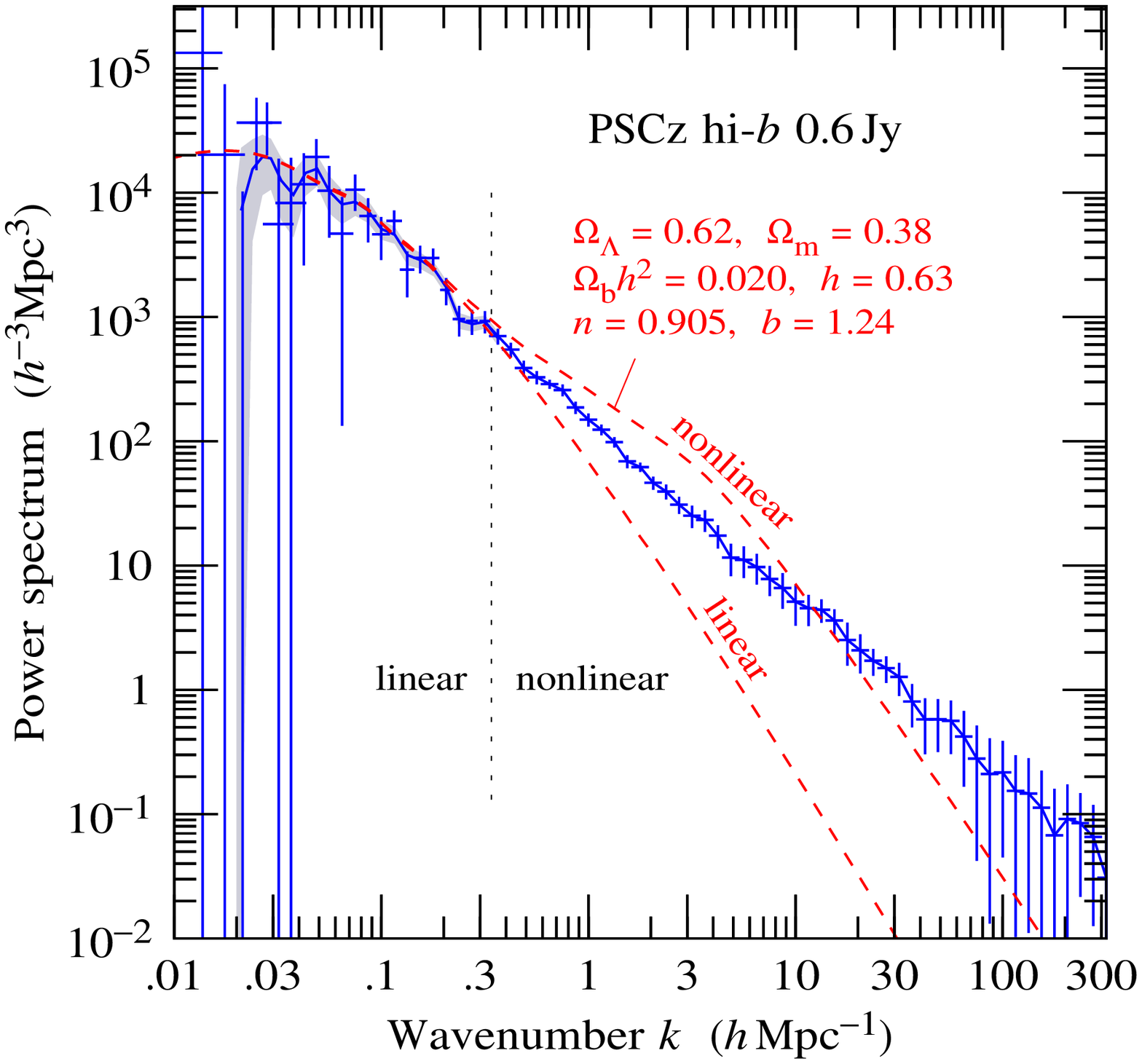}
    \end{center}
    \caption[1]{\small
Real space galaxy-galaxy power spectrum
measured from the PSCz 0.6~Jy survey with the high latitude angular mask.
To the left of the vertical dashed line
is the linear measurement from Hamilton et al.\ (2000),
while to the right
is the nonlinear measurement from the present paper.
The solid line is the correlated power spectrum.
In the linear regime (left of the vertical dashed line),
the shaded region is the $1 \sigma$ uncertainty in the correlated power spectrum,
and points with error bars constitute the decorrelated power spectrum
(Hamilton \& Tegmark 2000).
Each point of the decorrelated linear power spectrum is uncorrelated
with all other points.
It is not possible to decorrelate the nonlinear power spectrum,
so in the nonlinear regime (right of the vertical dashed line),
points with error bars are the errors in the correlated power spectrum.
The dashed lines are
the flat $\Lambda$CDM concordance model power spectrum
from Tegmark et al.\ (2001),
with parameters as indicated.
The lower dashed line is the linear model power spectrum,
the upper dashed line the model power spectrum
nonlinearly evolved according to the prescription of
Peacock \& Dodds (1996).
    \label{xik}
    }
    \end{minipage}
    \end{figure*}
}

\newcommand{\pairfig}{
    \begin{figure}
    \begin{center}
    \leavevmode
    \epsfxsize=2in	% normal
    \epsfbox{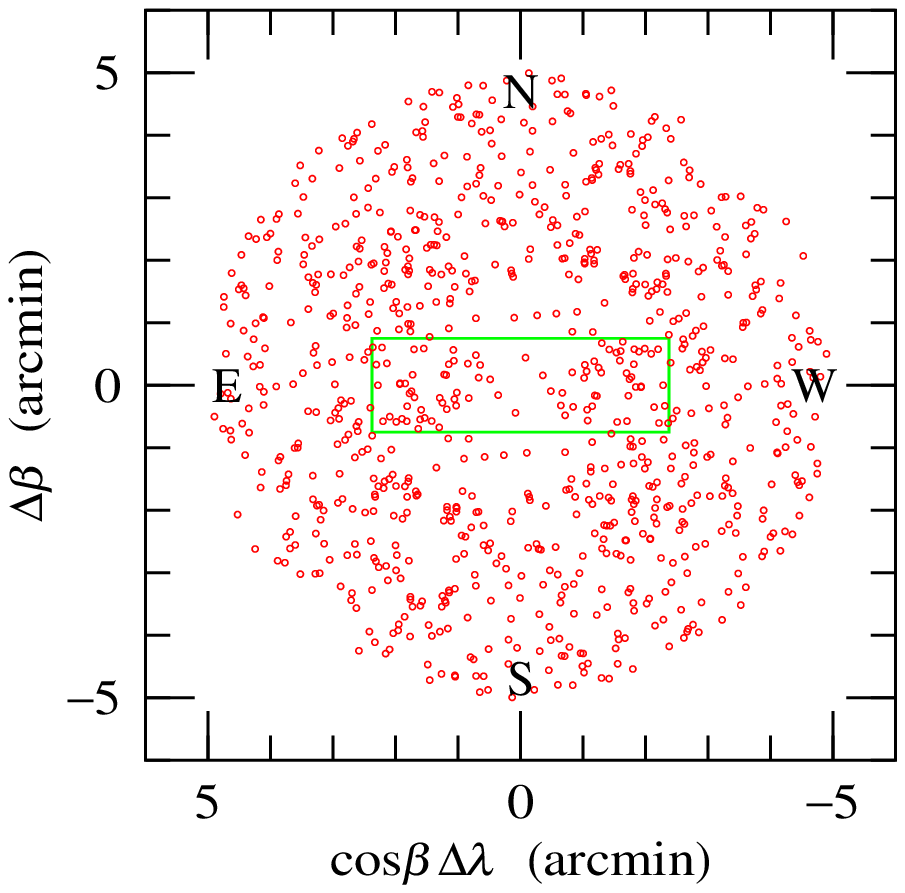}
    \end{center}
    \caption[1]{\small
The distribution of the 446 distinct pairs
closer than $10^\prime$ on the sky,
relative to a frame aligned with local ecliptic coordinates
$\lambda, \beta$
whose origin is the barycenter of each pair.
Ecliptic north is up, ecliptic east to the left.
The distribution has parity symmetry through the origin
(equivalently, it has $180^\circ$ rotation symmetry about the origin).
The rectangle at the centre illustrates
{\it IRAS\/}'s $1 \farcm 5$ in-scan $\times$ $4 \farcm 75$ cross-scan beam.
The effective angular resolution is higher,
particularly in the cross-scan (horizontal) direction,
thanks to the PSC strategy of combining
several scans at neighbouring longitudes.
Bear in mind that pairs are correlated with each other,
so that the distribution about the origin is not completely random.
    \label{pair}
    }
    \end{figure}
}

\newcommand{\xiksmallfig}{
    \begin{figure}
    \begin{center}
    \leavevmode
    \epsfxsize=3in	% normal
    \epsfbox{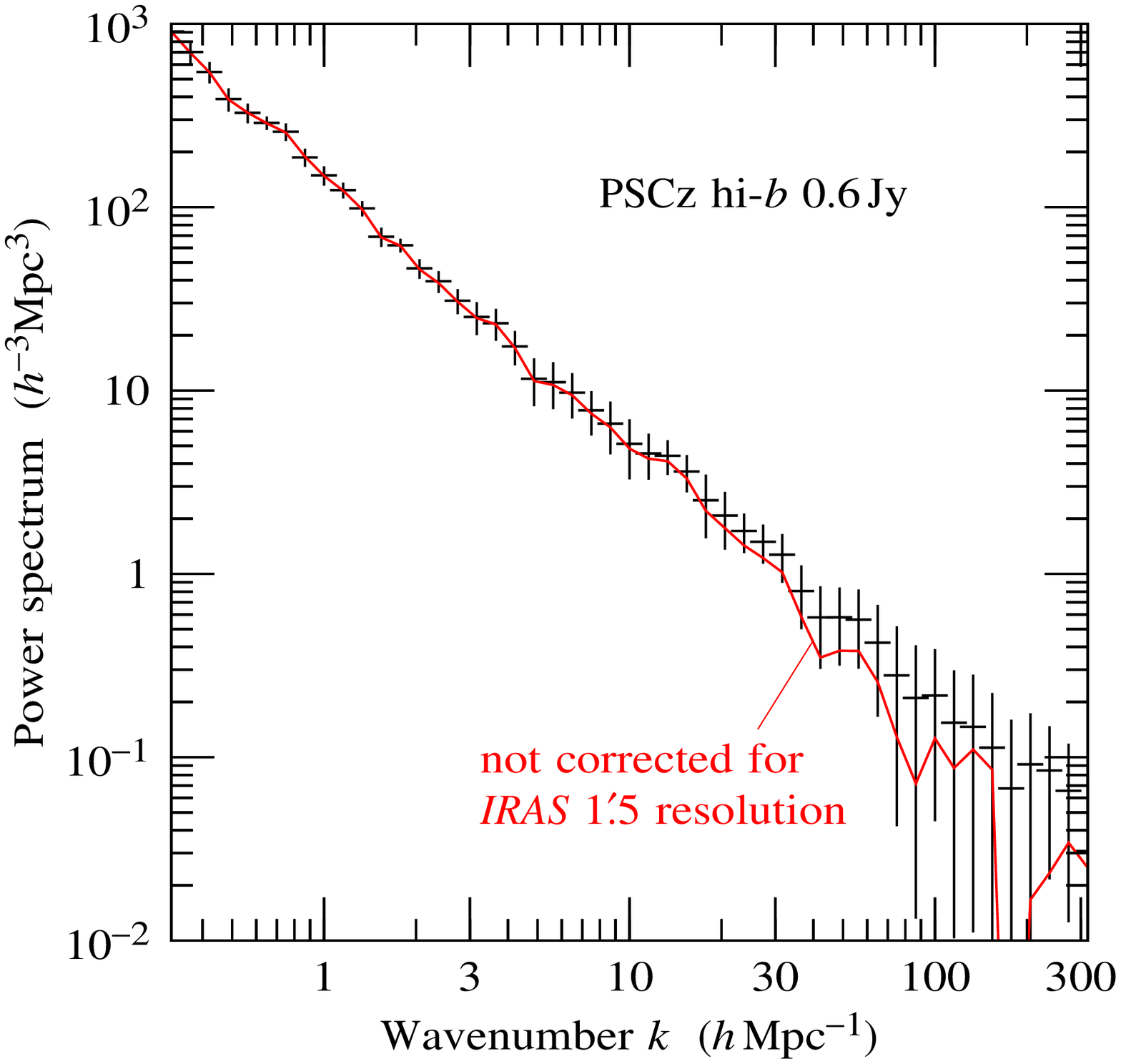}
    \end{center}
    \caption[1]{\small
{\it IRAS\/}'s $1 \farcm 5$ angular resolution
leads to a deficiency of pairs closer than $1 \farcm 5$ on the sky.
Points with error bars constitute the standard power spectrum from
Figure~\protect\ref{xik}, which takes this effect into account.
The solid line shows how the power spectrum is systematically depressed on the
smallest scales if the exclusion of close pairs is not taken into account.
    \label{xiksmall}
    }
    \end{figure}
}

\newcommand{\apmfig}{
    \begin{figure}
    \begin{center}
    \leavevmode
    \epsfxsize=3in	% normal
    \epsfbox{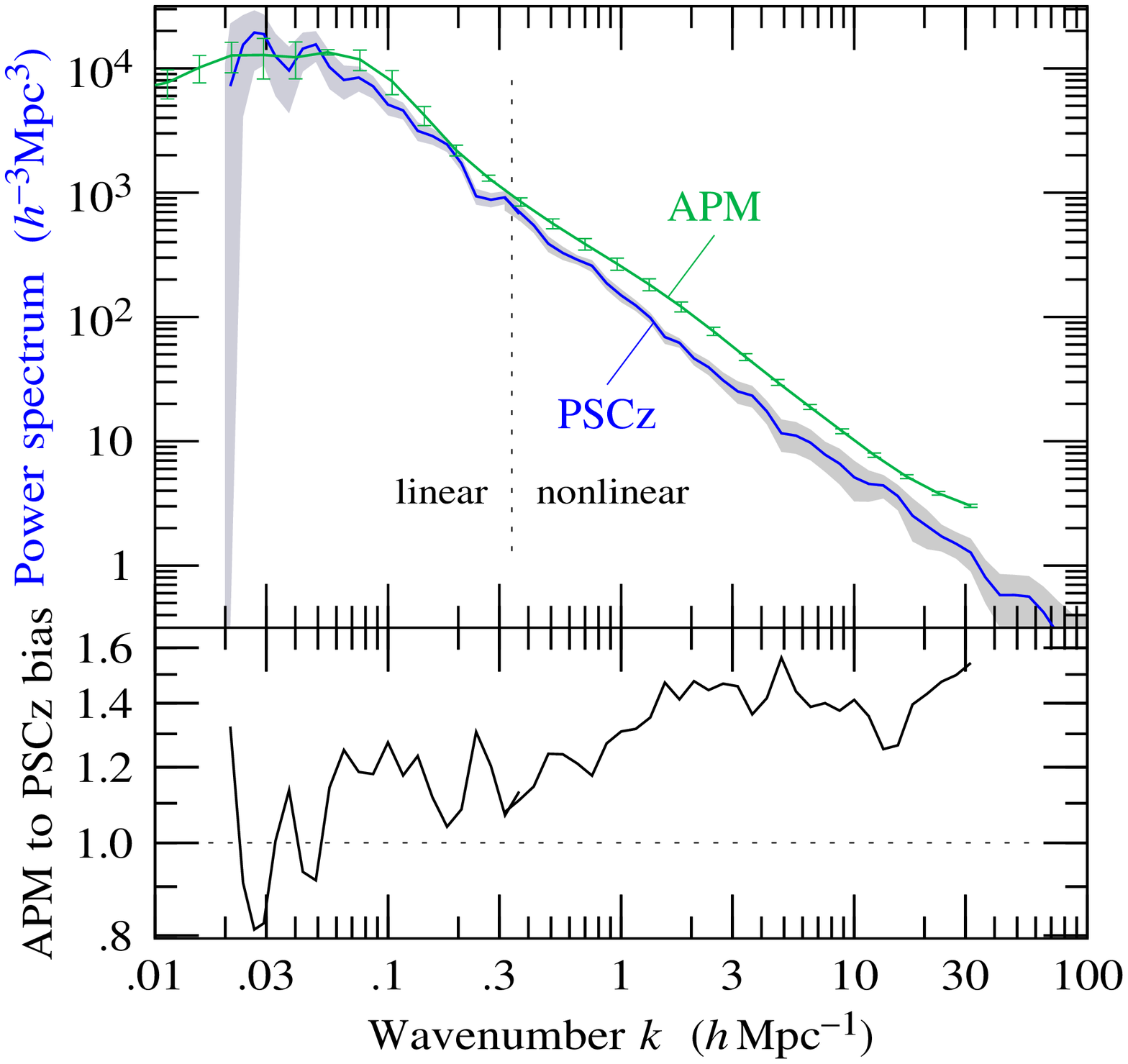}
    \end{center}
    \caption[1]{\small
Comparison of the real space power spectra
of the PSCz and APM (Gazta\~naga \& Baugh 1998) surveys.
The APM power has been renormalized upward by a factor $1.25$
(see text).
Shaded region is the $1 \sigma$ uncertainty in the
correlated power spectrum of PSCz.
The lower panel shows the ratio $b_{\rmn APM}/b_{\rmn PSCz}$
of the APM to PSCz bias,
the square root of the ratio of their power spectra.
The APM to PSCz bias is
$b_{\rmn APM}/b_{\rmn PSCz} \approx 1.15$
at linear scales,
$k \la 0.3 \, h \, \Mpc^{-1}$,
increasing to $b_{\rmn APM}/b_{\rmn PSCz} \approx 1.4$
at nonlinear scales,
$k \ga 1.5 \, h \, \Mpc^{-1}$.
Compare this Figure to Figure~2 of Peacock (1997).
    \label{apm}
    }
    \end{figure}
}

\newcommand{\xikcomfig}{
    \begin{figure}
    \begin{center}
    \leavevmode
    \epsfxsize=3in	% normal
    \epsfbox{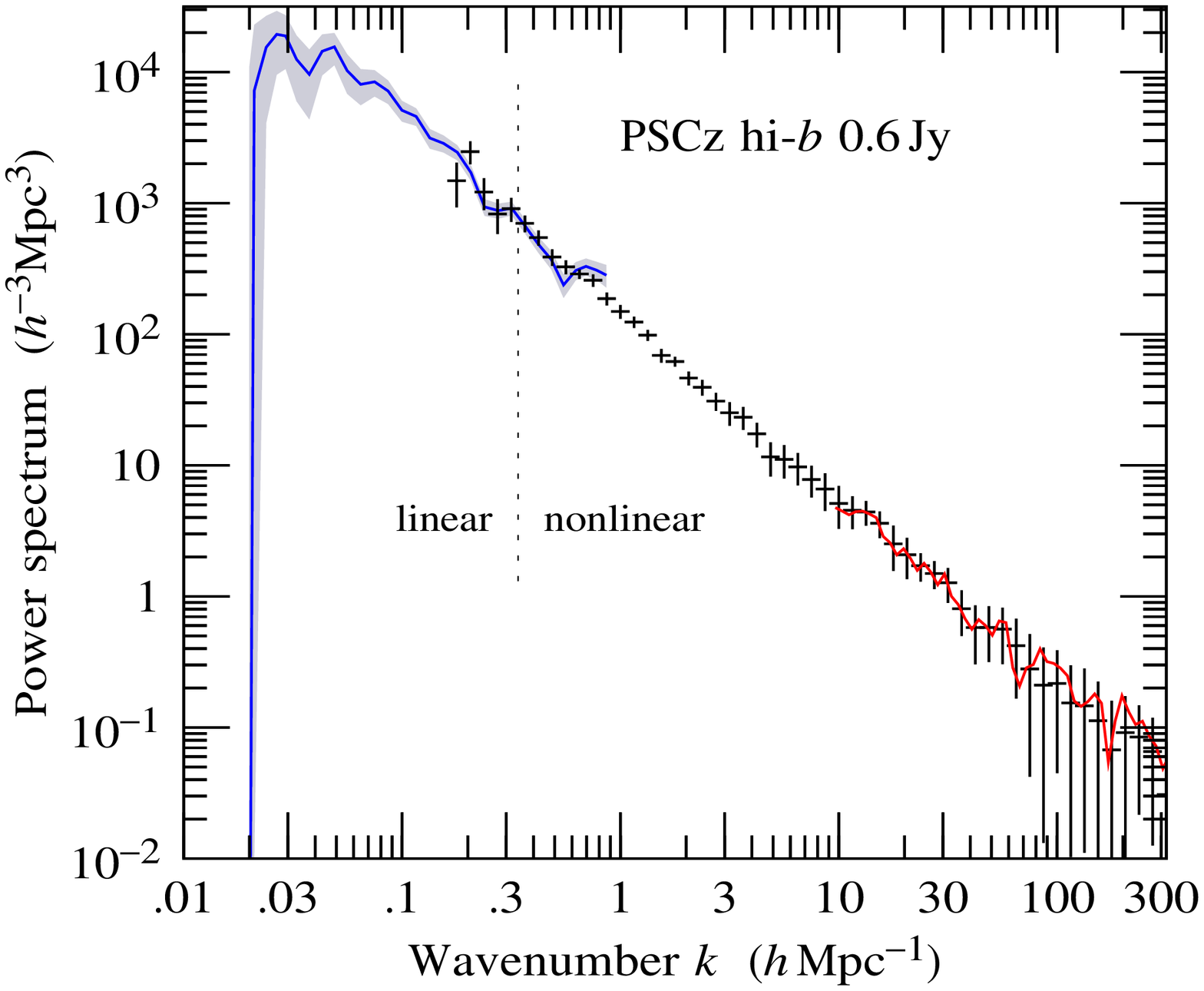}
    \end{center}
    \caption[1]{\small
Power spectrum of PSCz measured by different methods.
Solid line at large scales $k \la 1 \, h \, \Mpc^{-1}$
is the correlated power spectrum measured by the linear method,
and the shaded area its $1 \sigma$ limits.
Points with error bars
constitute the power spectrum measured by the nonlinear method
through band-power windows $\sim k^n \e^{-k^2}$ with $n = 72$.
Solid line at small scales $k \ga 10 \, h \, \Mpc^{-1}$
is the power through band-power windows with $n = 72 \times 4 = 288$.
The resolution of the $n = 288$ power spectrum is
$\Delta\log k \approx 1/24$ fwhm,
twice that of the $n = 72$ power spectrum.
For both $n = 72$ and $n = 288$,
the adopted maximum harmonic $\el_{\max}(k)$
is given by equation~(\ref{ladopt}),
with the additional constraint that $\el \le 72$ for $n = 72$.
Thus the $n = 288$ power spectrum uses more harmonics at
$k \ga 20 \, h \, \Mpc^{-1}$.
    \label{xikcom}
    }
    \end{figure}
}

\newcommand{\xikindfig}{
    \begin{figure}
    \begin{center}
    \leavevmode
    \epsfxsize=3in	% normal
    \epsfbox{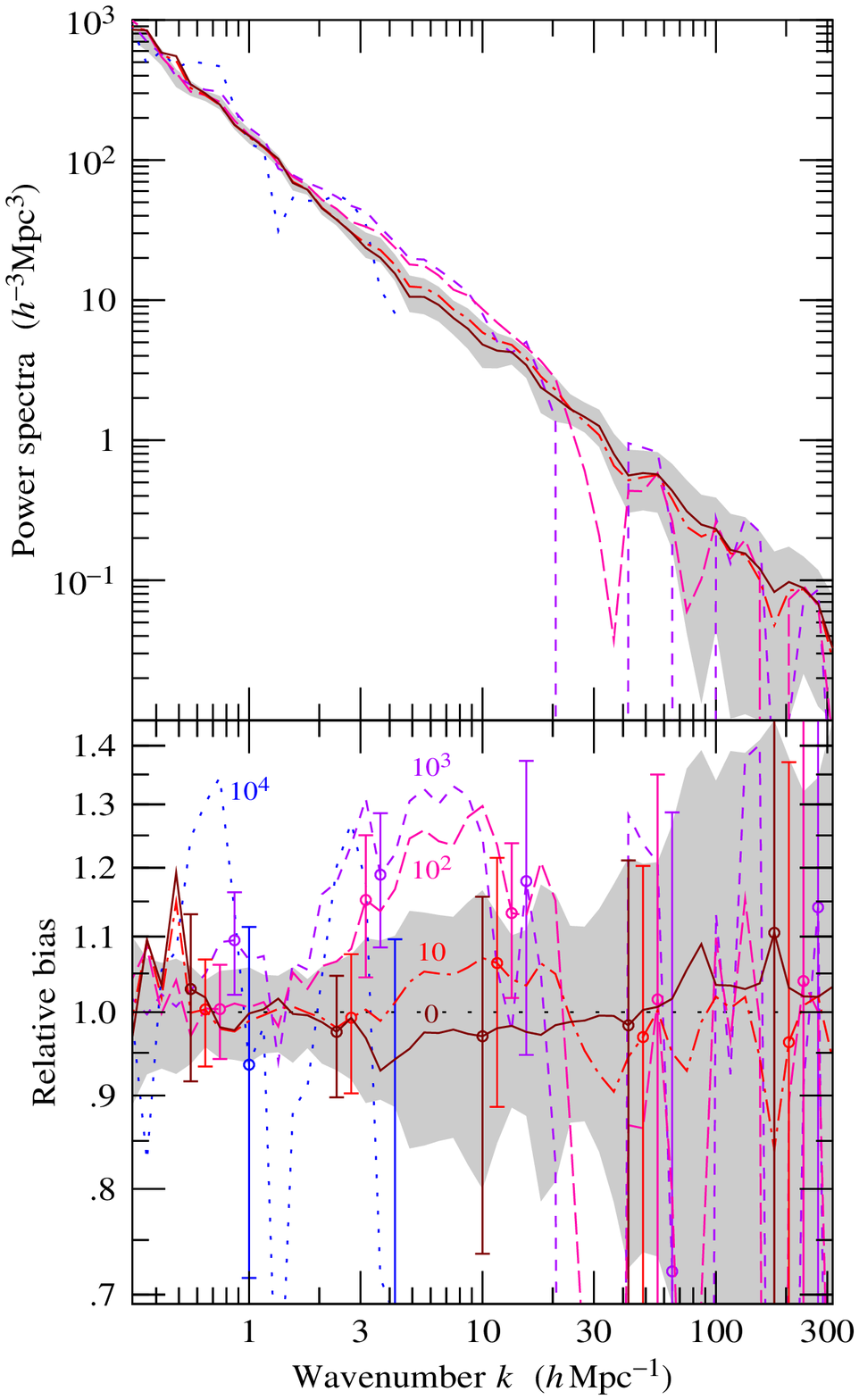}
    \end{center}
    \caption[1]{\small
(Upper panel)
Power spectra measured with fixed FKP constants.
(Lower panel)
Corresponding bias,
the square root of the ratio of the power spectrum
to the standard power spectrum of PSCz
plotted in Figure~\ref{xik}
and
tabulated in Table~\ref{xiktab}.
The shaded region represents the $1 \sigma$ uncertainty
in the standard power spectrum.
The different curves correspond to FKP constants
$J = 0$ (solid), $10$ (dot-dash), $10^2$ (long dash), $10^3$ (short dash),
and $10^4 \, h^{-3} \Mpc^3$ (dotted).
Larger FKP constants $J$ give relatively more weight to more distant
parts of the survey, i.e.\ to more luminous galaxies.
The curve with the largest FKP constant, $J = 10^4 \, h^{-3} \Mpc^3$ (dotted),
is plotted only up to $k \le 4 \, h \, \Mpc^{-1}$,
since its noisy criss-crossing confuses the plot at larger $k$.
A selection of $1 \sigma$ error bars is shown in the lower panel.
    \label{xikind}
    }
    \end{figure}
}

\newcommand{\xiknoprewhfig}{
    \begin{figure}
    \begin{center}
    \leavevmode
    \epsfxsize=3in	% normal
    \epsfbox{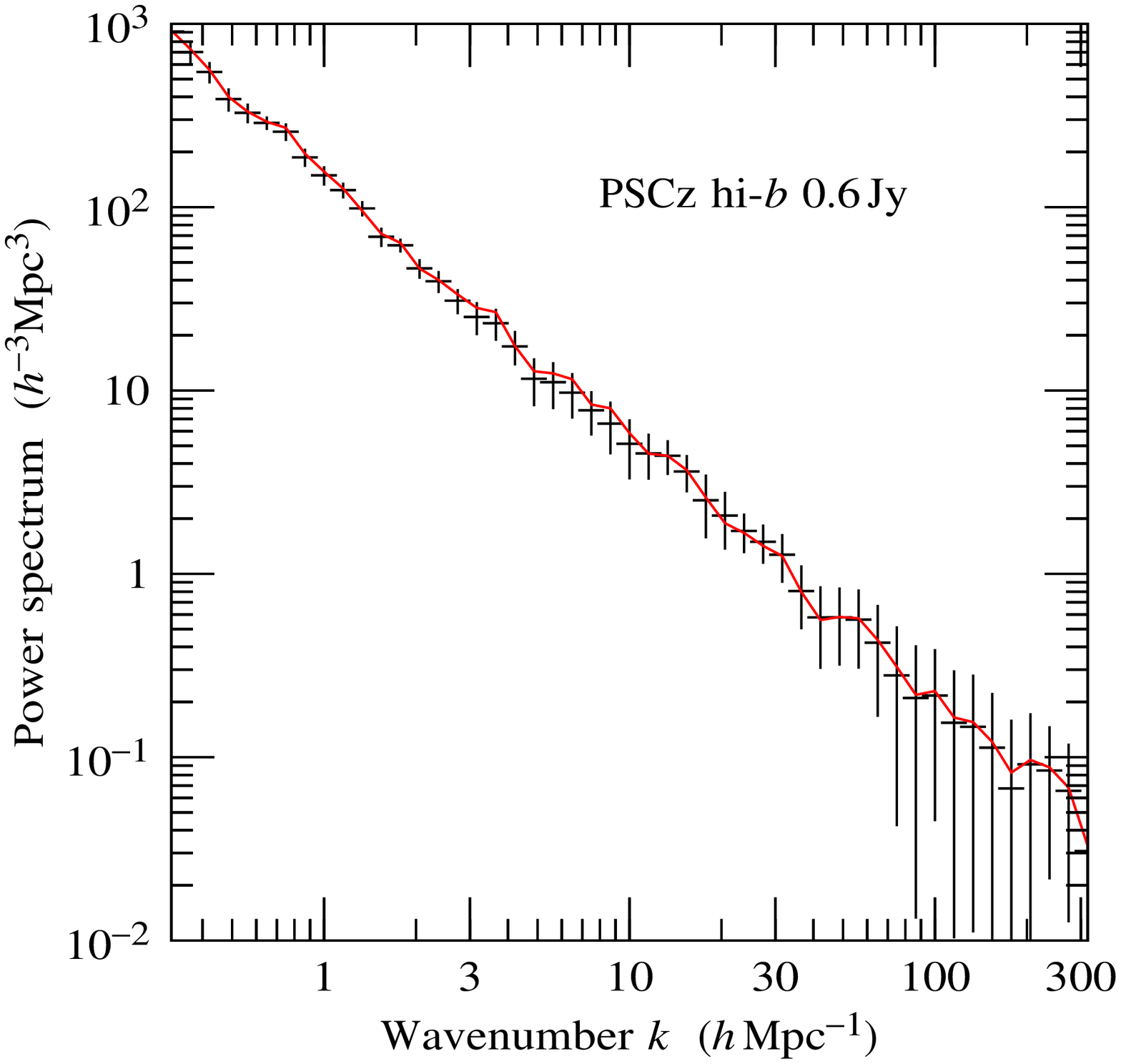}
    \end{center}
    \caption[1]{\small
Comparison of measured nonlinear power spectra
with and without prewhitening before FKP compression.
The points with error bars are the standard power spectrum
from Figure~\ref{xik},
in which the five FKP-weighted estimates of each band-power
are first prewhitened, then compressed, then unprewhitened.
The solid line is the power spectrum obtained from compressing
the five FKP-weighted estimates directly,
without the prewhiten-unprewhiten cycle.
The two power spectra are in good agreement.
    \label{xiknoprewh}
    }
    \end{figure}
}

\newcommand{\xirfig}{
    \begin{figure}
    \begin{center}
    \leavevmode
    \epsfxsize=3in	% normal
    \epsfbox{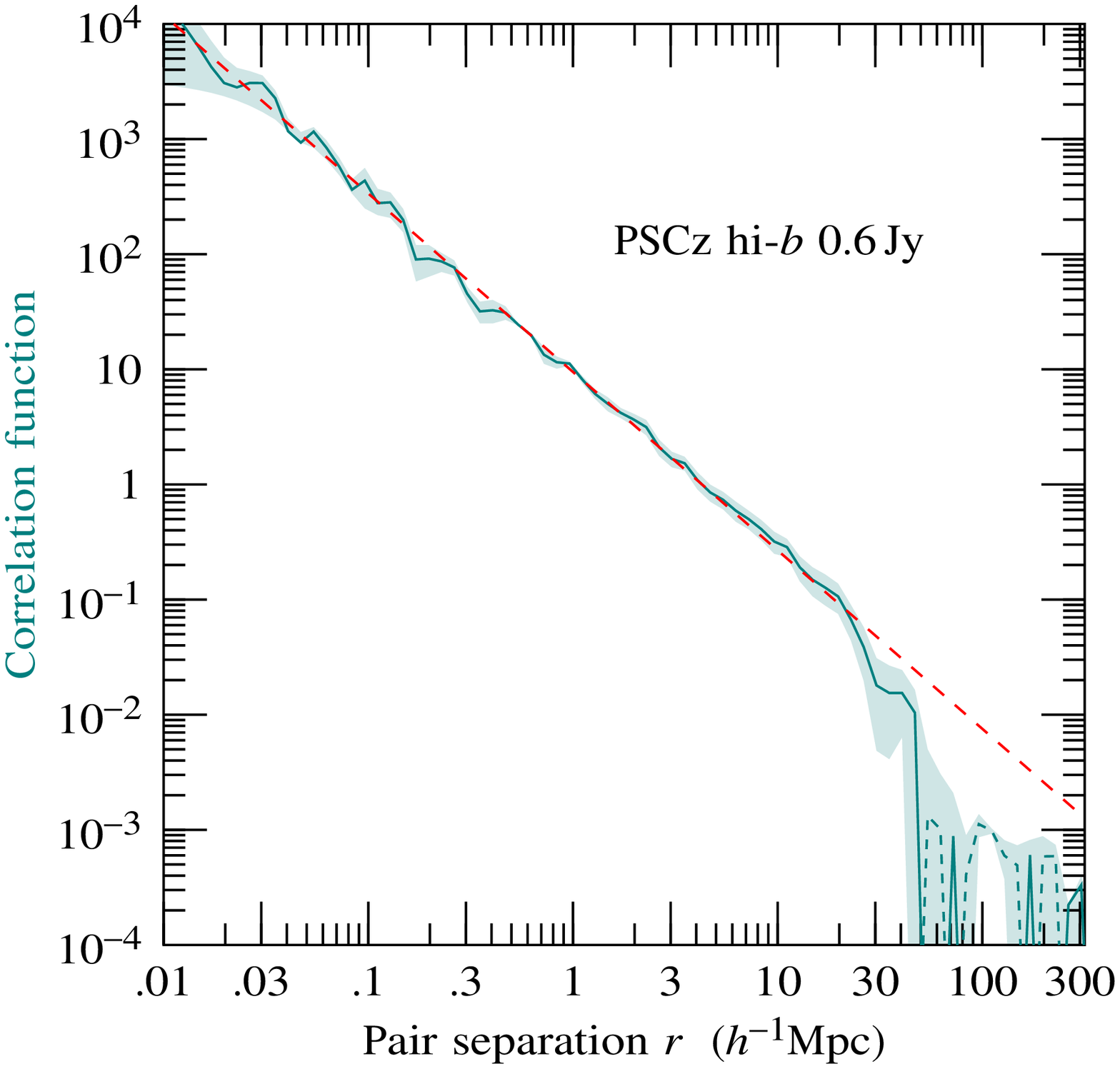}
    \end{center}
    \caption[1]{\small
Real space correlation function $\xi(r)$ of PSCz,
obtained by Fourier transforming the real space power spectrum.
The line is dashed where it is negative,
at pair separations $\approx 50$--$250 \, h^{-1} \Mpc$.
The shaded region is not the $1 \sigma$ uncertainty in $\xi(r)$,
but rather the envelope defined by the Fourier transforms of the
correlated power spectrum and its $\pm 1 \sigma$ extremes.
The dashed line is a power law $(r / 4.27 \, h^{-1} \Mpc)^{-1.55}$.
    \label{xir}
    }
    \end{figure}
}

\newcommand{\xiksfig}{
    \begin{figure}
    \begin{center}
    \leavevmode
    \epsfxsize=3in	% normal
    \epsfbox{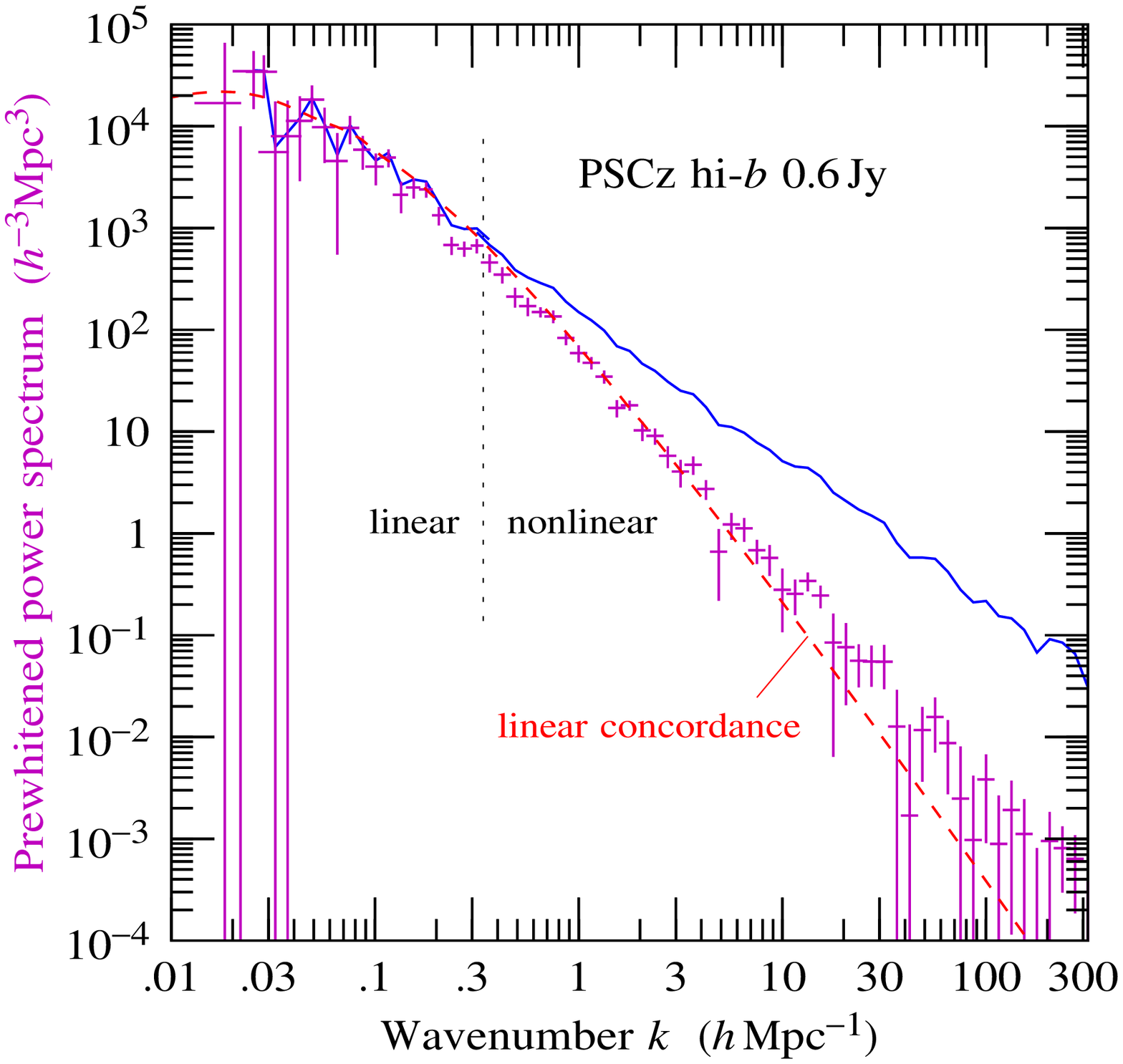}
    \end{center}
    \caption[1]{\small
Points with error bars constitute the prewhitened power spectrum of PSCz.
At linear scales, the points have been explicitly decorrelated.
At nonlinear scales, the points are somewhat correlated,
but less so than the (unprewhitened) power spectrum,
as illustrated in Figure~\ref{ccw}.
The solid line is the power spectrum which, when prewhitened,
equals the plotted prewhitened power spectrum.
The dashed line is the linear (not nonlinear) concordance
model power spectrum from Figure~\protect\ref{xik}.
    \label{xiks}
    }
    \end{figure}
}

\newcommand{\ccwfig}{
    \begin{figure}
    \begin{center}
    \leavevmode
    \epsfxsize=2in	% normal
    \epsfbox{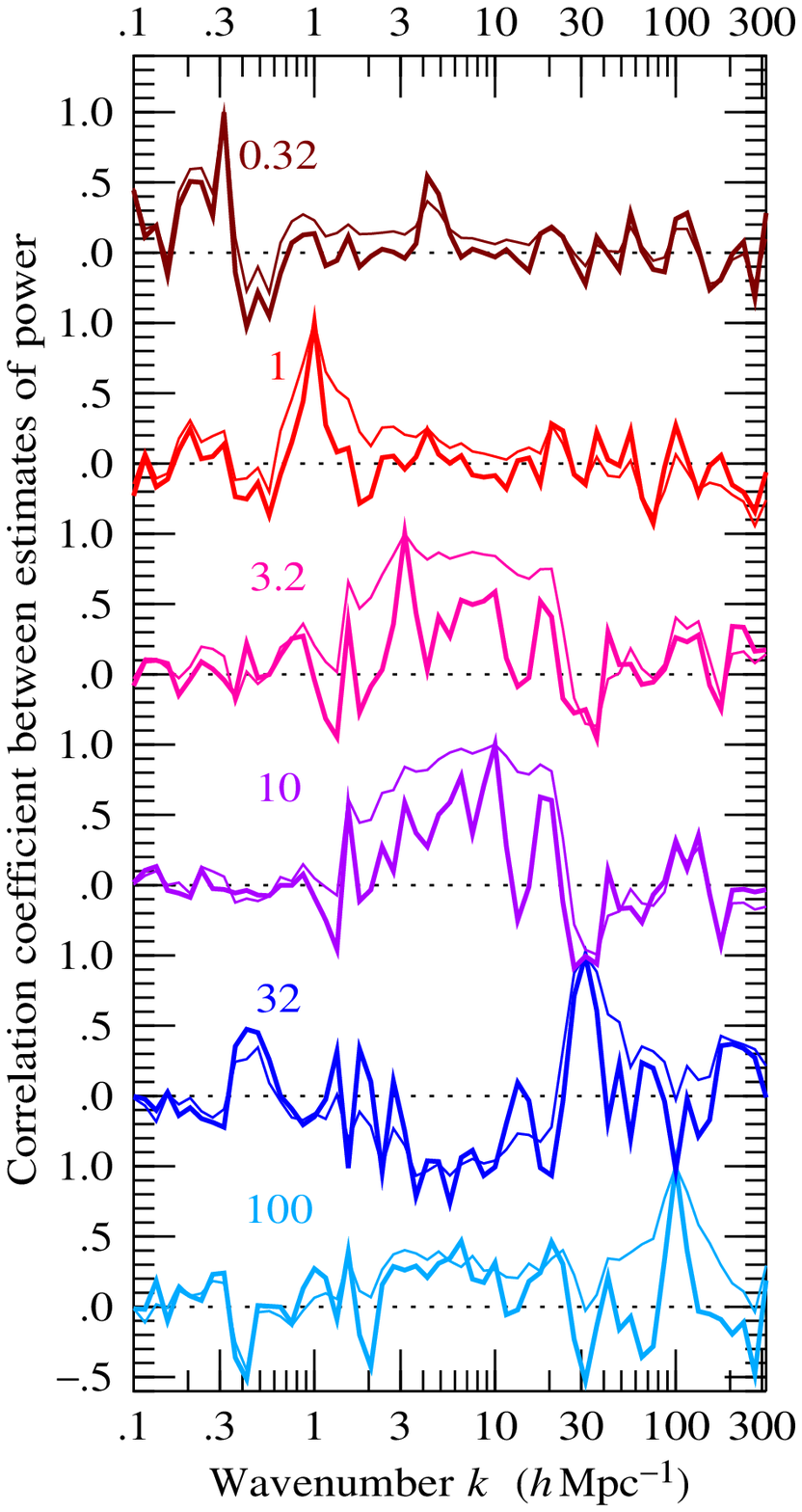}
    \end{center}
    \caption[1]{\small
Correlation coefficient $\mxC_{kk'}/(\mxC_{kk}^{1/2} \mxC_{k'k'}^{1/2})$
of estimates
of power (thin line)
and of prewhitened power (thick line)
in the PSCz survey.
The six plots are the correlation coefficients between the power at
$k' = 0.32$, $1$, $3.2$, $10$, $32$, and $100 \, h \, \Mpc^{-1}$, as labelled,
and the power at other wavenumbers $k$, as specified on the horizontal axis.
By construction, the correlation coefficient is unity at $k = k'$.
The Schwarz inequality requires that
the correlation coefficient lie between $1$ (perfect correlation)
and $-1$ (perfect anti-correlation).
The covariance of power is near diagonal both at large, linear scales,
where fluctuations are near-Gaussian,
and at small, highly nonlinear scales,
where shot noise dominates.
At intermediate scales, notably at $k' = 3.2$ and $10 \, h \, \Mpc^{-1}$,
the power is highly correlated,
whereas the prewhitened power is less so.
    \label{ccw}
    }
    \end{figure}
}

\newcommand{\Xsfig}{
    \begin{figure}
    \begin{center}
    \leavevmode
    \epsfxsize=2.4in	% normal
    \epsfbox{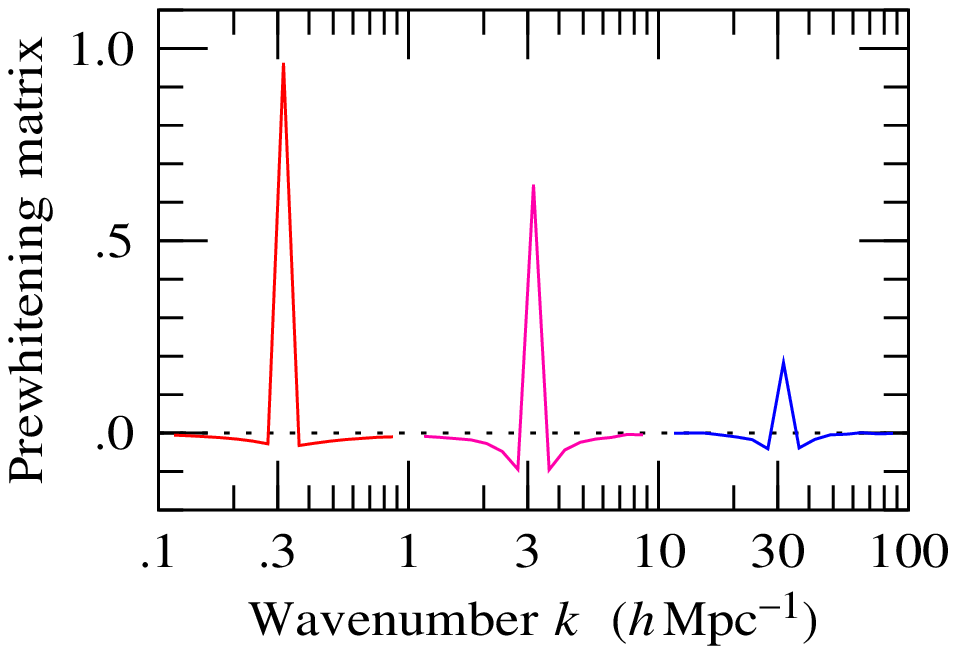}
    \end{center}
    \caption[1]{\small
Representative rows (or columns, since it is symmetric)
of the prewhitening matrix,
equation~(\ref{prewhmx}), in Fourier space,
appropriately discretized.
At linear scales the prewhitening matrix goes over to the unit matrix.
At nonlinear scales the prewhitening matrix looks like a high-pass filter.
There is a sharp peak along the diagonal, superimposed on a valley
that is deepest immediately adjacent to the diagonal.
The amplitude of the diagonal peak appears to decline
at larger wavenumbers because, in the discretized matrix,
the peak is being cancelled by a deeper valley.
    \label{Xs}
    }
    \end{figure}
}

\newcommand{\xiktable}{
    \begin{table*}
    \begin{minipage}{175mm}
    \caption{Correlated Power Spectrum}
    \label{xiktab}
    \begin{tabular}{@{}rrrrrrrrrrrrrrrrr}
\multicolumn{1}{c}{$k$} & \multicolumn{1}{c}{$k_-$} & \multicolumn{1}{c}{$k_+$} & \multicolumn{1}{c}{$P(k)$} & \multicolumn{1}{c}{$\Delta P(k)$} &  & 	\multicolumn{1}{c}{$k$} & \multicolumn{1}{c}{$k_-$} & \multicolumn{1}{c}{$k_+$} & \multicolumn{1}{c}{$P(k)$} & \multicolumn{1}{c}{$\Delta P(k)$} &  & 	\multicolumn{1}{c}{$k$} & \multicolumn{1}{c}{$k_-$} & \multicolumn{1}{c}{$k_+$} & \multicolumn{1}{c}{$P(k)$} & \multicolumn{1}{c}{$\Delta P(k)$} \\
\multicolumn{3}{c}{($h \, \Mpc^{-1}$)} & \multicolumn{2}{c}{($h^{-3} \Mpc^3$)} &  & 	\multicolumn{3}{c}{($h \, \Mpc^{-1}$)} & \multicolumn{2}{c}{($h^{-3} \Mpc^3$)} &  & 	\multicolumn{3}{c}{($h \, \Mpc^{-1}$)} & \multicolumn{2}{c}{($h^{-3} \Mpc^3$)} \\
$0.0210$ &  $0.0153$ &  $0.0269$ &  $7200.$ &  $15800.$ & &	$0.487$ &   $0.441$ &   $0.536$ &    $388.$ &     $56.$ & &	$13.3$ &   $12.1$ &   $14.7$ &    $4.41$ &     $0.94$ \\
$0.0239$ &  $0.0176$ &  $0.0298$ & $15500.$ &  $11400.$ & &	$0.562$ &   $0.510$ &   $0.619$ &    $327.$ &     $40.$ & &	$15.4$ &   $14.0$ &   $16.9$ &    $3.62$ &     $0.83$ \\
$0.0267$ &  $0.0203$ &  $0.0325$ & $19400.$ &   $9860.$ & &	$0.649$ &   $0.588$ &   $0.715$ &    $288.$ &     $24.$ & &	$17.8$ &   $16.1$ &   $19.6$ &    $2.52$ &     $0.96$ \\
$0.0293$ &  $0.0228$ &  $0.0355$ & $18900.$ &   $8300.$ & &	$0.750$ &   $0.679$ &   $0.825$ &    $258.$ &     $28.$ & &	$20.5$ &   $18.6$ &   $22.6$ &    $2.08$ &     $0.72$ \\
$0.0329$ &  $0.0257$ &  $0.0403$ & $12500.$ &   $6510.$ & &	$0.866$ &   $0.785$ &   $0.953$ &    $187.$ &     $21.$ & &	$23.7$ &   $21.5$ &   $26.1$ &    $1.72$ &     $0.42$ \\
$0.0376$ &  $0.0292$ &  $0.0467$ &  $9610.$ &   $5260.$ & &	$1.00$ &   $0.906$ &   $1.10$ &    $149.$ &     $18.$ & &	$27.4$ &   $24.8$ &   $30.1$ &    $1.50$ &     $0.36$ \\
$0.0431$ &  $0.0350$ &  $0.0518$ & $14400.$ &   $4970.$ & &	$1.15$ &   $1.05$ &   $1.27$ &    $124.$ &     $12.$ & &	$31.6$ &   $28.7$ &   $34.8$ &    $1.27$ &     $0.38$ \\
$0.0490$ &  $0.0406$ &  $0.0583$ & $15600.$ &   $4300.$ & &	$1.33$ &   $1.21$ &   $1.47$ &    $98.5$ &     $9.5$ & &	$36.5$ &   $33.1$ &   $40.2$ &    $0.805$ &    $0.307$ \\
$0.0560$ &  $0.0467$ &  $0.0666$ & $10200.$ &   $3420.$ & &	$1.54$ &   $1.40$ &   $1.69$ &    $69.0$ &     $8.3$ & &	$42.2$ &   $38.2$ &   $46.4$ &    $0.579$ &    $0.276$ \\
$0.0646$ &  $0.0536$ &  $0.0776$ &  $8060.$ &   $2480.$ & &	$1.78$ &   $1.61$ &   $1.96$ &    $62.0$ &     $5.4$ & &	$48.7$ &   $44.1$ &   $53.6$ &    $0.579$ &    $0.263$ \\
$0.0748$ &  $0.0626$ &  $0.0888$ &  $8430.$ &   $1920.$ & &	$2.05$ &   $1.86$ &   $2.26$ &    $46.4$ &     $5.6$ & &	$56.2$ &   $51.0$ &   $61.9$ &    $0.563$ &    $0.259$ \\
$0.0862$ &  $0.0728$ &  $0.101$ &   $7180.$ &   $1460.$ & &	$2.37$ &   $2.15$ &   $2.61$ &    $39.4$ &     $5.4$ & &	$64.9$ &   $58.8$ &   $71.5$ &    $0.421$ &    $0.255$ \\
$0.0998$ &  $0.0831$ &  $0.119$ &   $5110.$ &    $927.$ & &	$2.74$ &   $2.48$ &   $3.01$ &    $30.9$ &     $4.8$ & &	$75.0$ &   $67.9$ &   $82.5$ &    $0.280$ &    $0.238$ \\
$0.116$ &   $0.0973$ &  $0.137$ &   $4590.$ &    $703.$ & &	$3.16$ &   $2.87$ &   $3.48$ &    $25.2$ &     $5.2$ & &	$86.6$ &   $78.5$ &   $95.3$ &    $0.210$ &    $0.197$ \\
$0.134$ &   $0.113$ &   $0.158$ &   $3140.$ &    $538.$ & &	$3.65$ &   $3.31$ &   $4.02$ &    $23.3$ &     $4.6$ & &	$100.$ &   $90.6$ &   $110.$ &    $0.217$ &    $0.172$ \\
$0.155$ &   $0.131$ &   $0.182$ &   $2860.$ &    $425.$ & &	$4.22$ &   $3.82$ &   $4.64$ &    $17.4$ &     $3.7$ & &	$115.$ &   $105.$ &   $127.$ &    $0.154$ &    $0.144$ \\
$0.179$ &   $0.151$ &   $0.210$ &   $2440.$ &    $321.$ & &	$4.87$ &   $4.41$ &   $5.36$ &    $11.6$ &     $3.4$ & &	$133.$ &   $121.$ &   $147.$ &    $0.146$ &    $0.135$ \\
$0.207$ &   $0.175$ &   $0.240$ &   $1710.$ &    $233.$ & &	$5.62$ &   $5.10$ &   $6.19$ &    $11.1$ &     $3.2$ & &	$154.$ &   $140.$ &   $169.$ &    $0.113$ &    $0.111$ \\
$0.239$ &   $0.198$ &   $0.286$ &    $936.$ &    $136.$ & &	$6.49$ &   $5.88$ &   $7.15$ &    $9.74$ &    $2.70$ & &	$178.$ &   $161.$ &   $196.$ &    $0.068$ &    $0.093$ \\
$0.276$ &   $0.231$ &   $0.329$ &    $877.$ &    $115.$ & &	$7.50$ &   $6.79$ &   $8.25$ &    $7.80$ &    $2.12$ & &	$205.$ &   $186.$ &   $226.$ &    $0.091$ &    $0.082$ \\
$0.317$ &   $0.268$ &   $0.375$ &    $917.$ &    $109.$ & &	$8.66$ &   $7.85$ &   $9.53$ &    $6.60$ &    $2.11$ & &	$237.$ &   $215.$ &   $261.$ &    $0.085$ &    $0.063$ \\
$0.365$ &   $0.331$ &   $0.402$ &    $702.$ &    $102.$ & &	$10.0$ &   $9.06$ &   $11.0$ &    $5.13$ &    $1.85$ & &	$274.$ &   $248.$ &   $301.$ &    $0.066$ &    $0.053$ \\
$0.422$ &   $0.382$ &   $0.464$ &    $546.$ &     $72.$ & &	$11.5$ &   $10.5$ &   $12.7$ &    $4.54$ &    $1.28$ & &	$316.$ &   $287.$ &   $348.$ &    $0.031$ &    $0.047$ \\
    \end{tabular}
\medskip
\\
$k$ is the median wavenumber of the band-power window,
and $k_-$ and $k_+$ the wavenumbers
where the band-power window falls to half its maximum.
At linear scales, $k < 0.33 \, h \, \Mpc^{-1}$,
the median and half-maximum points are those of the scaled
and discretized band-power windows as defined in Hamilton \& Tegmark (2000).
At nonlinear scales, $k > 0.33 \, h \, \Mpc^{-1}$,
the band-powers have the power law times Gaussian form
detailed in Section~\ref{bandpowers}.
$P(k)$ is the estimated power in the band-power,
and $\Delta P(k)$ its $1 \sigma$ uncertainty.
    \end{minipage}
    \end{table*}
}
\newcommand{\xikdtable}{
    \begin{table*}
    \begin{minipage}{175mm}
    \caption{Decorrelated Linear Power Spectrum}
    \label{xikdtab}
    \begin{tabular}{@{}rrrrrrrrrrrrrrrrr}
\multicolumn{1}{c}{$k$} & \multicolumn{1}{c}{$k_-$} & \multicolumn{1}{c}{$k_+$} & \multicolumn{1}{c}{$P(k)$} & \multicolumn{1}{c}{$\Delta P(k)$} & & 	\multicolumn{1}{c}{$k$} & \multicolumn{1}{c}{$k_-$} & \multicolumn{1}{c}{$k_+$} & \multicolumn{1}{c}{$P(k)$} & \multicolumn{1}{c}{$\Delta P(k)$} \\
\multicolumn{3}{c}{($h \, \Mpc^{-1}$)} & \multicolumn{2}{c}{($h^{-3} \Mpc^3$)} & & 	\multicolumn{3}{c}{($h \, \Mpc^{-1}$)} & \multicolumn{2}{c}{($h^{-3} \Mpc^3$)} \\
$0.0137$ & $0.0097$ & $0.0171$ & $133000.$ & $920000.$ & & 		$0.0747$ & $0.0670$ & $0.0833$ & $10600.$ & $3400.$ \\
$0.0175$ & $0.0130$ & $0.0219$ & $20200.$ & $54200.$ & & 		$0.0863$ & $0.0783$ & $0.0947$ & $6490.$ & $2520.$ \\
$0.0214$ & $0.0165$ & $0.0264$ & $-11100.$ & $21300.$ & & 		$0.0998$ & $0.0902$ & $0.110$ & $4630.$ & $1750.$ \\
$0.0249$ & $0.0200$ & $0.0297$ & $36600.$ & $21400.$ & & 		$0.115$ & $0.106$ & $0.126$ & $5930.$ & $1270.$ \\
$0.0280$ & $0.0232$ & $0.0330$ & $36600.$ & $16600.$ & & 		$0.133$ & $0.123$ & $0.144$ & $2400.$ & $970.$ \\
$0.0319$ & $0.0268$ & $0.0376$ & $5580.$ & $13200.$ & & 		$0.154$ & $0.143$ & $0.165$ & $2990.$ & $750.$ \\
$0.0366$ & $0.0308$ & $0.0434$ & $8250.$ & $10800.$ & & 		$0.178$ & $0.166$ & $0.190$ & $2980.$ & $570.$ \\
$0.0422$ & $0.0365$ & $0.0492$ & $11700.$ & $9100.$ & & 		$0.205$ & $0.192$ & $0.219$ & $1650.$ & $410.$ \\
$0.0485$ & $0.0423$ & $0.0561$ & $19400.$ & $7600.$ & & 		$0.237$ & $0.221$ & $0.254$ & $963.$ & $266.$ \\
$0.0560$ & $0.0491$ & $0.0635$ & $10400.$ & $6000.$ & & 		$0.274$ & $0.257$ & $0.292$ & $929.$ & $211.$ \\
$0.0646$ & $0.0569$ & $0.0731$ & $4680.$ & $4550.$ & & 		$0.316$ & $0.298$ & $0.335$ & $927.$ & $189.$ \\
    \end{tabular}
\medskip
\\
See footnote to Table~\ref{xiktab}.
When fitting to theoretical models at linear scales,
this decorrelated power spectrum is to be preferred over the correlated
power spectrum of Table~\ref{xiktab},
since the decorrelated estimates can be treated as uncorrelated.
    \end{minipage}
    \end{table*}
}
\newcommand{\xirtable}{
    \begin{table*}
    \begin{minipage}{175mm}
    \caption{Correlation Function}
    \label{xirtab}
    \begin{tabular}{@{}crrrrcrrrrcrrr}
\multicolumn{1}{c}{$r$} & \multicolumn{1}{r}{$\xi\ \ \ $} & \multicolumn{1}{r}{$\xi_-\ \ $} & \multicolumn{1}{r}{$\xi_+\ \ $} & & 	\multicolumn{1}{c}{$r$} & \multicolumn{1}{r}{$\xi\ \ \ $} & \multicolumn{1}{r}{$\xi_-\ \ $} & \multicolumn{1}{r}{$\xi_+\ \ $} & & 	\multicolumn{1}{c}{$r$} & \multicolumn{1}{r}{$\xi\ \ \ $} & \multicolumn{1}{r}{$\xi_-\ \ $} & \multicolumn{1}{r}{$\xi_+\ \ $} \\
\multicolumn{1}{c}{($h^{-1} \Mpc$)} & & & & & 	\multicolumn{1}{c}{($h^{-1} \Mpc$)} & & & & & 	\multicolumn{1}{c}{($h^{-1} \Mpc$)} & & & \\
$0.00961$ &  $16200.$ &     $2980.$ &    $26000.$ & &$0.351$ &      $31.9$ &      $25.1$ &      $38.7$ & & $12.8$ &     $0.191$ &     $0.144$ &     $0.238$ \\
$0.0111$ &   $12500.$ &     $2890.$ &    $19800.$ & &$0.405$ &      $32.6$ &      $25.1$ &      $40.0$ & & $14.8$ &     $0.149$ &     $0.107$ &     $0.191$ \\
$0.0128$ &    $9130.$ &     $2790.$ &    $14500.$ & &$0.468$ &      $31.1$ &      $26.9$ &      $35.3$ & & $17.1$ &     $0.127$ &     $0.0886$ &    $0.165$ \\
$0.0148$ &    $6320.$ &     $2670.$ &    $10300.$ & &$0.541$ &      $24.4$ &      $24.7$ &      $24.0$ & & $19.7$ &     $0.106$ &     $0.0751$ &    $0.138$ \\
$0.0171$ &    $4240.$ &     $2520.$ &     $7130.$ & &$0.624$ &      $19.8$ &      $19.1$ &      $20.6$ & & $22.8$ &     $0.0670$ &    $0.0441$ &    $0.0899$ \\
$0.0197$ &    $3070.$ &     $2360.$ &     $5100.$ & &$0.721$ &      $13.4$ &      $11.2$ &      $15.6$ & & $26.3$ &     $0.0387$ &    $0.0196$ &    $0.0576$ \\
$0.0228$ &    $2810.$ &     $2170.$ &     $4150.$ & &$0.833$ &      $11.5$ &      $10.1$ &      $12.9$ & & $30.4$ &     $0.0180$ &    $4.87_{-3}$ &   $0.0310$ \\
$0.0263$ &    $3070.$ &     $1950.$ &     $3890.$ & &$0.961$ &      $11.2$ &      $10.7$ &      $11.7$ & & $35.1$ &     $0.0154$ &    $4.11_{-3}$ &   $0.0267$ \\
$0.0304$ &    $3070.$ &     $1720.$ &     $3570.$ & & $1.11$ &      $8.18$ &      $7.60$ &      $8.76$ & & $40.5$ &     $0.0154$ &    $6.36_{-3}$ &   $0.0245$ \\
$0.0351$ &    $2260.$ &     $1480.$ &     $2650.$ & & $1.28$ &      $6.12$ &      $5.54$ &      $6.70$ & & $46.8$ &     $0.0104$ &    $4.33_{-3}$ &   $0.0164$ \\
$0.0405$ &    $1170.$ &     $1250.$ &     $1510.$ & & $1.48$ &      $5.01$ &      $4.31$ &      $5.71$ & & $54.1$ &    $-1.32_{-3}$ &  $-5.00_{-3}$ &   $2.34_{-3}$ \\
$0.0468$ &     $931.$ &     $1030.$ &     $1150.$ & & $1.71$ &      $4.19$ &      $3.77$ &      $4.62$ & & $62.4$ &    $-1.01_{-3}$ &  $-3.06_{-3}$ &   $1.02_{-3}$ \\
$0.0541$ &    $1160.$ &      $841.$ &     $1270.$ & & $1.97$ &      $3.67$ &      $3.23$ &      $4.12$ & & $72.1$ &     $8.83_{-4}$ & $-3.26_{-4}$ &  $2.09_{-3}$ \\
$0.0624$ &     $841.$ &      $660.$ &      $974.$ & & $2.28$ &      $3.15$ &      $2.67$ &      $3.63$ & & $83.3$ &    $-4.06_{-4}$ & $-8.97_{-4}$ &  $8.9_{-5}$ \\
$0.0721$ &     $577.$ &      $486.$ &      $687.$ & & $2.63$ &      $2.11$ &      $1.77$ &      $2.45$ & & $96.1$ &    $-1.12_{-3}$ &  $-1.37_{-3}$ &  $-8.63_{-4}$ \\
$0.0833$ &     $364.$ &      $336.$ &      $447.$ & & $3.04$ &      $1.67$ &      $1.41$ &      $1.92$ & & $111.$ &    $-9.82_{-4}$ & $-9.27_{-4}$ & $-1.03_{-3}$ \\
$0.0961$ &     $435.$ &      $250.$ &      $562.$ & & $3.51$ &      $1.53$ &      $1.32$ &      $1.74$ & & $128.$ &    $-5.97_{-4}$ & $-3.73_{-4}$ & $-8.13_{-4}$ \\
$0.111$ &      $278.$ &      $219.$ &      $370.$ & & $4.05$ &      $1.09$ &     $0.905$ &      $1.28$ & & $148.$ &    $-4.91_{-4}$ & $-2.47_{-4}$ & $-7.34_{-4}$ \\
$0.128$ &      $282.$ &      $207.$ &      $343.$ & & $4.68$ &     $0.853$ &     $0.709$ &     $0.996$ & & $171.$ &     $6.05_{-4}$ &  $8.17_{-4}$ &  $3.87_{-4}$ \\
$0.148$ &      $198.$ &      $155.$ &      $247.$ & & $5.41$ &     $0.736$ &     $0.609$ &     $0.864$ & & $197.$ &    $-5.87_{-4}$ & $-2.99_{-4}$ & $-8.81_{-4}$ \\
$0.171$ &      $90.0$ &      $58.0$ &      $120.$ & & $6.24$ &     $0.592$ &     $0.474$ &     $0.709$ & & $228.$ &    $-5.91_{-4}$ & $-4.41_{-4}$ & $-7.41_{-4}$ \\
$0.197$ &      $91.5$ &      $63.7$ &      $120.$ & & $7.21$ &     $0.501$ &     $0.408$ &     $0.593$ & & $263.$ &     $2.23_{-4}$ &  $2.22_{-4}$ &  $2.27_{-4}$ \\
$0.228$ &      $86.5$ &      $70.1$ &      $103.$ & & $8.33$ &     $0.409$ &     $0.327$ &     $0.490$ & & $304.$ &     $3.31_{-4}$ &  $2.65_{-4}$ &  $3.95_{-4}$ \\
$0.263$ &      $76.6$ &      $65.2$ &      $88.2$ & & $9.61$ &     $0.319$ &     $0.250$ &     $0.389$ & & $351.$ &    $-1.53_{-4}$ & $-1.84_{-4}$ & $-1.25_{-4}$ \\
$0.304$ &      $45.3$ &      $38.3$ &      $52.3$ & & $11.1$ &     $0.286$ &     $0.235$ &     $0.337$ & &  &  &  &  \\
    \end{tabular}
\medskip
\\
$r$ is the pair separation,
$\xi$ the correlation function.
$\xi_-$ and $\xi_+$
are {\em not\/} $1 \sigma$ limits, but rather they are
the Fourier transforms of the $\pm 1 \sigma$ extremes $P(k) \pm \Delta P(k)$
of the correlated power from Table~\ref{xiktab}.
Notice that $\xi_-$ is not always less than $\xi_+$,
and that $\xi_-$ and $\xi_+$
do not necessarily encompass the central value $\xi$.
    \end{minipage}
    \end{table*}
}
\newcommand{\xikstable}{
    \begin{table*}
    \begin{minipage}{175mm}
    \caption{Prewhitened Power Spectrum}
    \label{xikstab}
    \begin{tabular}{@{}rrrrrrrrrrrrrrrrr}
\multicolumn{1}{c}{$k$} & \multicolumn{1}{c}{$k_-$} & \multicolumn{1}{c}{$k_+$} & \multicolumn{1}{c}{$P(k)$} & \multicolumn{1}{c}{$\Delta P(k)$} & & 	\multicolumn{1}{c}{$k$} & \multicolumn{1}{c}{$k_-$} & \multicolumn{1}{c}{$k_+$} & \multicolumn{1}{c}{$P(k)$} & \multicolumn{1}{c}{$\Delta P(k)$} & & 	\multicolumn{1}{c}{$k$} & \multicolumn{1}{c}{$k_-$} & \multicolumn{1}{c}{$k_+$} & \multicolumn{1}{c}{$P(k)$} & \multicolumn{1}{c}{$\Delta P(k)$} \\
\multicolumn{3}{c}{($h \, \Mpc^{-1}$)} & \multicolumn{2}{c}{($h^{-3} \Mpc^3$)} & & 	\multicolumn{3}{c}{($h \, \Mpc^{-1}$)} & \multicolumn{2}{c}{($h^{-3} \Mpc^3$)} & & 	\multicolumn{3}{c}{($h \, \Mpc^{-1}$)} & \multicolumn{2}{c}{($h^{-3} \Mpc^3$)} \\
$0.0183$ &  $0.0130$ &  $0.0220$ & $16900.$ &   $49000.$ & &$0.487$ &   $0.441$ &   $0.536$ &    $212.$ &      $47.$ & &$13.3$ &   $12.1$ &   $14.7$ &    $0.342$ &      $0.071$ \\
$0.0219$ &  $0.0165$ &  $0.0264$ & $-9780.$ &   $19800.$ & &$0.562$ &   $0.510$ &   $0.619$ &    $171.$ &      $35.$ & &$15.4$ &   $14.0$ &   $16.9$ &    $0.245$ &      $0.061$ \\
$0.0254$ &  $0.0200$ &  $0.0298$ & $34700.$ &   $20000.$ & &$0.649$ &   $0.588$ &   $0.715$ &    $149.$ &      $17.$ & &$17.8$ &   $16.1$ &   $19.6$ &    $0.0847$ &     $0.0784$ \\
$0.0284$ &  $0.0232$ &  $0.0330$ & $34300.$ &   $15400.$ & &$0.750$ &   $0.679$ &   $0.825$ &    $135.$ &      $19.$ & &$20.5$ &   $18.6$ &   $22.6$ &    $0.0762$ &     $0.0556$ \\
$0.0324$ &  $0.0268$ &  $0.0377$ &  $5570.$ &   $12000.$ & &$0.866$ &   $0.785$ &   $0.953$ &    $83.3$ &     $12.6$ & &$23.7$ &   $21.5$ &   $26.1$ &    $0.0562$ &     $0.0254$ \\
$0.0372$ &  $0.0308$ &  $0.0435$ &  $7980.$ &    $9870.$ & &$1.00$ &   $0.906$ &   $1.10$ &    $59.0$ &     $11.3$ & &$27.4$ &   $24.8$ &   $30.1$ &    $0.0552$ &     $0.0239$ \\
$0.0427$ &  $0.0365$ &  $0.0492$ & $11300.$ &    $8400.$ & &$1.15$ &   $1.05$ &   $1.27$ &    $47.4$ &      $6.6$ & &$31.6$ &   $28.7$ &   $34.8$ &    $0.0549$ &     $0.0254$ \\
$0.0490$ &  $0.0423$ &  $0.0563$ & $18300.$ &    $6940.$ & &$1.33$ &   $1.21$ &   $1.47$ &    $34.7$ &      $5.1$ & &$36.5$ &   $33.1$ &   $40.2$ &    $0.0127$ &     $0.0164$ \\
$0.0565$ &  $0.0490$ &  $0.0637$ &  $9780.$ &    $5420.$ & &$1.54$ &   $1.40$ &   $1.69$ &    $17.1$ &      $3.3$ & &$42.2$ &   $38.2$ &   $46.4$ &    $1.69_{-3}$ & $0.0116$ \\
$0.0653$ &  $0.0567$ &  $0.0734$ &  $4550.$ &    $4000.$ & &$1.78$ &   $1.61$ &   $1.96$ &    $18.1$ &      $2.1$ & &$48.7$ &   $44.1$ &   $53.6$ &    $0.0117$ &     $8.1_{-3}$ \\
$0.0754$ &  $0.0668$ &  $0.0836$ &  $9640.$ &    $2990.$ & &$2.05$ &   $1.86$ &   $2.26$ &    $10.3$ &      $2.2$ & &$56.2$ &   $51.0$ &   $61.9$ &    $0.0157$ &     $8.7_{-3}$ \\
$0.0871$ &  $0.0781$ &  $0.0950$ &  $5880.$ &    $2130.$ & &$2.37$ &   $2.15$ &   $2.61$ &    $9.08$ &     $1.65$ & &$64.9$ &   $58.8$ &   $71.5$ &    $8.72_{-3}$ & $5.98_{-3}$ \\
$0.101$ &   $0.0897$ &  $0.110$ &   $4020.$ &    $1380.$ & &$2.74$ &   $2.48$ &   $3.01$ &    $5.79$ &     $1.40$ & &$75.0$ &   $67.9$ &   $82.5$ &    $2.48_{-3}$ & $5.59_{-3}$ \\
$0.116$ &   $0.105$ &   $0.126$ &   $4930.$ &     $984.$ & &$3.16$ &   $2.87$ &   $3.48$ &    $4.06$ &     $1.23$ & &$86.6$ &   $78.5$ &   $95.3$ &    $9.73_{-4}$ & $3.20_{-3}$ \\
$0.134$ &   $0.122$ &   $0.145$ &   $2120.$ &     $725.$ & &$3.65$ &   $3.31$ &   $4.02$ &    $4.73$ &      $0.97$ & &$100.$ &   $90.6$ &   $110.$ &    $3.83_{-3}$ & $2.92_{-3}$ \\
$0.155$ &   $0.142$ &   $0.167$ &   $2500.$ &     $554.$ & &$4.22$ &   $3.82$ &   $4.64$ &    $2.74$ &      $0.60$ & &$115.$ &   $105.$ &   $127.$ &    $8.92_{-4}$ & $1.78_{-3}$ \\
$0.178$ &   $0.165$ &   $0.192$ &   $2400.$ &     $409.$ & &$4.87$ &   $4.41$ &   $5.36$ &    $0.661$ &      $0.443$ & &$133.$ &   $121.$ &   $147.$ &    $1.92_{-3}$ & $1.80_{-3}$ \\
$0.206$ &   $0.190$ &   $0.220$ &   $1330.$ &     $276.$ & &$5.62$ &   $5.10$ &   $6.19$ &    $1.23$ &      $0.36$ & &$154.$ &   $140.$ &   $169.$ &    $1.12_{-3}$ & $1.34_{-3}$ \\
$0.238$ &   $0.216$ &   $0.258$ &    $682.$ &     $137.$ & &$6.49$ &   $5.88$ &   $7.15$ &    $1.12$ &      $0.29$ & &$178.$ &   $161.$ &   $196.$ &   $-3.06_{-4}$ & $1.12_{-3}$ \\
$0.274$ &   $0.253$ &   $0.299$ &    $628.$ &     $106.$ & &$7.50$ &   $6.79$ &   $8.25$ &    $0.683$ &      $0.182$ & &$205.$ &   $186.$ &   $226.$ &    $9.53_{-4}$ & $8.85_{-4}$ \\
$0.316$ &   $0.295$ &   $0.341$ &    $671.$ &     $109.$ & &$8.66$ &   $7.85$ &   $9.53$ &    $0.576$ &      $0.193$ & &$237.$ &   $215.$ &   $261.$ &    $8.10_{-4}$ & $5.14_{-4}$ \\
$0.365$ &   $0.331$ &   $0.402$ &    $459.$ &      $92.$ & &$10.0$ &   $9.06$ &   $11.0$ &    $0.279$ &      $0.172$ & &$274.$ &   $248.$ &   $301.$ &    $6.36_{-4}$ & $4.51_{-4}$ \\
$0.422$ &   $0.382$ &   $0.464$ &    $348.$ &      $63.$ & &$11.5$ &   $10.5$ &   $12.7$ &    $0.255$ &      $0.098$ & &$316.$ &   $287.$ &   $348.$ &    $5._{-6}$ &    $3.74_{-4}$ \\
    \end{tabular}
\medskip
\\
See footnote to Table~\ref{xiktab}.
At linear scales $k < 0.33 \, h \, \Mpc^{-1}$
the estimates of prewhitened power have been decorrelated.
At nonlinear scales $k > 0.33 \, h \, \Mpc^{-1}$
inaccuracies in the covariance matrix prevent full decorrelation,
but it would not be unreasonable to treat the estimates of prewhitened
power as uncorrelated or nearly so.
    \end{minipage}
    \end{table*}
}
\begin{document}

\maketitle

\begin{abstract}
We report a measurement of the real space
(not redshift space) power spectrum of galaxies
over four and a half decades of wavenumber,
$0.01$ to $300 \, h \, \Mpc^{-1}$,
from the {\it IRAS} Point Source Catalog Redshift Survey (PSCz).
Since estimates of power are highly correlated in the nonlinear regime,
we also report results for the prewhitened power spectrum,
which is less correlated.
The inferred bias between optically-selected APM and {\it IRAS}-selected PSCz
galaxies is about $1.15$ at linear scales
$\la 0.3 \, h \, \Mpc^{-1}$,
increasing to about $1.4$ at nonlinear scales
$\ga 1 \, h \, \Mpc^{-1}$.
The nonlinear power spectrum of PSCz shows a near power-law behaviour
to the smallest scales measured, with possible mild upward curvature 
in the broad vicinity of $k \sim 2 \, h \, \Mpc^{-1}$.
Contrary to the prediction of unbiased Dark Matter models,
there is no prominent inflection at the linear-nonlinear transition scale,
and no turnover at the transition to the virialized regime.
The nonlinear power spectrum of PSCz requires scale-dependent bias:
all Dark Matter models without scale-dependent bias are ruled out
with high confidence.
\end{abstract}


%\clearpage

\section{Introduction}
\label{intro}

The power spectrum of galaxies can set
powerful constraints on cosmological parameters
(Eisenstein, Hu \& Tegmark 1988;
Tegmark, Zaldarriaga \& Hamilton 2001).
But while the cleanest information lies at large, linear scales,
most of the data is at smaller, nonlinear scales.
Potentially, there is much to be gained by pushing to smaller scales.

The galaxy power spectrum is complicated by
nonlinearity, redshift distortions, and galaxy-to-mass bias.
%Other possible complications include
%luminosity-dependent bias,
%evolution with redshift,
%and observational systematics.
Even without bias,
nonlinear redshift distortions pose a problem.
Whereas linear redshift distortions are well understood
(Kaiser 1987; Hamilton 1998),
%and the nonlinear real space (unredshifted) mass power spectrum
%is well modelled
%(Hamilton et al.\ 1991;
%Peacock \& Dodds 1994, 1996;
%Jain et al.\ 1995;
%Ma 1998;
%Ma et al.\ 1999),		% 9906174
nonlinear redshift distortions are not
(Hatton \& Cole 1997, 1999).
%Galaxy-to-mass bias only thickens the fog, of course.
Nonlinear redshift distortions are of considerable interest
in their own right
(Kepner, Summers \& Strauss 1997;	% 9607097
Davis, Miller \& White 1997;		% 9705224
Jing, Mo \& B\"orner 1998;		% 9707106
Strauss, Ostriker \& Cen 1998;		% 9707248
Landy, Szalay \& Broadhurst 1998;	% 9711045
Baker et al.\ 2000),			% 9909030
but they muddy interpretation of the power spectrum
observed in redshift space.

%Nonlinear redshift distortions are of considerable interest
%in their own right
%(Baker et al.\ 2000),	% 9909030
%but they have a detrimental influence on the information
%content of the galaxy power spectrum observed in redshift space.
%In Cold Dark Matter (CDM) Universes and their variants,
%Universes with larger matter density $\Omega_{\rmn m}$
%have relatively more real power on small scales,
%but also a higher velocity dispersion,
%depressing the apparent power spectrum observed in redshift space.
%As a result,
%the redshift power spectra of high and low $\Omega_{\rmn m}$ Universes
%can appear coincidentally similar
%(Suto \& Suginohara 1991;
%Brainerd et al.\ 1996).
%%see also:
%%Bahcall, Cen \& Gramann 1993;
%%Gramann, Cen \& Bahcall 1993;
%%Brainerd \& Villumsen  1993, 1994).

Fortunately,
the effect of redshift distortions,
linear or nonlinear, biased or not,
can be practically eliminated.
Because redshift distortions displace galaxies only in the radial direction,
the power spectrum in directions transverse to the line-of-sight
is unaffected by redshift distortions.
The fact that the angular clustering of galaxies is unaffected by redshift
distortions has been used by many authors
to deduce the real space correlation function or power spectrum
(Groth \& Peebles 1977;				% Lick
Davis \& Peebles 1983;				% CfA 1
Saunders, Rowan-Robinson \& Lawrence 1992;	% QDOT-QIGC
Fisher et al.\ 1994a;				% 1.2 Jy
Loveday et al.\ 1995;				% Stromlo-APM
Baugh 1996;					% APM xi(r), 9512011
Gazta\~naga \& Baugh 1998;			% APM P(k), 9704246
Ratcliffe et al.\ 1998;				% Durham/UKST, 9702228
Jing, Mo \& B\"orner 1998;			% LCRS, 9707106
Guzzo et al.\ 2000;				% ESP xi(r), 9901378
Dodelson \& Gazta\~naga 2000;			% APM P(k), 9906289
Eisenstein \& Zaldarriaga 2001).		% APM P(k), 9912149

\xikcontsfig

While large angular surveys, such as
%the Lick Survey
%(Shane \& Wirtanen 1967),
the Automatic Plate Measuring survey (APM)
(Maddox et al.\ 1990a,b, 1996),
or
the Edinburgh/Durham Southern Galaxy Catalogue (EDSGC)
(Nichol, Collins \& Lumsden 2001),
%or
%The Palomar Digital Sky Survey (DPOSS)
%(Djorgovski et al.\ 1999),
might seem to offer the most natural data sets for measuring the real space
power spectrum, redshift surveys contain additional information
-- the redshifts of galaxies --
that can be exploited to great effect.
% in determining the real space power spectrum.
That is,
even if the redshift of a galaxy does not determine its precise distance,
it nevertheless constrains that distance within narrow limits.
The additional redshift information allows
the real space power spectrum to be measured from a redshift survey
with accuracy comparable to that from an angular survey many times larger
%This is discussed in Section~\ref{infoz}.
(\S\ref{infoz}).

%Connoly et al.\ for photometric results.

The goal of the present paper is to measure the
real space power spectrum of the recently published
{\it IRAS\/} Point Source Catalog Redshift Survey (PSCz)
(Saunders et al.\ 2000).
Large volume and careful attention to uniformity of selection make the PSCz
the finest publicly available redshift survey for this purpose.

%Nonlinear inflection seen by APM
%(Gazta\~naga \& Juszkiewicz 2001).

The analysis is described
in Sections~\ref{analysisl} (linear) and \ref{analysisn} (nonlinear),
and results are presented in Section~\ref{results}.
Section~\ref{conclusions} summarizes the conclusions.
Tables of measurements are collected in an Appendix,
and are also available from
http:/$\!$/\discretionary{}{}{}casa\discretionary{}{}{}.colorado\discretionary{}{}{}.edu/\discretionary{}{}{}$\sim$ajsh/\discretionary{}{}{}pscz/.

\section{Analysis -- Linear Regime}
\label{analysisl}

At linear scales,
$k \la 0.3 \, h \, \Mpc^{-1}$,
we adopt the real space power spectrum of the PSCz survey measured by
Hamilton, Tegmark \& Padmanabhan (2000, hereafter HTP).
This measurement assumes that density fluctuations are Gaussian,
and that redshift distortions conform to the linear model
(Kaiser 1987).
The linear measurement yields three separate power spectra,
the galaxy-galaxy, galaxy-velocity, and velocity-velocity power spectra.
In the present paper we use only the galaxy-galaxy power spectrum,
which is the real space power spectrum,
redshift distortions having been isolated into the other two power spectra.

The linear measurement can lay claim to being optimal when
the prior assumptions are true,
but it becomes suboptimal, and eventually fails, at nonlinear scales.
This is not merely because the linear model of redshift distortions fails,
as of course it does,
nonlinear redshift distortions being dominated by fingers-of-god,
not by coherent infall toward large scale overdensities.
More fundamentally,
the assumption of Gaussian density fluctuations fails.
In particular,
the linear measurement seriously underestimates the variance of power
in the nonlinear regime,
by a factor $\sim ( 1 + \xi )$, where $\xi$ is the correlation function.

Thus an entirely different strategy is called for at nonlinear scales.

\section{Analysis -- Nonlinear Regime}
\label{analysisn}

At nonlinear scales,
$k \ga 0.3 \, h \, \Mpc^{-1}$,
a major simplifying assumption can be made,
that redshift distortions are plane-parallel
(the `distant observer' approximation).
The plane-parallel approximation fails at large scales,
so the nonlinear method breaks down at linear scales,
just as the linear method breaks down at nonlinear scales.

%two major simplifying assumptions can be made.
%The first is that redshift distortions are plane-parallel
%(the `distant observer' approximation).
%The second is that the survey contains enough effectively independent
%volumes that the Central Limit Theorem applies,
%so that estimates of power should be Gaussianly distributed
%about their true values.
%
%To exploit the data in the PSCz survey to best advantage,
%both of these assumptions must be refined,
%as discussed in the next two subsections.

\subsection{Real power is transverse power}

In the plane-parallel approximation,
the redshift space power spectrum $P^s(k_\perp,k_\parallel)$
(the superscript $s$ denotes quantities in redshift space)
at wavenumbers $k_\perp$ and $k_\parallel$
perpendicular and parallel to the line-of-sight
is the Fourier transform of the redshift space correlation function
$\xi^s(r_\perp,r_\parallel)$ at redshift separations
$r_\perp$ and $r_\parallel$
perpendicular and parallel to the line-of-sight:
\begin{equation}
\label{Psxis}
  P^s(k_\perp, k_\parallel) =
    \int \e^{\im \k_\perp.\r_\perp + \im k_\parallel r_\parallel}
    \xi^s(r_\perp, r_\parallel) \, \ddt r_\perp \dd r_\parallel
  \ .
\end{equation}
Redshift distortions affect only separations $r_\parallel$
in the line-of-sight direction.
Equation~(\ref{Psxis})
shows that the redshift power spectrum in the transverse direction,
where $k_\parallel = 0$,
involves an integral of the redshift space correlation function
over the line-of-sight separation $r_\parallel$.
Since redshift distortions displace galaxies along the line-of-sight,
but neither create nor destroy them,
the integral along the line-of-sight is left unchanged by redshift distortions.
%Thus for $k_\parallel = 0$,
%the redshift correlation function $\xi^s(\r)$ in the integrand of
%equation~(\ref{Psxis}) can be replaced by the real correlation function
%$\xi(\r)$.
It follows that
the redshift space power spectrum in the transverse direction
is equal to the real space power spectrum
\begin{equation}
\label{Ps0}
  P^s(k_\perp{=}k, k_\parallel{=}0) = P(k)
  \ .
\end{equation}
Thus the problem of measuring the real space power spectrum
reduces to that of measuring the redshift space power spectrum
in the transverse direction.

Figure~\ref{xikconts}
shows a contour plot of the redshift space power spectrum
$P^s(k_\perp,k_\parallel)$
of PSCz.
The redshift power shown in Figure~\ref{xikconts} is measured from the
harmonics of the redshift space power spectrum,
as explained in detail in the remainder of this Section.
%~\ref{analysisn}.
The nonlinear real space power spectrum reported in this paper
is equal to the redshift space power spectrum along the transverse axis
in Figure~\ref{xikconts}.
%but it is useful to see at the outset what the beast looks like.
%One of the features of Bayesian statistics,
%not to mention common sense,
%is that one is permitted to peek at the answer
%(allow the data to inform the prior)
%before deciding the best way to measure it.

\subsection{Information from galaxy redshifts}
\label{infoz}

Measuring real power from the redshift power at exactly $k_\parallel = 0$,
as specified by equation~(\ref{Ps0}),
is liable to lead to a rather noisy estimate.
A more precise estimate of real power could be obtained
by `averaging' (in some sense) the values of redshift power
in some interval about $k_\parallel = 0$.

Using redshift power at $k_\parallel \neq 0$
is equivalent to exploiting information from galaxy redshifts.
Suppose that velocity dispersion (or perhaps some other influence)
causes galaxy distances to be known only to an accuracy of $\sigma$.
Then the observed redshift power spectrum is the true power spectrum
multiplied by a window that looks like a 2-dimensional sheet
transverse to the line-of-sight, a horizontal ridge
of width $\Delta k_\parallel \sim 1/\sigma$
about $k_\parallel = 0$.
% in the radial direction.
It follows that redshift power within $\sim 1/\sigma$ of $k_\parallel = 0$
provides potentially useful information about real power.

If redshift information were discarded,
then the uncertainty in galaxy distances would increase to the
depth $\sim R$ of the survey,
and the window through which the power spectrum is observed
would thin to $\Delta k_\parallel \sim 1/R$.

Thus with galaxy redshifts
there is
$\sim R/\sigma$
times as much exploitable $k$-space
as without.
In the PSCz survey,
the central two quartiles in depth, containing half the galaxies,
run from $50$ to $150 \, h^{-1} \Mpc$.
The effective uncertainty in the distance of a galaxy
without a redshift can be taken to be half this,
$R \sim 50 \, h^{-1} \Mpc$.
%In PSCz, the median depth is $R \sim 85 \, h^{-1} \Mpc$,
The velocity dispersion is $\sigma \sim 3 \, h^{-1} \Mpc$.
Thus PSCz with redshifts is in a sense comparable to a no-redshift survey
some $50/3 \sim 16$ times larger.
The errors on the real space power spectrum of PSCz with redshifts
might be expected to be roughly $(50/3)^{1/2} \sim 4$
times smaller than PSCz without redshifts.
Evidently the gain in having redshift information may be considerable.

\subsection{Distance indicator versus true distance}

It is worth pointing out an important distinctive feature of
a redshift survey versus an angular or photometric survey.
In a redshift survey,
the relation between the distance indicator -- the redshift distance --
and the true distance is independent of depth
(at least to the extent that cosmological evolution of the power spectrum
can be neglected).
In an angular or photometric survey, by contrast,
the relation between distance indicator
-- apparent brightness in the angular survey,
or photometric distance in the photometric survey --
depends on depth.

The existence of a well-defined redshift space power spectrum $P^s(\k)$,
as in equation~(\ref{Psxis}),
depends implicitly on the assumption that
the relation between redshift distance and true distance
is independent of depth.

\subsection{Angular mask and selection function}
\label{mask}

We adopt the same angular mask and selection function as HTP.
The angular mask is the high-latitude mask of Saunders et al.\ (2000)
({\tt hibpsczmask.dat}, part of the PSCz package),
which leaves unmasked $9.0636 \, {\rmn str}$, or $72\%$ of the sky.
Measurement of the selection function is discussed below in
Section~\ref{selfn}.
The angular and radial cuts leave $12\,446$ galaxies
in the survey.

\xikcontsafig

\subsection{Approximating nonlinear redshift distortions by a finite sum of harmonics}

To exploit redshift information to best advantage,
it is necessary to have some model of nonlinear redshift distortions.
Since accurate a priori models of nonlinear redshift distortions are not
available (Hatton \& Cole 1997, 1999)
-- especially if nonlinear galaxy-to-mass bias is taken into account --
we resort to a semi-empirical approach,
motivated by a combination of theory and observation.
Our adopted solution is
to measure the harmonics of the redshift power spectrum,
and to assume that nonlinear redshift distortions
can be approximated by retaining only a finite number of harmonics,
the number of harmonics retained depending on $k$.
The procedure is analogous to the familiar one
of smoothing an image by eliminating high frequencies in Fourier space.

It is convenient to introduce the function $f(\k)$ defined to be
the ratio of redshift to real space power spectra
(cf.\ Landy, Szalay \& Broadhurst 1998)
\begin{equation}
\label{fk}
  f(\k) \equiv {P^s(\k) \over P(k)}
  \ .
\end{equation}
By construction, this ratio is unity in directions transverse
to the line-of-sight,
$f(k_\perp{=}k,k_\parallel{=}0) = 1$.

Figure~\ref{xikcontsa}
shows a contour plot of the ratio $f(\k)$ measured from the PSCz survey.
Naturally this plot represents our final, best measurement;
however, preliminary versions of this plot contributed to the
decision about the best way to measure it.
The final version of the plot is consistent with the preliminary
versions;
tweaking caused only minor adjustments in the contours,
with no significant systematic shifts.

In the linear regime,
$f(\k)$ is given by
Kaiser's (1987) famous formula for plane-parallel redshift distortions
\begin{equation}
\label{Kaiser}
  f(\k) = ( 1 + \beta \mu^2 )^2
\end{equation}
where $\mu \equiv k_\parallel/k$ is the cosine of the angle
between the wavevector $\k$ and the line-of-sight.
Here $f(\k)$ is a fourth order even polynomial in $\mu$.

In Eulerian second order perturbation theory,
$f(\k)$ becomes an eighth order even polynomial in $\mu$,
with coefficients that depend on the absolute value $k$ of the wavenumber
(Scoccimarro, Couchman \& Frieman 1999).

The precise behaviour of $f(\k)$ in the nonlinear regime is unknown.
A simple and widely used empirical approximation
is to assume that the redshift correlation function $\xi^s(\r)$
equals the real correlation function $\xi(r)$ modulated by
a random pairwise velocity distribution $f_v(v_\parallel)$
that is independent of pair separation
(note that $r$ has units of velocity:
$1 \, h^{-1} \Mpc = 100 \, {\rmn km} \, {\rmn s}^{-1}$)
\begin{equation}
  \xi^s(r_\perp,r_\parallel)
  = \int_{-\infty}^\infty \xi(r_\perp,r_\parallel-v_\parallel) f_v(v_\parallel)
    \, \dd v_\parallel
  \ .
\end{equation}
Most commonly,
the 1-dim\-en\-sional pairwise velocity distribution
$f_v(v_\parallel)$
is taken to be an exponential
\begin{equation}
\label{fexp}
  f_v(v_\parallel) = (2^{1/2} \sigma)^{-1} \exp(- 2^{1/2} |v_\parallel|/\sigma)
\end{equation}
with 1-dim\-en\-sional pairwise velocity dispersion $\sigma$.
The exponential pairwise velocity distribution
was first proposed by Peebles (1976),
and has continued to receive support from observations
(Davis \& Peebles 1983, CfA1;
Fisher et al.\ 1994b, 1.2~Jy survey;
Marzke et al.\ 1995, CfA2 + SSRS2;
%Lin 1995, LCRS		% PhD thesis
Landy, Szalay \& Broadhurst 1998, LCRS;
Jing, Mo \& B\"orner 1998, LCRS)
from $N$-body experiments
(Fisher et al.\ 1994b, Fig.~5;
Zurek et al.\ 1994, Fig.~7),
and from theoretical arguments
(Sheth 1996;
Diaferio \& Geller 1996;
Seto \& Yokoyama 1998;
Juszkiewicz, Fisher \& Szapudi 1998).

If the pairwise velocity distribution $f_v(v_\parallel)$
were indeed independent of scale,
then $f(\k)$ in equation~(\ref{fk}) would equal the 1-dim\-en\-sional
Fourier transform of $f_v(v_\parallel)$
\begin{equation}
  f(\k)
  = \int_{-\infty}^\infty f_v(v_\parallel) \e^{\im k_\parallel v_\parallel}
    \, \dd v_\parallel
\end{equation}
a function only of $k_\parallel = k \mu$.
For the exponential pairwise velocity distribution, equation~(\ref{fexp}),
$f(\k)$ would be a Lorentzian
\begin{equation}
\label{fkexp}
  f(\k) = {1 \over 1 + \frac{1}{2} (\sigma k_\parallel)^2}
  = {1 \over 1 + \frac{1}{2} (\sigma k \mu)^2}
  \ .
\end{equation}
%We do {\em not\/} use the prior form~(\ref{fkexp}) here,
%since the form is indicative rather than definitive,
%and unnecessarily restrictive for our purpose.
%However,
Equation~(\ref{fkexp})
is a specific example of the general expectation that
$f(\k)$ in the nonlinear regime should be a smooth function,
peaked at $k_\parallel = 0$,
with width $\Delta k_\parallel \sim 1/\sigma$.

Figure~\ref{xikcontsa}
shows that in reality the pairwise velocity dispersion $\sigma$
is not independent of scale.
Rather, the velocity dispersion reaches a maximum at
$k \approx 1.3 \, h \, \Mpc^{-1}$
(where the contours of $f(\k)$ crowd the horizontal axis most closely),
and decreases to smaller scales (larger $k$).
%where the contours are more widely spaced.
This decrease in velocity dispersion to smaller scales
is qualitatively (though not necessarily quantitatively)
consistent with the expectation from the virial theorem that
$\sigma^2 \sim r^2 \xi(r) \sim k P(k)$
(Davis \& Peebles 1977; Peebles 1980, \S75),
which with $P(k) \simpropto k^{-1.5}$
(as found in \S\ref{results})
would predict $\sigma \simpropto k^{-0.25}$.

Jing \& B\"orner (2001)
find in $N$-body simulations of CDM variants
that $f(\k)$ falls somewhat faster than the Lorentzian model,
equation~(\ref{fkexp}),
at large $\sigma k \mu$.
They find that a better fit is
\begin{equation}
  f(\k)
  = {(1 + \beta \mu^2)^2
    \over 1 + \frac{1}{2} (\sigma k \mu)^2 + \eta (\sigma k \mu)^4}
\end{equation}
with $\sigma$ a function of $k$,
and $\eta$ a fitting parameter.

%The above examples indicate that $f(\k)$
%is a smooth function, an even polynomial in $\mu$
%that starts at 4th order in the linear regime,
%grows to 8th order in the quasi-linear regime,
%and then to infinite order (a Taylor expansion)
%in the fully nonlinear regime.

The above examples suggest the idea of approximating
$f(\k)$ as an even order polynomial in $\mu \equiv k_\parallel/k$,
or equivalently as a finite sum of even harmonics,
\begin{equation}
\label{fkl}
  f(\k) = \sum_{\el = 0}^{\el_{\max}(k)} f_\el(k) {\cal P}_\el(\mu)
\end{equation}
where ${\cal P}_\el(\mu)$ denotes a Legendre polynomial,
with maximum harmonic $\el_{\max}(k)$ depending on wavenumber $k$.
Of course the Lorentzian example, equation~(\ref{fkexp}),
is not a finite polynomial
(nor even a convergent Taylor series, if
$\frac{1}{2} (\sigma k \mu)^2 \ge 1$);
but evidently it could be approximated as such.
The principal advantages of the description in terms of harmonics are
(1) its flexibility, and
(2) fitting to a linear combination of even harmonics
(i.e.\ a polynomial in $\mu^2$) is far easier
than nonlinear fitting to, for example, a rational function of $\mu^2$.

A key question is how many harmonics to include in the sum~(\ref{fkl}).
Too many harmonics will yield an unnecessarily noisy estimate;
too few harmonics will fail to resolve the hill at $\mu = 0$,
and will tend to bias the measurement low.

At linear scales, the maximum harmonic should be $\el_{\max}(k) = 4$,
in accordance with Kaiser's formula~(\ref{Kaiser}).
At nonlinear scales,
it is necessary to resolve radial wavenumbers
comparable to the inverse pairwise velocity dispersion,
$1/\sigma$,
in accordance with the arguments in Section~\ref{infoz}.
Harmonics up to $\el$ can resolve angles $\sim \upi/\el$,
hence radial wavenumbers
$\Delta k_\parallel \sim k \upi/\el$.
Thus resolving $\Delta k_\parallel \sim 1/\sigma$
requires harmonics up to
\begin{equation}
\label{lapprox}
  \el_{\max}(k) \sim \upi \sigma k
  \ .
\end{equation}
If the velocity dispersion is $\sigma \sim 3 \, h^{-1} \Mpc$,
then equation~(\ref{lapprox}) suggests
$\el_{\max} \sim 10$ at $k \sim 1 \, h \, \Mpc^{-1}$.
The linear and nonlinear estimates together thus suggest, provisionally,
\begin{equation}
\label{ladoptp}
  \el_{\max}(k) = 4 + 6 \, k
\end{equation}
with $k$ measured in $h \, \Mpc^{-1}$.

The maximum harmonic specified by
equation~(\ref{ladoptp})
was our original choice,
and we carried out a complete set of measurements with it.
The preliminary measurements indicated
that redshift power was possibly slightly under-resolved at
$k \sim 1 \, h \, \Mpc^{-1}$,
but over-resolved at large $k$.
This can be seen in Figure~\ref{xikcontsa},
which shows that the ridge of redshift power along the transverse axis
reaches its narrowest point at
$k \approx 1.3 \, h \, \Mpc^{-1}$,
where
$\Delta k_\parallel \approx 0.33 \, h \, \Mpc^{-1}$,
but broadens out at larger $k$.
The velocity dispersion
$\sigma \sim 1 / \Delta k_\parallel$
is thus comparable to
$3 \, h^{-1} \Mpc$
at
$k \sim 1 \, h \, \Mpc^{-1}$,
but is smaller at large $k$.
Consequently the maximum harmonic $\el_{\max}$ of equation~(\ref{ladoptp}),
which provisionally presumed that
$\sigma \sim 3 \, h^{-1} \Mpc$,
is about right at
$k \sim 1 \, h \, \Mpc^{-1}$
but unnecessarily large at large $k$.
On the basis of the preliminary measurements,
we revised the choice of maximum harmonic to (the nearest even integer to)
\begin{equation}
\label{ladopt}
  \el_{\max}(k) = 16 \, k^{1/2}
\end{equation}
again with $k$ measured in $h \, \Mpc^{-1}$.
The revised choice of maximum harmonic $\el_{\max}(k)$
is slightly larger than the provisional choice at
$k \sim 1 \, h \, \Mpc^{-1}$
(so as to be on the safe side),
but smaller at large $k$.
The milder increase of maximum harmonic with wavenumber,
$\el_{\max} \propto k^{1/2}$ instead of
$\el_{\max} \propto k$ of equation~(\ref{ladoptp}),
reflects not only the fact that the velocity dispersion $\sigma$
decreases at larger $k$, as seen in Figure~\ref{xikcontsa},
but also that the statistical uncertainties increase at larger $k$.
More harmonics means smaller systematic bias,
but larger statistical uncertainty.
The choice~(\ref{ladopt}) is intended
to make the statistical error as small as possible
while ensuring that the systematic bias is small compared to the
statistical error.
Note that the nonlinear measurements are limited to
$k \ge 0.1 \, h \, \Mpc^{-1}$,
and that equation~(\ref{ladopt}) gives $\el_{\max} = 6$
at the smallest wavenumber of the nonlinear range,
$k = 0.1 \, h \, \Mpc^{-1}$.

%In practice, the change of maximum harmonic from equation~(\ref{ladoptp})
%to (\ref{ladopt}) caused little discernible systematic change in the
%measured power.
%Any residual bias from the truncation at $\el_{\max}$
%appears to be, as intended, only a fraction of the statistical error.

Equation~(\ref{ladopt}) is our adopted final choice of maximum harmonic
$\el_{\max}(k)$.
For other reasons, to be discussed in Section~\ref{bandpowers},
we also limit the maximum harmonic to
\begin{equation}
\label{lmax}
  \el_{\max}(k) \le 72
  \ .
\end{equation}
Numerical experiment,
reported in Section~\ref{com},
indicates that the maximum harmonic specified by
equations~(\ref{ladopt}) and (\ref{lmax})
is large enough that any bias caused by using too few harmonics
is small compared to the statistical uncertainty.
In practice,
the measured power spectrum proves satisfyingly robust against changes
in the choice of maximum harmonic,
the changes being typically some fraction of $1\sigma$,
and random rather than systematic.

%Why not truncate the harmonics more gradually, rather than sharply?
%For example,
%why not truncate harmonics with a Gaussian smoothing window
%$\e^{-\alpha \el(\el+1)}$?
%Our concern here is that a gradual cutoff is more likely
%to smooth out the expected peak at $\mu = 0$,
%leading to a systematic underestimate of real space power.

\subsection{Measuring harmonics of band-powers}
\label{harmonics}

We measure harmonics of band-powers of the redshift space power spectrum
using essentially the same procedure as
Hamilton (1995, 1998; hereafter H95, H98),
which is a slightly refined version of the method of
Hamilton (1992, 1993; hereafter H92, H93).

A feature of this analysis is that,
although it is the power spectrum that is being measured,
all the calculations are done in real (redshift) space
rather than in Fourier space.
In measuring redshift distortions,
it is important to disentangle the true distortion from the
artificial distortion introduced by a non-uniform survey window.
In real (redshift) space, the observed galaxy density
is the product of the true density and the selection function.
In Fourier (redshift) space, this product becomes a convolution.
Thus the natural place to `deconvolve' observations from the selection function
is real space, where deconvolution reduces to division,
and where the observations exist in the first place.

Let $\tilde P^s_\el(\tilde k)$
denote the $\el$'th harmonic of the redshift
power spectrum folded through some band-power window
$W(\tilde k, k)$
(the tildes distinguish band-powers $\tilde P^s$
and their characteristic wavenumbers $\tilde k$
from the raw power spectrum $P^s(\k)$;
tildes are tacitly dropped in the Results Section~\ref{results},
even though the powers reported there are in fact band-powers):
\begin{equation}
\label{Pslk}
  \tilde P^s_\el(\tilde k)
  =
    \int W(\tilde k, k) (2 \el + 1) {\cal P}_\el(\mu) P^s(\k)
    \, \ddd k/(2\upi)^3
  \ .
\end{equation}
The band-power windows $W(\tilde k, k)$ will be chosen momentarily
(\S\ref{bandpowers})
to be strictly positive functions narrowly peaked about a
central wavenumber $\tilde k$,
but for the moment equation~(\ref{Pslk}) is entirely general.
The band-power $\tilde P^s_\el(\tilde k)$, equation~(\ref{Pslk}),
can be expressed as an integral
over the redshift space correlation function
(H98, \S5.2)
\begin{equation}
\label{Pslr}
  \tilde P^s_{\el}(\tilde k)
  =
    \int W_\el(\tilde k, r) (2 \el + 1) {\cal P}_\el(\mu_\r) \xi^s(\r)
    \, \ddd r
\end{equation}
where $W_\el(\tilde k, r)$ is a spherical Bessel transform of $W(\tilde k, k)$:
\begin{equation}
\label{Wl}
  W_\el(\tilde k, r)
  = \im^\el \int_0^\infty j_\el(k r) W(\tilde k, k)
    \, 4\upi k^2 \dd k /(2\upi)^3
  \ .
\end{equation}
Equation~(\ref{Pslr})
is the basic equation that allows
galaxy pair counts to be converted directly into band-powers.

The redshift correlation function $\xi^s(r,\mu_\r)$
at separation $r$ and cosine angle $\mu_\r = \hat\z.\hat\r$
to the line of sight $\z$ is estimated by the H93 estimator
(the hat on $\hat \xi^s$ in eq.~\ref{xiest} is a reminder that it is an
estimate, not the true value)
\begin{equation}
\label{xiest}
  1 + \hat\xi^s(r,\mu_\r) =
    {\langle D D \rangle \langle R R \rangle \over \langle D R \rangle^2}
\end{equation}
where, following the conventional notation of the literature,
$D$ signifies data, and $R$ signifies random background points
(although in practice all the background integrals here were done as integrals,
not as Monte-Carlo integrals).
The angle brackets $\langle \, \rangle$ in equation~(\ref{xiest}) represent
FKP-weighted (see \S\ref{FKP})
averages over pairs at separation $r$ and $\mu_\r$.
The line of sight $\z$ is defined separately for each pair
as the angular bisector of the pair.
To allow for {\it IRAS\/}'s $1 \farcm 5$ angular resolution,
only pairs further apart than $1 \farcm 5$ are retained
(see \S\ref{smallk} for further discussion of this important effect),
and to ensure the validity of the plane-parallel approximation,
only pairs closer than $50^\circ$ on the sky are retained.
Poisson sampling noise is removed by excluding self-pairs
(pairs consisting of a galaxy and itself).

We continue the tradition of H92--H98
in computing the angular part of the pair integrals
$\langle D R \rangle$ and $\langle R R \rangle$
analytically
(H93, Appendix),
which leaves a single numerical integral over the radial direction.
The procedure is faster and more accurate than Monte Carlo methods,
and eliminates the artificial problem of shot noise in the background counts.
We also continue the tradition of H92--H98 in explicitly subtracting
the shot noise contribution to $\langle D R \rangle^2$
that comes from the same galaxy contributing to $D$ in both factors
of $\langle D R \rangle$
(\S2c of H93),
eliminating the small bias that arises if that contribution is not subtracted.

\subsection{Band-power windows}
\label{bandpowers}

The resolution $\Delta k$ with which the power spectrum can be measured
is limited by the characteristic size $R$ of the survey
to $\Delta k \sim 1/R$.
At linear scales this size,
and indeed the detailed shape of the survey volume,
plays an essential role in constructing band-power windows,
but at nonlinear scales there is greater freedom to choose band-power windows
more arbitrarily.

Following H95, H98,
we adopt band-power windows that are power laws times a Gaussian,
$W \sim k^n \e^{-k^2}$,
suitably scaled and normalized
(see eq.~\ref{W} below).
The advantages of this choice are:
(1) the band-power windows are strictly positive,
preserving the intrinsic positivity of the power spectrum;
(2) they vanish at zero wavenumber
(provided that $n > 0$),
so immunizing the measurement of power against
uncertainty in the mean density
(which makes a delta-function contribution to power at zero wavenumber);
(3) they are analytically convenient;
(4) they yield Gaussian convergence as a function of pair separation $r$
in the corresponding real space windows $W_\el(\tilde k, r)$,
equation~(\ref{Wlag}),
for harmonics $\el \le n$,
provided that $n$ is chosen to be an even integer.

Amusingly,
a power law times Gaussian, $k^n \e^{-k^2}$,
is the lowest energy eigenstate
of a three-dimensional simple harmonic oscillator
with angular momentum $n$.
Thus there is a least-squares sense
in which the band-power window
yields a measurement of the $n$'th harmonic of the power spectrum
at the smallest possible wavenumber
with the smallest possible pair separations
(Tegmark 1995).

As a compromise between resolution and the size of error bars
(higher resolution means larger error bars),
we choose band-powers uniformly spaced at $\Delta\log k = 1/16$,
the same resolution adopted by HTP in the linear regime.
The resolution of the band-power windows
$k^n \e^{-k^2}$,
equation~(\ref{W}),
increases with the exponent $n$,
the full width at half maximum (fwhm) going approximately as
$\Delta\log k$ $\simpropto$ $n^{-1/2}$.
We choose $n = 72$,
which has a fwhm of
$\Delta\log k$ $\approx$ $1/12$,
slightly wider than the adopted band-power spacing of $\Delta\log k = 1/16$.

\wfig

The maximum measurable harmonic at $n = 72$ is $\el = 72$,
which explains the limit~(\ref{lmax}).
We also measured band-powers with
exponents $n$ = $72 \times 4$ = $288$,
whose fwhm is $1/2$ that of the $n=72$ band-powers,
and for $n$ = $72 \times 9$ = $648$,
whose fwhm is $1/3$ that of the $n=72$ band-powers.
Since the higher resolution measurements were consistent
with the lower resolution $n = 72$ measurement
(see \S\ref{individual}),
with little sign of any systematic offset caused by insufficient resolution,
we choose to report as standard the result from the lower resolution
$n = 72$ measurement, which has slightly smaller error bars
(after the higher resolution measurements are rebinned in $k$
to the lower resolution).

Suitably scaled,
and normalized so
$\int W(\tilde k,k)$\,\discretionary{}{}{}{}$\ddd k/(2\upi)^3$ = $1$,
the band-power windows are
\begin{equation}
\label{W}
  W(\tilde k, k) {\ddd k \over (2\upi)^3} \equiv
    {2 \, \e^{-q^2} q^{n+2} \, dq \over \Gamma[(n{+}3)/2]}
  \ , \quad
  q \equiv
    {\alpha k \over \tilde k}
  \ .
\end{equation}
The constant
$\alpha$
is chosen so that the band power window
$W(\tilde k, k)$ is centred at
$k \approx \tilde k$.
Following H95, H98,
we choose the constant
$\alpha = \left\{ \Gamma[(n{+}3)/2] \right.$/\discretionary{}{}{}$\left.\Gamma[(n{+}\gamma)/2] \right\}^{[1/(3{-}\gamma)]}$
so that the smoothed monopole power at wavenumber $\tilde k$
is equal to the unsmoothed monopole power at the same wavenumber,
$\tilde P^s_0(\tilde k) = P^s_0(\tilde k)$,
for the particular case where the power spectrum is a power law
$P^s_0(k) \propto k^{\gamma - 3}$
(corresponding to $\xi(r) \propto r^{-\gamma}$)
of index $\gamma = 1.5$,
that is, for $P^s_0(k) \propto k^{-1.5}$.
For the case $n = 72$ in the window~(\ref{W}),
this fixes $\alpha = 6.051$.

\mapfig

The harmonics $\tilde P^s_\el(\tilde k)$
of the redshift power spectrum folded through the window~(\ref{W}) are,
according to equation~(\ref{Pslr}),
equal to the harmonics of the redshift correlation function folded through
the corresponding windows $W_\el(\tilde k, r)$ given by equation~(\ref{Wl}):
\begin{equation}
\label{Wlag}
  W_\el ( \tilde k, r ) =
  {\im^\el [(n{-}\el)/2]! \over (3/2)_{(n/2)}} \:
  s^\el \e^{-s^2} L_{(n-\el)/2}^{\el+(1/2)} (s^2)
  \ , \quad
  s \equiv {\tilde k r \over 2 \alpha}
\end{equation}
(note that $W_0(\tilde k, 0) = 1$)
where $L_{\nu}^{\lambda}$ are Laguerre polynomials
(Abramowitz \& Stegun 1964)
and
$(3/2)_{(n/2)}$ = $\Gamma [(n{+}3)/2]$/\discretionary{}{}{}$\Gamma (3/2)$
is a Pochhammer symbol.

Figure~\ref{w}
illustrates both the Fourier band-power window $W(\tilde k, k)$,
equation~(\ref{W}),
and a selection of its real space counterparts $W_\el(\tilde k, r)$,
equation~(\ref{Wlag}),
for the case $n = 72$.
The Figure illustrates that
measuring higher harmonics of power requires finer resolution in Fourier space,
hence wider separations in real space.
At small separations $r$,
the real space windows $W_\el(\tilde k, r)$
alternate between being positive or negative, as $\el/2$ is even or odd,
thanks to the $\im^\el$ factor in equation~(\ref{Wlag}).

%\begin{equation}
%  \frac{1}{2}
%  \left( - \nabla^2 + \omega^2 r^2 - {\el(\el + 1) \over r^2} \right)
%    \phi_{n\el}(r)
%  = \left( \frac{3}{2} + n \right) \phi_{n\el}(r)
%\end{equation}

One of the features of the $k^n \e^{-k^2}$ band-power window
is that it vanishes at $k = 0$.
It follows that any constant contribution to the correlation function $\xi^s$,
equivalent to a delta-function contribution to power at $\k = 0$,
vanishes when folded through the windows $W_\el(\tilde k, r)$
given by equation~(\ref{Wlag}).
Thus in estimating $\tilde P^s_\el(\tilde k)$ by equation~(\ref{Pslr}),
the $\xi^s$ factor in the integrand can be replaced by $1 + \xi^s$:
it is unnecessary to subtract the $1$ part of the estimator
$1 + \hat\xi^s$ of equation~(\ref{xiest}).

\subsection{Covariance matrix}
\label{covariance}

Reliable error bars
on a measurement are as important as the measurement itself.
Indeed,
if precise comparison to theoretical models is to be made,
then a full covariance matrix is essential
(Eisenstein \& Zaldarriaga 2001;
Tegmark et al.\ 2001).

There are essentially three ways to determine uncertainties,
differing in how much prior information they invoke.
%(1) know the covariance matrix a priori;
%(2) estimate the covariance from the scatter in measurements
%from ensembles of mock catalogues;
%(3) estimate the covariance from the scatter in the data.

The ideal situation is to know a priori what the covariance matrix is,
or to know its form as a function of a modest number of parameters.
Precisely this situation obtains
for Gaussian fluctuations in the linear regime.
Unfortunately,
notwithstanding valuable progress
(Scoccimarro \& Frieman 1999;
Szapudi, Colombi \& Bernardeau 1999)
the covariance matrix of nonlinear power is not accurately known
(in either real or redshift space),
and indeed the simplest model, based on the hierarchical model
with constant hierarchical amplitudes, is known to be inconsistent,
because it violates the Schwarz inequality
(Scoccimarro, Zaldarriaga \& Hui 1999;
Hamilton 2000).

A second commonly used strategy
is to estimate the covariance from the scatter in measurements
from ensembles of mock catalogues constructed from $N$-body simulations
to resemble the survey as closely as possible
(e.g.\ Fisher et al.\ 1993;
Cole et al.\ 1998).

A third alternative is to measure the covariance
directly from the level of fluctuations observed in the survey itself
(H93; Szapudi 2000),
and here we follow this latter approach.
%The method makes minimal assumptions about the origin of fluctuations,
The approach takes full account of the correlated character of the
fluctuations in a survey.
Although the method is expected to break down at scales approaching
the size of the survey,
it should work fine at the nonlinear scales addressed here.

H93's method for measuring covariance works in essence as follows
(see \S4 of H93 for intricate details).
Let $\hat P$ be a quadratic estimator,
some integral of products of pairs of galaxy densities.
For example, $\hat P$ could be an estimate of $\tilde P^s_\el(\tilde k)$,
the $\el$'th harmonic of some band-power in redshift space,
equation~(\ref{Pslr}).
%Let $\hat\xi(r_\perp,r_\parallel) = W_{ij} \delta_i \delta_j$,
%a quadratic integral over galaxy overdensities $\delta_i$,
%be an unbiased estimate of the redshift space correlation function
%at transverse and line-of-sight separations $r_\perp$ and $r_\parallel$.
Divide the survey into a reasonably large number of subvolumes.
Here we choose 22 angular regions,
as shown in the inset to Figure~\ref{map},
times 10 radial shells, each 0.2 dex wide,
covering radial depths from $10^{0.625}$ to $10^{2.625} \, h^{-1} \Mpc$
(i.e.\ 4.2 to $420 \, h^{-1} \Mpc$).
Imagine attaching a weight $w_i$ to each of these
$22 \times 10 = 220$ subvolumes.
As each of these weights is varied,
the estimated value $\hat P$ changes.
Note that the estimator $\hat P$
is being supposed subject to an overall normalization condition
such that it remains an unbiased estimate of the thing being estimated,
as the weights $w_i$ are varied;
in other words, only the relative weights $w_i$ really matter.
Define the fluctuation $\Delta\hat P_i$ in $\hat P$
attributable to subvolume $i$ by
\begin{equation}
\label{DhatP}
  \Delta\hat P_i = \frac{1}{2} w_i {\upartial \hat P \over \upartial w_i}
\end{equation}
where the important factor $1/2$ arises because $\hat P$ depends quadratically
on galaxy density.
Then (H93) the variance of $\hat P$ is given by
a sum over pairs $ij$ of subvolumes
\begin{equation}
\label{DhatP2}
  \langle \Delta\hat P^2 \rangle
  = \sum_{ij} \Delta\hat P_i \Delta\hat P_j
  \ .
\end{equation}

The fluctuations $\Delta\hat P_i$ are subject to
a `pair-integral constraint' that their sum over all subvolumes
should be zero, $\sum_i \Delta\hat P_i = 0$.
This follows from the fact that changing all the weights $w_i$
by the same constant factor leaves the estimate $\hat P$ unchanged.
If all pairs $ij$ of subvolumes were included in
the sum on the right hand side of equation~(\ref{DhatP2}),
then the variance $\langle \Delta\hat P^2 \rangle$
would be zero,
because of the integral constraint $\sum_i \Delta\hat P_i = 0$.
Consider instead including in the sum only pairs $ij$ of subvolumes
closer than some given separation.
Characteristically, as this maximum separation between subvolumes increases,
the sum on the right hand side of equation~(\ref{DhatP2})
increases, reaches a maximum,
and then declines to exactly zero when all pairs of subvolumes are included.
We follow H93's proposal of
approximating the variance $\langle \Delta\hat P^2 \rangle$
by its maximum value attained as the maximum separation between subvolumes
is increased.
This approximation reflects on the one hand
the idea that it is nearby regions that are most correlated,
and on the other hand the desire to include
as much of the correlation between nearby regions as possible.

As discussed by H93,
the pair-integral constraint means that the variance
$\langle \Delta\hat P^2 \rangle$
is inevitably underestimated at scales approaching the size of the survey.
However,
this effect should be minor at the nonlinear scales addressed here.
Conversely,
there may be some tendency to overestimate the variance because
noise is liable to make
the measured maximum in the variance fluctuate above the true maximum.

The covariance between $\hat P$
and another any quadratic estimator $\hat P'$
is given by a generalization of equation~(\ref{DhatP2}),
\begin{equation}
\label{DhatPDhatP}
  \langle \Delta\hat P \Delta\hat P' \rangle
  = \sum_{ij} \Delta\hat P_i \Delta\hat P'_j
  \ .
\end{equation}
Again, if all pairs $ij$ of subvolumes were included in
the sum on the right hand side of equation~(\ref{DhatPDhatP}),
then the covariance would be zero,
because of the integral constraint $\sum_i \Delta\hat P_i = 0$.
In this case the strategy of approximating the covariance by the maximum
value attained, as pairs $ij$ of greater and greater separation are
included in the sum, fails.
The strategy fails
partly because covariances need not be positive,
and partly because choosing covariances to be large
is not necessarily a conservative approach
--
whereas increasing variances always reduces information content,
increasing covariances can actually increase information content,
because two highly correlated quantities contain information about each other.

Here we estimate the covariance
$\langle \Delta\hat P \Delta\hat P' \rangle$
as the average of the sums
$\sum_{ij} \Delta\hat P_i \Delta\hat P'_j$
evaluated at the two places where the variances
$\sum_{ij} \Delta\hat P_i^2$
and
$\sum_{ij} \Delta\hat P'^2_i$
reach a maximum.

\subsection{Prewhitened power}
\label{prewhitenedpower}

The term `prewhitening' comes from signal-processing,
and refers to the operation of transforming a signal in such a way
that the noise becomes white, or constant
(Blackman \& Tukey 1959, \S11).
The notion of prewhitening the power spectrum of galaxies
as a means of narrowing the covariance of estimates of power
at nonlinear scales was proposed by Hamilton (2000, hereafter H00).
Whereas at linear scales the covariance of estimates of power is (nearly)
diagonal, at nonlinear scales the covariance of estimates of power
is broadly correlated over different wavenumbers,
as emphasized by
Meiksin \& White (1999)
and
Scoccimarro, Zaldarriaga \& Hui (1999),
and as illustrated in Section~\ref{prewhiten} of the present paper.

H00 showed empirically that prewhitening the power spectrum
narrowed the covariance of power in a broad range of models.
As will be seen in Section~\ref{prewhiten},
the measured covariance of prewhitened power in PSCz
is indeed narrower than the covariance of power itself.

The prewhitened power spectrum is defined to be the Fourier transform,
$X(k) = \int_0^\infty \e^{\im \k.\r} X(r) \, \ddd r$,
of the prewhitened correlation function $X(r)$ defined by
(H00, \S5.1)
\begin{equation}
\label{Xr}
  X(r) \equiv {2 \, \xi(r) \over 1 + [1 + \xi(r)]^{1/2}}
  \ .
\end{equation}
Differentiating equation~(\ref{Xr}) gives, to lowest order,
\begin{equation}
\label{DXr}
  \Delta X(r) = {\Delta \xi(r) \over [1 + \xi(r)]^{1/2}}
\end{equation}
so that the covariance of estimates $\hat X(r)$
of the prewhitened correlation function is
(for small errors)
\begin{eqnarray}
\label{DXrDXr}
  \lefteqn{
  \langle \Delta\hat X(r) \Delta\hat X(r') \rangle
  }
  &&
  \nonumber
  \\
  &&
  = [1 + \xi(r)]^{-1/2}
    \langle \Delta\hat\xi(r) \Delta\hat\xi(r') \rangle
    [1 + \xi(r')]^{-1/2}
\end{eqnarray}
(note that $\langle \hat\xi(r) \rangle = \xi(r)$,
if $\hat\xi(r)$ is an unbiased estimator).
Since the shot noise contribution to
$\langle \Delta\hat\xi(r) \Delta\hat\xi(r') \rangle$,
i.e.\ the contribution that comes from the covariance between a pair
of galaxies and itself,
is in real space a diagonal matrix proportional to $1 + \xi(r)$
(H00, eq. 38),
it follows that
the prewhitened covariance, equation~(\ref{DXrDXr}), has the property
that the shot noise contribution to
$\langle \Delta\hat X(r) \Delta\hat X(r') \rangle$
is proportional to the unit matrix.

The covariance of estimates $\Delta\hat X(k)$ of prewhitened power is given by
the Fourier transform of equation~(\ref{DXrDXr}),
\begin{equation}
\label{DXkDXk}
  \langle \Delta\hat X(k) \Delta\hat X(k') \rangle
  = \mxH^{-1/2}
    \langle \Delta\hat P(k) \Delta\hat P(k') \rangle
    \mxH^{-1/2}
\end{equation}
where $\mxH$ is the Fourier transform of the matrix
which in real space is diagonal with diagonal entries
$1 + \xi(r)$.
The shot noise (self-pair) contribution to
$\langle \Delta\hat X(k) \Delta\hat X(k') \rangle$
is again proportional to the unit matrix,
since the unit matrix remains the unit matrix in any representation.

Some numerical issues concerning prewhitening are discussed in \S4.2 of H00,
and as an aid to the reader,
Appendix~\ref{howtoprewhiten}
contains practical instructions on how to prewhiten a power spectrum
numerically.

One slightly subtle issue is that the power spectrum is estimated
in discrete band-powers, not as a continuous function of wavenumber.
Our policy is to adhere to the definition~(\ref{Xr})
of the prewhitened correlation function
\begin{equation}
\label{hatXr}
  \hat X(r) \equiv {2 \, \hat\xi(r) \over 1 + [1 + \hat\xi(r)]^{1/2}}
\end{equation}
with $\hat\xi(r)$ in both numerator and denominator
being understood to be band-estimates,
Fourier transforms of the band-powers.

\subsection{FKP weightings}
\label{FKP}

In a seminal paper,
Feldman, Kaiser \& Peacock (1994, hereafter FKP)
showed that at wavelengths large enough to be Gaussian,
but still small compared to the scale of the survey,
the optimal weighting of pairs $ij$ of volume elements
for measuring the power spectrum $P(k)$ at wavenumber $k$ is
\begin{equation}
\label{wFKP}
  {\bar n(\r_i) \bar n(\r_j)
  \over [ 1 + \bar n(\r_i) P(k) ] [ 1 + \bar n(\r_j) P(k) ]}
  \ .
\end{equation}
The FKP weighting goes over to equal weighting of volumes where the
selection function $\bar n(\r)$ is large,
and equal weighting of galaxies where the
selection function is small,
which makes physical sense.

The FKP weighting is often referred to as `minimum variance'
(or more cautiously, `near minimum variance'),
yet the range of scales over which it is strictly valid is limited
(even non-existent).
Of course it is commonly, and correctly, argued
in defense of the more general use of the FKP weighting
that because the variance changes quadratically about its minimum,
a near minimum variance weighting should give a result not much worse
than the true minimum variance.

The simplicity of the FKP weighting, equation~(\ref{wFKP}),
springs from the fact that, for Gaussian fluctuations,
the covariance matrix $\langle \Delta\hat P(k) \Delta\hat P(k') \rangle$
of estimates of power
(including the shot noise contribution) is diagonal
(for Gaussian fluctuations, at wavelengths small compared to the survey).
Thus the inverse covariance matrix,
which determines the optimal weighting of pairs,
is similarly diagonal.
The eigenvalues of the inverse covariance constitute the FKP weights,
equation~(\ref{wFKP}).
By contrast, the covariance of estimates of the correlation function $\xi(r)$,
for example, is not diagonal,
and the optimal weighting of pairs is, strictly, a complicated matrix.

At nonlinear scales the covariance of power ceases to be diagonal,
and the FKP weighting ceases to be optimal.
However,
H00 showed that a weighting similar to the FKP weighting is valid
for the prewhitened power spectrum (\S\ref{prewhitenedpower})
to the extent that the covariance of prewhitened power is indeed
(nearly) diagonal.
The more general weighting differs from FKP in that $P(k)$ in the denominator
of the weighting is replaced by an `FKP constant' $J$,
whose value is model-dependent, but of order $\sim 1$--$3$ times the
(unprewhitened) power $P(k)$
(H00, Fig.~11):
\begin{equation}
\label{J}
  {\bar n(\r_i) \bar n(\r_j)
  \over [ 1 + \bar n(\r_i) J ] [ 1 + \bar n(\r_j) J ]}
  \ .
\end{equation}

The strategy of the present paper is to measure
band-powers using FKP weightings, equation~(\ref{J}),
with 5 values of the FKP constant,
$J = 0$, $10$, $10^2$, $10^3$, and $10^4 \, h^{-3} \Mpc^3$,
and then (cautiously) compress (\S\ref{compress})
the 5 measurements into a single best estimate of the band-power.

In accordance with the above arguments,
we compress not the band-powers themselves,
but rather the prewhitened band-powers.
In other words, to form the best estimate of the band-power,
we first first prewhiten
(\S\ref{prewhitenedpower})
the 5 FKP-weighted estimates,
which we then combine into a best estimate of prewhitened power,
which we then unprewhiten.

Why choose 5 particular values of the FKP constant,
rather than follow H00 and adopt, at each wavenumber $k$,
a single FKP constant $J$ equal to $1$--$3$ times the power $P(k)$?
The reasons are both practical and philosophical.
The practical reason is as follows.
We wish to make an estimate of the prewhitened power
in which the estimate $\hat\xi(r)$ in the denominator of equation~(\ref{hatXr})
is the same as the $\hat\xi(r)$ in the numerator, at every separation $r$.
But the best choice of FKP constant $J$ varies with $k$,
which has the consequence that the best estimate of prewhitened power
involves estimates of (unprewhitened) power at many $J$'s.
An alternative procedure that naturally suggests itself
might be to measure the power spectrum with a fixed $J$, prewhiten it,
and call that the best estimate of prewhitened power at a particular $k$.
However, the prewhitened power from the latter procedure does not
satisfy the desideratum that the estimates $\hat\xi(r)$ in the numerator
and denominator of equation~(\ref{hatXr}) are the same.
Our view is that it is better to impose the a priori requirement
that the $\hat\xi(r)$ in the numerator and denominator be the same,
than to discard that information.
Given that
it is necessary to measure the (unprewhitened) power $P(k)$ at many $J$'s,
for each wavenumber $k$,
one is also faced with
the necessity of measuring the covariances between powers
with different $J$'s and different $k$'s.
But limitations of computer power then constrain one to using just a
handful of $J$'s.
This is the practical reason behind the procedure adopted here.

%To ensure that the estimate $\hat\xi(r)$ in the denominator of
%equation~(\ref{hatXr}) is the same as the $\hat\xi(r)$ in the numerator
%at every separation $r$,
%we follow an iterative procedure in which
%we start by compressing the 5 estimates without prewhitening,
%derive the best estimate power, use that to prewhiten,
%rederive the best estimate power, and iterate to convergence.
%In practice only a few iterations are required.

The philosophical reason for measuring the power
by compressing estimates from a handful of $J$'s,
rather than adopting at each $k$ a single FKP constant $J$ equal to
$1$--$3$ times the power $P(k)$,
is that the factor of $1$--$3$
depends on the assumed model for the behaviour of higher order correlations,
and there is no assurance that the PSCz data conform to the model.
Indeed the model adopted  by H00
--- the hierarchical model with constant hierarchical amplitudes ---
is certainly wrong at some level,
because the resulting covariances of power violate the Schwarz inequality
unless the 4-point star amplitude is equal to minus the
4-point snake amplitude, $R_b = -R_a$, contrary to observation
(Scoccimarro, Zaldarriaga \& Hui 1999 \S3.3; H00).
Our preference is therefore to allow the PSCz data
to `choose' the best weighting.

\subsection{Cautious Fisher compression}
\label{compress}

At this point,
the data consist of
$5$ FKP weightings
%($J = 0$, $10$, $10^2$, $10^3$, and $10^4 \, h^{-3} \Mpc^3$)
of each of $37$ harmonics
(even harmonics up to $\el_{\max} = 72$)
of band-powers at each of $57$ wavenumbers
($k = 0.1$ to $316 \, h \, \Mpc^{-1}$ logarithmically spaced at
$\Delta\log k = 1/16$),
a total of $5 \times 37 \times 57 = 10\,545$ quantities.
Along with the data are their fluctuations,
equation~(\ref{DhatP}),
with respect to each of $220$ volume elements,
a total of $10\,545 \times 220 = 2\,319\,900$ fluctuations.
The $10\,545 \times 10\,545$ covariance matrix of the data
is constructed (or at least constructible)
from the fluctuations as described in Section~\ref{covariance}
(in effect, the fluctuations provide a convenient way to store in abbreviated
form the variances and covariances between all $10\,545$ quantities).

In principle,
the Fisher matrix formalism
(see Tegmark, Taylor \& Heavens 1997 for a review)
allows one to take the $10\,545$ data and use their Fisher matrix
-- their inverse covariance --
to compress them optimally into $57$ measurements of real space power.
Unfortunately,
errors in the measured covariance matrix
thwart so idealistic an enterprise.
We relegate this moral tale of failed ambition to its rightful place,
an Appendix.

A symptom of the difficulty with the covariance matrix
is that a good fraction of its eigenvalues are negative,
whereas in reality the covariance matrix should be positive definite,
with all positive eigenvalues.

If the only problem were negative eigenvalues,
then it would be easy to solve by Singular Value Decomposition.
%replacing negative eigenvalues of the Fisher matrix by zero.
The more serious problem is that the covariance matrix contains
positive eigenvalues some of which are evidently spuriously small.
A small positive eigenvalue can signify
either that a quantity is accurately measured,
or else that there is some highly correlated set of quantities.
Clearly one wants to retain a well-measured quantity;
on the other hand one might be inclined to discard some component
of a set of highly correlated quantities.

The problem is not that the covariance matrix is particularly badly measured.
In fact the level of fluctuations in the measured covariances,
such as can be seen in Figure~\ref{ccw},
suggests that the covariances are typically accurate to $\sim 20\%$.
Moreover there is general consistency with errors measured (HTP) by the
linear method.

Abandoning any grand compression scheme
(Appendix~\ref{failed}),
we revert to a simpler program,
to compress the 5 FKP-weighted estimates of each band-power into one.

We first form an estimate $\hat P(\tilde k)$ of the real space power
at each FKP weighting and each wavenumber
from the redshift space power in the transverse direction,
$\mu = 0$,
including only harmonics of redshift power up to
$\el_{\max}(k)$ given by equations~(\ref{ladopt}) and (\ref{lmax})
(the hat on $\hat P^s_\el(\tilde k)$ in the following equation
is a reminder that it is an estimate, not the true value,
of the band-power harmonic
$\tilde P^s_\el(\tilde k)$, eq.~\ref{Pslk}):
\begin{equation}
  \hat P(\tilde k) =
  \sum_{\el = 0}^{\el_{\max}(k)}
    \hat P^s_\el(\tilde k) {\cal P}_\el(\mu{=}0)
  \ .
\end{equation}
We compute the $5 \times 5$ covariance matrix of the
five FKP-weighted estimates $\hat P(\tilde k)$
from the fluctuations $\Delta \hat P(\tilde k)$,
equation~(\ref{DhatP}),
as described in Section~\ref{covariance}.
The resulting covariance matrix is consistent with
that computed less directly (hence presumably less accurately)
via the covariance matrix of harmonics.

We then prewhiten (\S\ref{prewhitenedpower})
each of the 5 FKP-weighted estimates
$\hat P(\tilde k)$,
and prewhiten their covariance matrix correspondingly.
Since prewhitening requires knowledge of the full power spectrum,
we start by compressing the 5 estimates without prewhitening,
derive the best estimate power, use that to prewhiten,
rederive the best estimate power, and iterate to convergence.
If $\hat X_i$ denotes the $i$'th of 5 estimates of prewhitened power,
then the overall best estimate $\hat X$ is that which minimizes $\chi^2$
\begin{equation}
\label{chi2X}
  \chi^2 = \sum_{ij}
    ( \hat X_i - \hat X )
    \mxC^{-1}_{ij}
    ( \hat X_j - \hat X )
\end{equation}
where $\mxC_{ij} \equiv \langle \Delta \hat X_i \Delta \hat X_j \rangle$
is the $5 \times 5$ covariance matrix of estimates of prewhitened power.
The minimum $\chi^2$ solution of equation~(\ref{chi2X}) is
\begin{equation}
\label{bestX}
  \hat X = \sum_i w_i \hat X_i
  \ , \quad
  w_i = {\sum_j \mxC^{-1}_{ij} \over \sum_{kl} \mxC^{-1}_{kl}}
  \ .
\end{equation}

Typically, the covariance matrix $\mxC_{ij}$
contains some small (sometimes negative) eigenvalues,
indicating that the 5 estimates are highly correlated
-- not a particularly surprising result.
However, $\chi^2$ minimization typically responds to the high correlation
by assigning one estimate a large positive weight,
and another an almost cancelling large negative weight.
Such behaviour is clearly spurious,
an artefact of errors in the covariance matrix
having random ill effects on small eigenvalues.

We solve the problem by requiring that the weights
that go into the best estimate, equation~(\ref{bestX}),
all be positive, $w_i \ge 0$.
We do this in a dumb way:
we find the minimum $\chi^2$ solution for each of the $2^5 - 1 = 31$
nontrivial ways in which each of the 5 weights is free or fixed at $0$,
and choose that positive weighting that has the smallest $\chi^2$.
Typically two or three of the 5 estimates have nonzero weights
in the best estimate.
The other estimates, having zero weight, are in effect discarded,
the least informative way of using those data.

The weightings for the full set of band-powers
show a plausible and expected pattern.
Band-powers at larger scales, where $P(k)$ is large,
prefer weightings with larger FKP constants $J$,
while band-powers at smaller scales prefer smaller $J$.

Finally, having obtained the best estimate prewhitened power $\hat X$,
we unprewhiten to obtain the best estimate power $\hat P$.
As commented above,
several iterations are needed to ensure that the power spectrum
used in (un)prewhitening is the same as the best estimate.

The main effect of prewhitening before compressing,
as opposed to compressing powers directly,
is to prefer smaller FKP constants $J$.
The consequences of this preference are commented on
in Section~\ref{individual}.

\subsection{Selection function}
\label{selfn}

\sffig

Since HTP give only a brief description of
the measurement of the selection function, we offer more details here.
We adopt three simplifying assumptions commonly made
in measuring the selection function of a flux-limited galaxy survey
(see e.g.\ the reviews by
Binggeli, Sandage \& Tammann 1988;
Willmer 1997;
Tresse 1999):
(1) that the luminosity function is independent of position;
(2)
that the survey is complete to the specified flux limit;
and
(3)
that distances and galaxy fluxes are measured with negligible error.
Undoubtedly all of these assumptions fail at some level.

If the above three assumptions are taken to be true,
then there is a unique exact solution
(modulo an overall normalization factor),
a solution for the luminosity function and radial density distribution
of galaxies that exactly reproduces the observed distribution
of luminosities and distances.
The exact solution is given by Lynden-Bell's (1971) $C^-$ method,
which coincides with Turner's (1979) method in the limit of infinitesimal bins,
and with Efstathiou, Ellis \& Peterson's (1988)
Stepwise Maximum Likelihood method in the limit of infinitesimal steps.
The exact solution is a sum of delta-functions:
the luminosity function is a sum of delta-functions at the observed
luminosities of the galaxies;
and the galaxy density is a sum of delta-functions at the observed
distances of the galaxies.
This is perhaps not too surprising given that the observations -- galaxies --
are themselves described as delta-functions in luminosity and distance.
The resulting selection function,
the integral of the luminosity function, is a step-function,
with a step at the limiting distance of each galaxy in the survey.
In practice we evaluate the selection function
using Turner's (1979) method adapted to the case of infinitesimal bins;
the algorithm has the merit of being exceedingly fast.

Figure~\ref{sf}
shows the resulting exact solution
for the selection function and the inferred galaxy density.

The selection function so computed is `exact'
only to the extent that the prior assumptions are valid.
Clearly,
the `exact' selection function, being a step-function,
does not incorporate the Bayesian prejudice
that the selection function is likely to be smooth.
For this reason it is usual to fit the selection function
to a smooth analytic function.
We use the maximum likelihood method of
Sandage, Tammann \& Yahil (1979),
and fit the selection function
$\bar n(r)$
to a function whose form is inspired by the Schechter (1976) function,
but with enough free parameters to yield a good fit,
also shown in Figure~\ref{sf}:
\begin{eqnarray}
\label{sffit}
  \lefteqn{
  \log_{10}(\bar n)
  =
  \mbox{} - 0.646 \, \log_{10}(r_{100})
  }
  &&
  \nonumber
  \\
  &&
  \mbox{} - \, {
    1.86 + 1.836 \, r_{100} + 0.3811 \, r_{100}^2 + 0.02074 \, r_{100}^3
  \over
    1 + 0.2073 \, r_{100} + 0.08386 \, r_{100}^2
  }
\end{eqnarray}
where $r_{100}$
is the comoving depth in units of $100 \, h^{-1} \Mpc$.
The assumed redshift-distance relation is that of a flat $\Lambda$CDM
model with $\Omega_{\rmn m} = 0.3$, $\Omega_\Lambda = 0.7$,
in which comoving distance $r$ (in velocity units)
is related to redshift $z$ by
\begin{equation}
  r = {c \over 3 \, \Omega_\Lambda^{1/6} \Omega_{\rm m}^{1/3}}
    \left[ B\left( \Omega_\Lambda(z), 1/6, 1/3 \right)
         - B\left( \Omega_\Lambda, 1/6, 1/3 \right) \right]
\end{equation}
where $c$ is the speed of light,
$B(x,a,b) \equiv \int_0^x t^{a-1} (1-t)^{b-1} \, dt$
is the incomplete Beta function,
and
$\Omega_\Lambda(z) = \Omega_\Lambda/[\Omega_\Lambda + \Omega_{\rm m} (1+z)^3]$
is the density of vacuum energy as a function of redshift.

Measurement of the selection function as described above
determines its shape, but not the overall normalization
(Binggeli, Sandage \& Tammann 1988).
The normalization factor is measured here as one of the parameters
of the linear method of HTP.
The fitting function~(\ref{sffit}) is thus maximum likelihood
not only with respect to the shape, but also with respect to the normalization.
The measured normalization depends mainly on the amplitude of the `mean mode'
(the mode whose angular shape is the cut monopole,
and whose radial shape is that of the selection function),
but it self-consistently incorporates information from the amplitudes of
all other linear modes.

A difficulty one encounters in implementing
a maximum likelihood fit to the selection function,
per Sandage et al. (1979),
is that there are many spurious non-smooth solutions
that wiggle fiercely and look awful,
but nevertheless have formally greater likelihood than the desired smooth
solutions.
This strange behaviour can be traced to the fact that the `exact' solution
for the luminosity function and galaxy density is a sum of delta-functions.
Formally,
the `exact' step-function solution has infinitely
greater likelihood than any smooth solution.
Increasing the number of parameters in the fitting function
increases the tendency for the maximum likelihood solution
to slide off into a spurious non-smooth solution.
To reduce this instability,
we start by carrying out a simplified least squares fit
to the `exact' selection function,
since least squares quickly finds an approximate fit without
serious problems of stability.
The resulting approximate values of the parameters of the fit provide
the starting point from which to search for the maximum likelihood solution.
Even so,
the maximum likelihood fitting becomes unstable with too many parameters.
The adopted fit~(\ref{sffit}) contains 7 free parameters,
and formally all of these are significant;
for example,
increasing the number of parameters from 5 to 7 increases
the log-likelihood by $\Delta \ln {\cal L} = 7$.
However, we could not increase the number of parameters beyond this
without the solution veering into instability.
We interpret this behaviour as suggesting that,
among functions of its form,
equation~(\ref{sffit})
almost exhausts the possibilities for finding a better smooth fit.

As found by Saunders et al.\ (1990)
in the case of the QDOT survey (the 1-in-6 precursor to PSCz),
measurement of the selection function yields evidence for what appears to be
strong evolution,
in the sense that galaxies used to be more numerous, or more luminous,
than they are now.
We choose to model evolution by pure luminosity evolution,
which is mathematically indistinguishable from a spectral $K$-correction.
Specifically, we adopt a luminosity-cum-spectral correction of the form
$K = (1+z)^\kappa$
in the relation
$F = K L / [ 4 \upi (1+z)^2 r^2 ]$
between the observed flux $F$,
luminosity $L$, redshift $z$, and comoving distance $r$
of a galaxy.
Figure~\ref{sf}
shows the observed number density of galaxies,
divided by the measured selection function,
both with evolution, $\kappa = 3.4$, and without, $\kappa = 0$.
The Figure shows that,
in the absence of an evolutionary correction,
the galaxy density appears to increase substantially with redshift.
The large degree of evolution is consistent with that
reported in QDOT by Saunders et al.\ (1990).
Actually,
a canonical {\it IRAS\/} galaxy spectrum $d L / d \nu \propto \nu^{-2}$
(Saunders et al.\ 1990)
would predict a spectral $K$-correction with $\kappa = -1$.
In that case,
the actual luminosity evolution would be one power steeper
than indicated in Figure~\ref{sf}.

The best fit value of the evolutionary exponent
$\kappa$ increases systematically as the flux limit is decreased,
from 1 at 1.2~Jy, to 2.9 at 0.75~Jy, to 3.4 at 0.60~Jy.
This suggests the possibility that at least part of the effect
may be caused not by evolution, but rather by Malmquist bias,
in which the increasing number of galaxies at fainter fluxes,
combined with random flux errors at the flux limit of the survey,
cause galaxies to fluctuate preferentially into rather than out of the survey.
Malmquist bias is expected to be most marked in more distant regions
of the survey, where the selection function is steepest.

Since galaxies which randomly fluctuate into the sample should
be clustered in the same way as galaxies which correctly belong
to the sample,
Malmquist bias should not bias measurement of the power spectrum,
so long as the bias is homogeneous over the sky.
As discussed in \S4.4 of Saunders et al.\ (2000),
Malmquist bias in the PSCz survey is probably inhomogeneous at some level,
notably because flux errors are higher in the 2HCON regions of the survey
than in the 3HCON regions.
However,
if inhomogeneous Malmquist bias were important,
then it should show up as an excess of angular power over radial power
at the largest scales.
The investigations of HTP reveal no strong excess of
angular power at large scales in the redshift distortions
either of the correlation function, Fig.~2 of HTP,
or of the power spectrum, Fig.~4 of HTP.
We tentatively conclude that inhomogeneous Malmquist bias
is not a major problem in the PSCz survey.

Besides evolution,
Figure~\ref{sf}
also suggests growing incompleteness at the greatest depths.
This may be presumed to be the incompleteness at high redshift
described in \S4.2 of Saunders et al.\ (2000),
associated with the policy
not to pursue redshifts of galaxies optically fainter than $b_J = 19.5^{\rm m}$.
Since this incompleteness is greater in regions of higher
optical extinction, and is systematic rather than random over the sky
(Fig~4. of Saunders et al.\ 2000),
we choose to cut the survey at
$10^{2.625} \, h^{-1} \Mpc \approx 420 \, h^{-1} \Mpc$,
as previously did HTP.
But whereas HTP set the lower depth limit at $1 \, h^{-1} \Mpc$,
here we choose the slightly more conservative lower limit of
$10^{0.625} \, h^{-1} \Mpc \approx 4.2 \, h^{-1} \Mpc$,
about a correlation length, to reduce `local bias' resulting from the fact
that we, sitting in a galaxy, the Milky Way, are not at a random location.

The angular and radial cuts leave $12\,446$ galaxies
(out of an original
%$15\,411$
$14\,677$ galaxies with redshifts) in the survey.

\xikfig

\section{Results}
\label{results}

\subsection{Real space power spectrum}
\label{realpk}

Figure~\ref{xik}
shows the real space power spectrum of the PSCz 0.6~Jy survey
with the high-latitude angular mask.
The values at linear scales are from HTP,
while those at nonlinear scales are measured as described in
Section~\ref{analysisn}.
The plotted values are tabulated in
Tables~\ref{xiktab} and \ref{xikdtab}.

At linear scales
Figure~\ref{xik}
shows both correlated and decorrelated power spectra,
as measured by HTP,
tabulated separately in Tables~\ref{xiktab} and \ref{xikdtab}.
The correlated power spectrum
is the one that emerges most directly from the data,
and in essence represents the power spectrum smoothed through
the Fourier transform of the optimally weighted survey window.
The errors in the correlated power spectrum are correlated.
The decorrelated power spectrum is partially deconvolved in such
a way that estimates of power at different wavenumbers are
uncorrelated with each other (Hamilton \& Tegmark 2000).
The decorrelated power spectrum is to be preferred,
if one wants to compare a model power spectrum to the PSCz data
at linear scales.

At nonlinear scales the power spectrum cannot be decorrelated sensibly
(unless it is first prewhitened -- see \S\ref{prewhiten})
so Table~\ref{xikdtab} lists the decorrelated power only at linear scales.
If one attempted to decorrelate the nonlinear power spectrum into a set
of uncorrelated band-powers, then the band-power windows would be so broad,
with almost cancelling positive and negative parts,
that it would be hard to interpret the band-powers as
representing the power spectrum in any meaningful way.

Integration over the (decorrelated) power spectrum yields an rms fluctuation
\begin{equation}
  \sigma_r \equiv
  \left\{ \int_0^\infty \left[ 3 j_1( k r ) \over k r \right]^2 P(k)
    \, {4\upi k^2 \dd k \over (2\upi)^3} \right\}^{1/2}
\end{equation}
in $r = 8 \, h^{-1} \Mpc$ radius spheres of
\begin{equation}
  \sigma_8 = 0.80 \pm 0.05
  \ .
\end{equation}

Figure~\ref{xik}
also shows the concordance model power spectrum of Tegmark et al.\ (2001),
nonlinearly evolved by the method of Peacock \& Dodds (1996).
Although the concordance model fits well at linear scales,
it evidently fails dismally at nonlinear scales.

In fact all Dark Matter (DM) models with constant galaxy-to-mass bias
-- to be precise, all DM models in the
Eisenstein \& Hu (1998, 1999) suites,
nonlinearly evolved by the method of Peacock \& Dodds (1996),
and all the Cold+Hot DM models of Ma (1998a,b),
all arbitrarily normalized
--
fail at nonlinear scales, with high confidence.

The concordance model illustrated in Figure~\ref{xik}
shows two characteristic features of all DM power spectra:
an inflection
(Gazta\~naga \& Juszkiewicz 2001)
at the linear-nonlinear transition scale
(here $k \sim 0.3 \,h \, \Mpc^{-1}$),
%where $\Delta(k)^2 \equiv [4\upi k^3 P(k)]/(2\upi)^3 \approx 1$,
and a turnover at the transition between
the nonlinear collapse and virialized regimes
(in the model at $k \sim 3 \,h \, \Mpc^{-1}$).
Instead, the observed PSCz power spectrum shows a near power law behaviour
$P(k) \sim k^{-1.5}$
over virtually the entire observed range.
The power law is not exact:
visually there appears to be a mild upward curvature of power
in the broad vicinity of $k \sim 2 \, h \, \Mpc^{-1}$.
But there is no prominent nonlinear inflection,
as there is in APM
(Gazta\~naga \& Baugh 1998;
Gazta\~naga \& Juszkiewicz 2001).

These conclusions are essentially the same as those previously arrived at by
Peacock (1997) and Jenkins et al.\ (1998).

While the disagreement between theory and observation may presage
a drastic failure of DM models, or of the Peacock-Dodds or Ma transformations,
it seems more likely that scale-dependent galaxy-to-mass bias is responsible.

To make theory and observation agree requires antibias at intermediate scales,
and positive bias at small scales,
as can be seen in Figure~\ref{xik}.
Remarkably,
precisely this type of behaviour is reproduced in some $N$-body experiments
(Kravtsov \& Klypin 1999;
Col\'{\i}n et al.\ 1999;		% 9809202
Benson et al.\ 2000),			% 9903343
and there is already vigorous theoretical effort to understand it
in terms of the way galaxies populate dark matter haloes
(Ma \& Fry 2000a,b; 			% 0001347, 0003343
Seljak 2000, 2001;			% 0001493, 0009016
Peacock and Smith 2000;			% 0005010
Scoccimarro et al.\ 2001).		% 0006319

%It should be cautioned that most of the signal at the smallest scales
%comes from the nearest part of the survey.
%Tripling the lower limit on depth from $10^{0.625} \approx 4.2 \, h^{-1} \Mpc$
%to $10^{1.125} \approx 13 \, h^{-1} \Mpc$,
%removes 455 galaxies and
%eliminates the signal at $k \ga 150 \, h \, \Mpc^{-1}$. % Big deal

We admit some frustration over our failure,
documented in Section~\ref{compress} and Appendix~\ref{failed},
to measure a positive definite covariance matrix
for the nonlinear power spectrum.
Without such a matrix,
and given the broad covariance of power in the nonlinear regime,
it is impossible to assess rigorously the statistical significance
of the tentative mild upward curvature of power near
$k \sim 2 \, h \, \Mpc^{-1}$.
If the off-diagonal elements of the covariance matrix are simply discarded
-- an inadmissible procedure, but no better option presents itself --
then the best single power law fit over the range
$k = 0.05$--$300 \, h \, \Mpc^{-1}$
is
(with $k$ measured in $h \, \Mpc^{-1}$)
\begin{equation}
\label{onepowerlaw}
  P(k) \approx 150 \ k^{-1.46} \, h^{-3} \Mpc^3
\end{equation}
with $\chi^2 = 25$ for 59 nominal degrees of freedom.
The low $\chi^2$ per degree of freedom is indicative of the high degree
of correlation of the nonlinear estimates of power,
not of the excellence of the fit.
The best fit to a sum of two power laws over
$k = 0.05$--$300 \, h \, \Mpc^{-1}$
is
(with $k$ measured in $h \, \Mpc^{-1}$)
\begin{equation}
\label{twopowerlaw}
  P(k) \approx
  \left( 72 \ k^{-1.72}
  +
  74 \ k^{-1.28} \right) \, h^{-3} \Mpc^3
\end{equation}
with $\chi^2 = 19$ for 57 nominal degrees of freedom.
The reduction of $\chi^2$ by 6 for 2 additional parameters
can by no means be construed as implying that the upward curvature of power is
statistically significant;
but there is a possibility that it may be statistically significant.
The exponents $-1.72$ and $-1.28$
in the two power law fit, equation~(\ref{twopowerlaw}),
may exaggerate slightly the asymptotic slopes of the power spectrum
at large and small scales:
the best fitting exponents to single power laws
at large,
$k = 0.05$--$2 \, h \, \Mpc^{-1}$,
and small,
$k = 2$--$300 \, h \, \Mpc^{-1}$,
scales
are $-1.53$ and $-1.37$
respectively.

\pairfig

\subsection{Power at the smallest scales}
\label{smallk}

How reliable are the measurements of power at the smallest scales,
$k \approx 300 \, h \, \Mpc^{-1}$?
Such scales correspond to separations of the order
of a galaxy size,
$\upi/k \approx 10 \, h^{-1} \kpc$.

Beyond the minimum depth of
$10^{0.625} \, h^{-1} \Mpc \approx 4.2 \, h^{-1} \Mpc$
considered in this paper,
there are 7 distinct pairs of galaxies with transverse separations
closer than
$10 \, h^{-1} \kpc$
(and redshift separations small enough that they are probably
physically associated),
mostly near the plane of the Local Supercluster.
%with a median depth of $8 \, h^{-1} \Mpc$.
There are a further 34 distinct pairs with transverse separations
in the interval
$10$--$30 \, h^{-1} \kpc$,
variously distributed over the sky
(3 of the 34 pairs actually live in 3 distinct triple systems).
%with a median depth of $19 \, h^{-1} \Mpc$.
The number of close pairs, though not large,
appears to be enough to provide a statistically significant sample.
%And although the sample of close pairs is somewhat
%concentrated toward the local overdensity,
%The fact that the close pairs are fairly well spread out,
%rather than being concentrated in one or two small clumps,
%suggests that the sample is not hopelessly biased.

%Nevertheless, one should bear in mind that the small scale power plotted
%in Figure~\ref{xik}
%is based on a relatively modest volume of the nearby Universe,
%which may not be representative of the Universe as a whole.

An important systematic effect arises from
{\it IRAS\/}'s
$\sim 1 \farcm 5$
angular resolution,
which is expected to lead to a deficiency of galaxy pairs
at small angular separations.
{\it IRAS\/} scanned roughly along lines of constant ecliptic longitude
(see e.g.
http:/$\!$/\discretionary{}{}{}www\discretionary{}{}{}.ipac\discretionary{}{}{}.caltech\discretionary{}{}{}.edu/\discretionary{}{}{}Outreach/\discretionary{}{}{}Gallery/\discretionary{}{}{}IRAS/\discretionary{}{}{}allsky.html),
and the angular resolution for a single scan was typically
$\sim 1 \farcm 5$ in-scan
by $\sim 4 \farcm 75$ cross-scan
(\S2.3 of Saunders et al.\ 2000).
As described in the {\it IRAS\/} Explanatory Supplement
(Beichman et al.\ 1988, \S V.H),
the resolution of the Point Source Catalog (PSC)
was improved
by combining several scans at neighbouring longitudes.
The selection rules for the PSC
impose an absolute lower limit on pair separation of
$0 \farcm 5$ in-scan
by $1 \farcm 5$ cross-scan,
although this limit is occasionally violated
because of variations in processing.
In the PSCz sample considered in this paper,
there are in practice 5 distinct pairs closer than $1 \farcm 5$,
though none closer than $0 \farcm 75$.

Figure~\ref{pair}
shows the distribution of close pairs
relative to a frame aligned with local ecliptic coordinates.
The Figure
shows that the effective resolution in the cross-scan direction
is substantially higher than the $4 \farcm 75$ single-beam resolution,
indicating that the PSC strategy of combining scans from neighbouring longitudes
was particularly effective in the cross-scan direction.
Indeed,
the Figure suggests that the resolution in the cross-scan (horizontal)
direction is if anything slightly higher than the resolution
in the in-scan (vertical) direction.
We have also checked the distribution of close pairs on the sky,
and find no tendency for close pairs to lie preferentially
near the ecliptic poles, where scans cross,
and where the angular resolution might be expected to be high
in all directions.

Given the evidence of Figure~\ref{pair},
we assume that the {\it IRAS\/} beam is effectively isotropic,
with an angular resolution of $1 \farcm 5$.

\xiksmallfig

Independent of any a priori knowledge about the {\it IRAS\/} beam,
the fact that there is a deficiency of close pairs in PSCz can be demonstrated
by looking at the distribution of pairs as a function of depth.
If the sample were complete,
then the distribution of pairs with depth
would be independent of pair separation,
at least at pair separations small compared to the scale of the survey.
Specifically,
the expected number of pairs in some prescribed volume $V$ of the survey is,
at pair separations small compared to the scale of the survey,
\begin{equation}
\label{pairdistr}
  \mbox{expected number of pairs} =
  C \sum_{\mbox{\small {gals $i$}}} \bar n(\r_i)
\end{equation}
where the sum is over all galaxies $i$ in the volume $V$,
the quantity
$\bar n(\r_i)$
is the selection function
at the position $\r_i$ of galaxy $i$,
and
$C = \int [1 + \xi(r)] \, \ddd r$
is an integral over pair separations in the interval of interest.
The distribution of pairs with depth is determined entirely by the factor
$\sum_i \bar n(\r_i)$,
which is independent of pair separation,
the factor $C$ being a constant for any specified interval of pair separations.

For the sample used in this paper,
the PSCz high latitude sample at comoving depths
$4.2$--$420 \, h^{-1} \Mpc$,
formula~(\ref{pairdistr}) predicts that the
10, 25, 50 (median), 75, and 90 percentile
depths of close pairs
% lnl.hist
should be
7.5, 11, 19, 36, and
$56 \, h^{-1} \Mpc$
respectively.
By comparison,
the median and maximum depths of the 7 pairs with transverse separation
$\le 10 \, h^{-1} \kpc$
are
8 and
$19 \, h^{-1} \Mpc$,
indicating a significant deficiency of pairs,
with of order 30 to 50 percent completeness,
while the median and maximum depths
of the 34 pairs with transverse separations
$10$--$30 \, h^{-1} \kpc$
are
18
and $56\, h^{-1} \Mpc$,
consistent with little or no deficiency,
of order 90 percent completeness.

The distribution of close pairs with depth is consistent
with the hypothesis that there is a cutoff at $\sim 1 \farcm 5$.
This angular separation corresponds to transverse separations of
8 and
$25 \, h^{-1} \kpc$
at the 50 and 90 percentile depths
19 and
$56 \, h^{-1} \Mpc$
of the survey.
Thus if pairs closer than $1 \farcm 5$ are missing, then
pairs at tranverse separation $8 \, h^{-1} \kpc$ should be 50 percent complete,
and pairs at transverse separation $25 \, h^{-1} \kpc$ should be 90 percent
complete.
These levels of completeness are consistent with those inferred for
observed pairs in the
$\le 10 \, h^{-1} \kpc$
and
$10$--$30 \, h^{-1} \kpc$
ranges of separation.

We choose to deal with the incompleteness
by imposing a sharp lower limit of $1 \farcm 5$
in the angular separation of pairs,
in both real and `background' pair counts.
Figure~\protect\ref{xiksmall}
compares the power spectra measured
with and without the $1 \farcm 5$ cutoff.
At the smallest scales,
the power spectrum without the cutoff is systematically lower
than the canonical power spectrum with the cutoff.

We caution that there is expected to be at least some incompleteness
in pairs at angular separations $\sim 1 \farcm 5$--$5^\prime$,
so our estimate of the power spectrum at the smallest scales
may be systematically underestimated.
We hesitate to attempt to correct for this residual incompleteness,
given the uncertainty in {\it IRAS\/}'s effective beam.

Whether the small scale power spectrum of PSCz galaxies
is systematically underestimated or not,
it demonstrates dramatically
that the power spectrum continues to small scales
with no hint of any turnover such as expected in the matter power spectrum.

\subsection{Comparison of methods}
\label{com}

\xikcomfig

Figure~\ref{xikcom}
compares the power spectrum
measured by the linear and nonlinear methods separately,
demonstrating good agreement between the two methods where they overlap,
around $k \sim 0.3 \, h \, \Mpc^{-1}$.
This agreement constitutes a powerful end-to-end test of both methods,
since they involve completely different approximations and computational
approaches.

Quantitative comparison is complicated by the fact that
the band-power windows have somewhat different shapes for the
linear and nonlinear methods.
Moreover the nonlinear method assumes a weaker prior,
since it allows higher harmonics of redshift power,
so the errors on the nonlinear estimates might be expected to be slightly
larger where both methods work well.
However, the agreement
is encouraging despite these differences.
For example,
the results for the two band-powers
adjacent to the linear-nonlinear boundary
are as follows.
For the band-power centred at
$k = 0.317 \, h \, \Mpc^{-1}$,
the linear and nonlinear methods yield
$P(k) = 917 \pm 109 \, h^{-3} \Mpc^3$
and
$P(k) = 908 \pm 190 \, h^{-3} \Mpc^3$
respectively,
a $1\%$ mismatch in power and a $74\%$ larger error for the nonlinear case.
Similarly,
for the band-power centred at
$k = 0.365 \, h \, \Mpc^{-1}$,
the linear and nonlinear methods yield
$P(k) = 674 \pm 85 \, h^{-3} \Mpc^3$
and
$P(k) = 702 \pm 102 \, h^{-3} \Mpc^3$
respectively,
a $4\%$ mismatch in power, and a $20\%$ larger error for the nonlinear case.
Tightening the nonlinear prior by reducing the maximum number
$\el_{\max}$ of harmonics, equation~(\ref{ladopt}),
reduces the error bars in the nonlinear case,
bringing them into closer agreement with the linear method.

At linear scales the nonlinear method breaks down,
in part because the plane-parallel approximation breaks down,
but also because the band-power window we have used at nonlinear scales,
$\sim k^n \e^{-k^2}$ with $n = 72$,
which has a fwhm of $\Delta\log k \approx 1/12$,
becomes too narrow in low wavenumber band-powers
to be resolved by the survey.
We assess the problem quantitatively
by introducing an explicit maximum pair separation of
$\approx 270 \, h^{-1} \Mpc$,
and computing the neglected contribution
to monopole power from separations exceeding the limit.
The neglected contribution increases with exponential rapidity
at large scales,
from a fractional correction of
$\sim 10^{-10}$ to the band-power at $\approx 0.3 \, h \, \Mpc^{-1}$,
to $\sim 10^{-3}$ at $\approx 0.2 \, h \, \Mpc^{-1}$,
to overwhelmingly dominant at $\approx 0.1 \, h \, \Mpc^{-1}$.
This explains why the power computed by the nonlinear method is
plotted only at $k \ga 0.2 \, h \, \Mpc^{-1}$ in Figure~\ref{xikcom}.

At nonlinear scales the linear method breaks down,
in part because both the assumption of Gaussian density fluctuations
and the linear model of redshift distortions fail,
but also because the number 4096 of Karhunen-Lo\`eve modes used by HTP is,
by design, sufficient to achieve good coverage of $k$-space only
up to $k \la 0.3 \, h \, \Mpc^{-1}$.
At larger wavenumbers the coverage of $k$-space becomes increasingly sparse.
This explains why the power computed by the linear method appears
to become noisier at $k \ga 0.5 \, h \, \Mpc^{-1}$,
and why it is plotted only to
$k \la 0.9 \, h \, \Mpc^{-1}$ in Figure~\ref{xikcom}.

Figure~\ref{xikcom}
also compares the power spectrum
measured by the nonlinear method using two different band-power windows,
$\sim k^n \e^{-k^2}$ with $n = 72$ and $n = 288$.
The high resolution band-powers, $n = 288$,
have resolution $\Delta\log k \approx 1/24$ fwhm twice
that of the low resolution band-powers, $n = 72$.
Evidently the two sets of band-powers yield results in good agreement.
We also experimented with $n = 648$,
which has three times the resolution of $n = 72$;
again the results were in good agreement.

We also computed
a power spectrum using the nonlinear method with $n = 72$
but with twice as many harmonics,
$\el_{\max} = 32 \, (k / 1 \, h \, \Mpc^{-1})^{1/2}$,
as the adopted maximum, equation~(\ref{ladopt}).
The power spectrum agrees well with the original calculation,
but we choose to omit it from Figure~\ref{xikcom} to avoid confusing the plot.

The maximum harmonic measurable
with a band-power $\sim k^n \e^{-k^2}$ is $\el = n$.
The concern with the low resolution band-powers, $n = 72$, is that
at large wavenumbers
there are not enough harmonics to resolve the expected hill
in the redshift power at $\mu = 0$,
the all-important place where redshift power equals real power.
In fact equation~(\ref{ladopt}) would suggest that,
in order to resolve redshift power satisfactorily,
harmonics
$\el > 72$ are required at $k \ga 20 \, h \, \Mpc^{-1}$,
with $\el \approx 284$ required
at $k = 316 \, h \, \Mpc^{-1}$.
One might anticipate that too few harmonics would tend to smooth out the hill,
hence bias the estimate of real power systematically low.
However,
Figure~\ref{xikcom} shows little sign that the lower resolution band-powers
with $n = 72$
are biased low compared to the higher resolution band-powers
with $n = 288$.
Some bias surely remains,
but it is apparently small compared to the statistical uncertainty.
Since the low resolution band-powers have smaller error bars
than the high resolution band-powers binned to the same resolution,
we prefer the low resolution $n = 72$ band-powers at all nonlinear scales.

\apmfig

\subsection{Comparison to APM}
\label{compareapm}

To date the best published measurement of the real space galaxy
power spectrum is that of the APM survey\footnote{
The APM power spectrum in the present paper is taken from Table~2
of Gazta\~naga \& Baugh (1998),
who state that their tabulated numbers are
essentially the same as those of Baugh \& Efstathiou (1993).
Eisenstein \& Zaldarriaga (2000)
have critiqued the error bars of Baugh \& Efstathiou (1993, 1994),
and to a lesser extent those of Dodelson \& Gazta\~naga (2000),
as overly optimistic,
mainly because of the neglect of covariances.
Unfortunately Eisenstein \& Zaldarriaga limit their analysis
to $k \la 0.8 \, h \, \Mpc^{-1}$,
so in the present paper we choose to quote
the power spectrum of Gazta\~naga \& Baugh (1998).
The Eisenstein \& Zaldarriaga power spectrum has factor of 2 larger
error bars, and scatters about more, than the Gazta\~naga \& Baugh spectrum,
but the two measurements are otherwise consistent with each other.
}
(Baugh \& Efstathiou 1993, 1994;
Maddox et al.\ 1996;
Gazta\~naga \& Baugh 1998, Table~2;
Dodelson \& Gazta\~naga 2000;
Eisenstein \& Zaldarriaga 2000).

As discussed by Baugh \& Efstathiou (1993)
and
Eisenstein \& Zaldarriaga (2000),
the APM survey has a median depth in redshift of $z \approx 0.11$,
and transforming the power spectrum to zero redshift depends on cosmology.
The main effect is that the redshift-distance relation is
different in different cosmologies.
The canonical APM power spectrum quoted by
Baugh \& Efstathiou, Gazta\~naga \& Baugh,
and Eisenstein \& Zaldarriaga
assumes a flat matter-dominated cosmology, $\Omega_{\rmn m} = 1$.
In a $\Lambda$CDM cosmology, $\Omega_{\rmn m} = 0.3$, $\Omega_\Lambda = 0.7$
(as assumed for the redshift-distance relation in our PSCz measurements),
the power spectrum would be $\sim 20\%$ higher.
Following Peacock (1997),
we renormalize the APM power spectrum upward by a factor $1.25$,
which according to Peacock brings it into agreement with the
real space correlation function of the APM-Stromlo survey
(Loveday et al.\ 1995).

Figure~\ref{apm}
compares the real space power spectrum of PSCz to that of APM.
The relative bias between APM and PSCz,
defined as the square root of the ratio of their power spectra,
reveals a suggestively simple pattern.
At linear scales
$k \la 0.3 \, h \, \Mpc^{-1}$
the relative bias is approximately constant,
$b_{\rmn APM}/b_{\rmn PSCz} \approx 1.15$.
At transition scales
$k \sim 0.3$--$1.5 \, h \, \Mpc^{-1}$
the APM to PSCz bias increases,
settling down at nonlinear scales
$k \ga 1.5 \, h \, \Mpc^{-1}$
to another constant,
$b_{\rmn APM}/b_{\rmn PSCz} \approx 1.4$.

Intriguingly, the APM to PSCz bias would have been close to unity
at linear scales if we had {\em not\/} renormalized the APM power spectrum
by Peacock's factor $1.25$.
However, we are persuaded that it is correct to renormalize.

\xikindfig

The fact that APM to PSCz bias is consistent with being constant at linear
scales is an encouraging confirmation of the prediction of local bias models,
that bias at large, linear scales should be scale-independent
(Coles 1993;
Fry and Gazta\~naga 1993;
Scherrer \& Weinberg 1998;
Coles, Melott \& Munshi 1999;
Heavens, Matarrese \& Verde 1999).
Scale-independence of bias at linear scales
is also a feature of $N$-body experiments
(Kravtsov \& Klypin 1999;
Col\'{\i}n et al.\ 1999;		% 9809202
Narayanan, Berlind \& Weinberg 2000;	% 9812002
Benson et al.\ 2000).			% 9903343

\subsection{Power spectra from individual FKP weightings}
\label{individual}

Figure~\ref{xikind}
compares the power spectra
measured from the five individual FKP pair-weightings
(\S\ref{FKP}),
with FKP constants $J = 0$, $10$, $10^2$, $10^3$,
and $10^4 \, h^{-3} \Mpc^3$,
equation~(\ref{J}).
To show more detail, the lower panel of Figure~\ref{xikind} shows
the bias of the power spectra,
defined here to be the square root of the ratio of the
power spectrum to the standard power spectrum of PSCz
plotted in Figure~\ref{xik} and
tabulated in Table~\ref{xiktab}.
Figure~\ref{xikind}
demonstrates that there is a general consistency between the power spectra
measured with different pair-weightings.

\xiknoprewhfig

Larger FKP constants $J$ give greater effective weight to more distant
regions of the survey, hence to more luminous galaxies.
Figure~\ref{xikind}
gives weak indication that power spectra measured with larger FKP constants
have higher bias over the range $\sim 2$--$20 \, h \, \Mpc^{-1}$,
which in turns suggests weakly that {\it IRAS}-luminous galaxies may be more
clustered than less luminous galaxies at these scales.
If this is correct,
then it would suggest that the power spectrum of the more luminous
{\it IRAS\/} galaxies may be similar to the power spectrum of APM galaxies,
Figure~\ref{apm}.

One should be careful not to overinterpret Figure~\ref{xikind}.
The fact that measurements for $J = 10^2$ and $10^3 \, h^{-3} \Mpc^3$
appear systematically high, at the $1$--$2 \sigma$ level,
over the range $k \sim 3$--$10 \, h \, \Mpc^{-1}$,
might suggest that the difference is statistically significant.
However,
the power spectrum is highly correlated over this range,
as seen in Figure~\ref{ccw} below,
and the significance is more marginal than it appears.

Figure~\ref{xikind}
also gives some suggestion that power spectra with larger FKP constants $J$
may switch to being biased low at smaller scales, $k \ga 20 \, h \, \Mpc^{-1}$.
%suggesting that {\it IRAS}-luminous galaxies are actually less
%clustered at tiny scales.
However, as is evident from the errors bars in the lower panel of
Figure~\ref{xikind},
the noise is really too great to tell.
%What happens is that
%the individual power spectra with larger $J$ appear to die to zero
%at small scales, at the same time becoming rather noisy.
%Despite the noise, this may be a real effect,
%caused by exclusion of close pairs, either physical or instrumental.
%The smallest wavenumber measured here, $k \sim 300 \, h \, \Mpc^{-1}$,
%corresponds to about the physical size of a galaxy,
%$\sim \upi/(300 \, h \, \Mpc^{-1}) \approx 10 \, h^{-1} \kpc$,
%so physical exclusion may affect power at the smallest scales.
%On the instrumental side,
%{\it IRAS\/}'s $\sim 1 \farcm 5$ angular resolution
%(Saunders et al.\ 2000, \S2.3)
%should be able to resolve pair separations
%$10 \, h^{-1} \kpc$
%to a distance of $\sim 20 \, h^{-1} \Mpc$.
%%$\upi/( 20 \, h \, \Mpc^{-1} ) \approx 0.15 \, h^{-1} \Mpc$
%%to a distance of $\sim 300 \, h^{-1} \Mpc$.
%It seems plausible that some merging of {\it IRAS\/} images
%occurs in more distant, luminous galaxies,
%which might contribute to the observed exclusion effect.
%We will return to this issue in \S\ref{smallk}.

As discussed in Section~\ref{compress},
instead of compressing the five FKP-weighted estimates of each band-power
directly, we first prewhiten the power, then compress, then unprewhiten,
since in theory it is better to apply an FKP-like weighting
to almost uncorrelated measure like the prewhitened power (H00).
The general effect of prewhitening before compressing
is to prefer smaller FKP constants $J$,
i.e.\ to give relatively more weight to nearer, less luminous galaxies.
Figure~\ref{xiknoprewh}
shows the power spectra measured
both with and without prewhitening before compression.
The consequence on the power spectrum is for the most part small.
The most noticeable effect is what might be expected
on the basis of Figure~\ref{xikind}:
prewhitening before compressing
decreases power by $\sim 10\%$ over the range
$k \sim 3$--$10 \, h \, \Mpc^{-1}$.

Perhaps the greatest concern over luminosity-dependent bias
is that it could bias the estimation of cosmological parameters.
If more luminous galaxies are more clustered,
then estimates of power at large scales,
which depend more on distant, luminous galaxies,
would be biased upward,
giving the power spectrum a false red tilt.
Encouragingly,
Figure~\ref{xikind} shows no evidence of significant luminosity bias at scales
$k \la 1 \, h \, \Mpc^{-1}$.
Although these measurements are restricted to the nonlinear regime,
they do suggest that luminosity bias is probably not a major effect
on the cosmological parameter analysis of
Tegmark et al.\ (2001),
which used PSCz data only at linear scales
$k < 0.3\, h \, \Mpc^{-1}$.

Three recent studies,
by
Beisbart \& Kerscher (2000), % 0003358
Szapudi et al.\ (2000),	% 0007243
and
Hawkins et al.\ (2001),
have found no evidence of significant difference between the clustering of
luminous and faint galaxies in the PSCz survey.
Our results,
while not constituting a formal study of differential biasing with luminosity,
are consistent with the conclusions of these authors.

\subsection{Real space correlation function}
\label{corrfn}

\xirfig

The correlation function $\xi(r)$ remains one of the most
popular statistics for characterizing large scale structure
(Peebles 1980).

Figure~\ref{xir}
shows the real space correlation function of PSCz,
obtained as the Fourier transform of the real space power spectrum
shown in Figure~\ref{xik}.
The covariance properties of the correlation function $\xi(r)$ are less
than ideal,
since there are broad correlations between estimates
at different pair separations $r$.
We make no attempt at a rigorous treatment of errors,
and instead simply show in Figure~\ref{xir}
the envelope defined by the Fourier transforms of the
correlated power spectrum and its $\pm 1 \sigma$ extremes.

Table~\ref{xirtab}
tabulates the correlation function $\xi(r)$,
the Fourier transform of the correlated power $P(k)$ from Table~\ref{xiktab},
and the correlation functions $\xi_-(r)$ and $\xi_+(r)$ which are
the Fourier transforms of the $\pm 1 \sigma$ extremes $P(k) \pm \Delta P(k)$
of the correlated power from Table~\ref{xiktab}.
Notice that $\xi_-$ is not always less than $\xi_+$,
and that $\xi_-$ and $\xi_+$
do not necessarily encompass the central value $\xi$.

We Fourier transform the power spectrum to the correlation function
using the fast, logarithmically-spaced
Fourier-Hankel method of Talman (1978),
as implemented in the \FFTLog\ code described in Appendix~B of
H00, and available at
http:/$\!$/\discretionary{}{}{}casa\discretionary{}{}{}.colorado\discretionary{}{}{}.edu/\discretionary{}{}{}$\sim$ajsh/\discretionary{}{}{}FFTLog/.
Besides being able to cover a broader range of scales,
the logarithmic FFT has the advantage
that it does not suffer from the serious problem of ringing that
afflicts the normal FFT when applied to cosmological power spectra
(H00, Fig.~12).

To avoid artefacts
arising from the periodicity in log space assumed by \FFTLog,
we padded the power spectrum with a power law at each end
to quadruple (double would have sufficed) the logarithmic interval,
$P(k) \propto k$ to $k = 10^{-9} \, h \, \Mpc^{-1}$,
and
$P(k) \propto k^{-1.4}$ to $k = 10^{9} \, h \, \Mpc^{-1}$.
We then applied the most straightforward version of the \FFTLog\ transform,
i.e.\ no power-law bias ($q = 0$), and a low-ringing value of $kr$.
Finally, we retained only the central part of the correlation function
$\xi(r)$, from $r = 0.01$ to $300 \, h^{-1} \Mpc$.

A by-eye fit of the resulting correlation function
to a power-law yields
$\xi(r) \approx (r/r_0)^{-\gamma}$
with correlation length
$r_0 = 4.27 \, h^{-1} \Mpc$
and index
$\gamma = 1.55$
over the range $r = 0.01$--$20 \, h^{-1} \Mpc$.
The fit is illustrated in Figure~\ref{xir}.
The correlation function is a factor $\approx 1.2$ higher than,
but has about the same slope as,
the correlation function measured by
Saunders et al.\ (1992),
who found
$r_0 = 3.79 \pm 0.14 \, h^{-1} \Mpc$
and
$\gamma = 1.57 \pm 0.03$
over pair separations $r = 0.1$--$20 \, h^{-1} \Mpc$
from a power law fit to the projected cross-correlation function
between the QDOT survey (the 1-in-6 precursor to PSCz)
and its parent QIGC angular catalogue.
Our power-law fit is also higher, but slightly shallower, than that of
Fisher et al.\ (1994a),
who inferred
$r_0 = 3.76^{+0.20}_{-0.23} \, h^{-1} \Mpc$
and
$\gamma = 1.66^{+0.12}_{-0.09}$
over $r = 1$--$20 \, h^{-1} \Mpc$
from a power law fit to the projected correlation function
of the {\it IRAS} 1.2~Jy survey.

Fitting by eye is not satisfactory,
but as in the case of the power spectrum,
discussed at the end of Section~\ref{realpk},
our attempt to carry out rigorous fits is thwarted
by the fact that the covariance matrix measured at nonlinear scales
is not positive definite
(see \S\ref{compress} and Appendix~\ref{failed}).
The best that we have been able to do in terms of rigorous
fitting at nonlinear scales is discussed in the following subsection,
on the prewhitened power spectrum.

\subsection{Prewhitened power spectrum}
\label{prewhiten}

\xiksfig

\ccwfig

Nonlinear evolution induces broad correlations
between estimates of power at different wavenumbers
(Meiksin \& White 1999;
Scoccimarro, Zaldarriaga \& Hui 1999;
H00).
In effect, nonlinear evolution blurs whatever information
may have been present in the linear power spectrum,
such as baryonic wiggles
(Meiksin, White \& Peacock 1999).

H00 showed that
prewhitening (\S\ref{prewhitenedpower}) the nonlinear power spectrum
-- transforming the power in such a way that the shot noise contribution
to the covariance is proportional to the unit matrix --
appears empirically to narrow the covariance of power substantially.
The extent to which the prewhitened nonlinear power spectrum
may be a better carrier of information than the nonlinear power itself
remains to be explored,
but whatever the case,
the prewhitened power spectrum is less correlated,
and therefore should offer
better control of errors in fitting to cosmological models.

Figure~\ref{xiks}
shows the prewhitened power spectrum of PSCz,
and Table~\ref{xikstab}
tabulates the corresponding values.
Figure~\ref{xiks}
also shows the linear (not nonlinear) concordance model power spectrum
from Figure~\ref{xik}.
As remarked by H00,
the prewhitened nonlinear power spectrum appears intriguingly similar to
the underlying linear power spectrum, for realistic power spectra.
It is not clear whether the similarity has some physical cause,
or whether is is merely coincidental.

At linear scales,
the prewhitened power plotted in Figure~\ref{xiks}
has been explicitly decorrelated (Hamilton \& Tegmark 2000),
so that each point is uncorrelated with every other.
The (unprewhitened) power spectrum shown in Figure~\ref{xiks}
is the one that, when prewhitened, yields the plotted decorrelated
prewhitened spectrum.
The (unprewhitened) power in Figure~\ref{xiks} is not the same as either
the correlated or uncorrelated powers shown in Figure~\ref{xik};
rather, it is that power which becomes decorrelated after being prewhitened.

We also tried decorrelating the prewhitened power at nonlinear scales,
but the measured prewhitened covariance matrix proved too noisy
to admit believable decorrelation band-powers
(\S\ref{compress}).
While the prewhitened powers at nonlinear scales are therefore
somewhat correlated,
it would be not unreasonable to treat them
as being uncorrelated, or nearly so, in fitting to theoretical models.

With the points treated as uncorrelated,
a power law fit to the prewhitened power spectrum at
%fully nonlinear scales,
%$k = 1$--$300 \, h \, \Mpc^{-1}$,
nonlinear scales,
$k = 0.3$--$300 \, h \, \Mpc^{-1}$,
yields
\begin{equation}
\label{Xfit}
  X(k) =
%   ( 59.0 \pm 4.1 ) \, (k/1 \, h \, \Mpc^{-1})^{-2.22 \pm 0.05}
%   ( 6.47 \pm 0.31 ) \, (k/2.7 \, h \, \Mpc^{-1})^{-2.225 \pm 0.05}
    ( 18.0 \pm 0.7 ) \, (k/1.7 \, h \, \Mpc^{-1})^{-2.16 \pm 0.04}
    h^{-3} \Mpc^3
\end{equation}
%with $\chi^2 = 35.8$ for $39$ degrees of freedom.
%The pivot point $k = 2.7 \, h \, \Mpc^{-1}$ of the fit
with $\chi^2 = 34.8$ for $46$ degrees of freedom.
The pivot point $k = 1.7 \, h \, \Mpc^{-1}$ of the fit
in equation~(\ref{Xfit})
is chosen so that the error bars on the amplitude
and exponent of the fit are essentially uncorrelated.
The $\chi^2$ per degree of freedom of $35/46$ is closer to one
than the $25/59$ for the power law fit to the power spectrum reported
in Section~\ref{realpk},
but the $\chi^2$ remains lower than expected for uncorrelated points,
suggesting that there remains some residual correlation
in the estimates of prewhitened power.

The fitted nonlinear slope
%$-2.225 \pm 0.05$
$-2.16 \pm 0.04$
of the prewhitened power
would predict that the prewhitened correlation function
would have a nonlinear slope of
%$3 + (- 2.225 \pm 0.05) = 0.775 \pm 0.05$.
$3 + (- 2.16 \pm 0.04) = 0.84 \pm 0.04$.
According to the defining equation~(\ref{Xr}),
the (unprewhitened) correlation function would then have a nonlinear slope of
%$\gamma = 2 \times ( 0.775 \pm 0.05 ) = 1.55 \pm 0.10$,
$\gamma = 2 \times ( 0.84 \pm 0.04 ) = 1.68 \pm 0.08$.
%consistent with the by-eye fit
%$\gamma \approx 1.52$
This is slightly steeper than the by-eye slope of
$\gamma \approx 1.55$
fitted to the correlation function
in Section~\ref{corrfn}.
Similarly,
the nonlinear slope of the prewhitened power would predict that
the power spectrum would have a nonlinear slope of
%$(1.55 \pm 0.10) - 3 = - 1.45 \pm 0.10$.
$(1.68 \pm 0.08) - 3 = - 1.32 \pm 0.08$,
somewhat shallower than the slope $-1.46$ fitted directly to the power spectrum,
equation~(\ref{onepowerlaw}).
However,
as discussed in Section~\ref{realpk},
there is some suggestion that the power spectrum flattens to smaller scales,
and the shallower slope predicted by the prewhitened power spectrum
is consistent with such a flattening.

Figure~\ref{ccw}
shows the correlations between estimates of power,
and between estimates of prewhitened power,
measured in the PSCz survey.
The plotted quantity is the correlation coefficient
$\mxC_{kk'}/(\mxC_{kk}^{1/2} \mxC_{k'k'}^{1/2})$,
which the Schwarz inequality implies must lie between
$-1$ (perfect anti-correlation)
and
$1$ (perfect correlation).
The covariances $\mxC_{kk'}$ of power estimates are measured
from the fluctuations in the PSCz data themselves (\S\ref{covariance}),
and are essentially free from prior assumption.
The measurements properly take into account the correlation
between different subregions of the survey.

Figure~\ref{ccw} confirms that prewhitening the power spectrum
narrows its covariance.
However,
the narrowing is not as good as found in analytic models by H00,
and we confess some disappointment at the result.
One unexpected feature of the covariance plotted in Figure~\ref{ccw}
is that the power at
$k' = 32 \, h \, \Mpc^{-1}$
appears somewhat anti-correlated with power at $\sim 5 \, h \, \Mpc^{-1}$.
We have no explanation for this.

\section{Conclusions}
\label{conclusions}

\subsection{What we have done}

The paper combines two separate measurements at linear and nonlinear scales
to yield a measurement of the real space power spectrum
of the {\it IRAS} PSCz 0.6~Jy survey
(Saunders et al.\ 2000)
over four and a half decades of wavenumber.
The linear measurement
(HTP)
assumes Gaussian fluctuations and that redshift distortions
conform to the linear model,
while the nonlinear measurement assumes the plane-parallel approximation,
and infers the real space power spectrum
from the redshift space power spectrum in the transverse direction.
The measurements are tabulated in an Appendix.

At nonlinear scales the power spectrum
is broadly correlated over different wavenumbers,
which not only blurs the information content of the power spectrum,
but also complicates rigorous comparison to cosmological models.
We therefore also report a measurement of the prewhitened power spectrum
of PSCz, which is less correlated than the (nonlinear) power
spectrum itself.
To assist the reader,
Appendix~\ref{howtoprewhiten}
contains practical instructions on how to prewhiten a power spectrum.

\subsection{Methodology}

We have shown how to exploit galaxy redshifts to measure the real space
power spectrum with accuracy comparable to that attainable
from an angular survey many times larger.
%Absent a reliable theory of nonlinear redshift distortions,
%we have adopted a weak prior assumption that the redshift space power
%$P^s(\k)$ is, at each wavenumber $k$,
%a smooth function of the cosine $\mu \equiv\hat\k.\hat\z$
%between the wavevector $\k$ and the line-of-sight $\z$,
%and we have used the measured redshift power to inform
%the appropriate level of smoothing.

We have successfully applied H00's proposal
to reduce the degree of correlation of the nonlinear power spectrum
by prewhitening it.
Statistical uncertainties in the covariance matrix of power estimates
prevented complete decorrelation of the prewhitened nonlinear power spectrum.
More reliable models, coupled with more precise measurements,
of nonlinear covariance could permit full decorrelation in future analyses.

By combining separate methods at linear and nonlinear scales, 
the present work completes the two-pronged program envisaged by 
Tegmark et al.\ (1998).
%and further extends it to 
%incorporate redshift distortions and prewhitening.
The fact that there is a range of scales where the two methods
overlap and agree well suggests that this two-pronged
approach should be fruitful for ongoing projects such as
the 2dF Survey and the Sloan Digital Sky Survey.

\subsection{What the results show}

The relative bias between optically-selected APM galaxies
and {\it IRAS}-selected PSCz galaxies
is consistent with being constant at linear scales,
with $b_{\rmn APM}/b_{\rmn PSCz} \approx 1.15$.
The relative bias then rises
to a second plateau
$b_{\rmn APM}/b_{\rmn PSCz} \approx 1.4$
at nonlinear scales $k \ga 1.5 \, h \, \Mpc^{-1}$.
This is essentially the same behaviour as found by Peacock (1997).

All Dark Matter models
predict an inflection in the matter power spectrum at the transition
between the linear and nonlinear regimes at $k \sim 0.3 \, h \, \Mpc^{-1}$,
and a turnover at the transition from nonlinear collapse to
the virialized regime at $k \sim 3 \, h \, \Mpc^{-1}$.
The PSCz galaxy power spectrum shows neither of these features,
but instead displays a near power-law behaviour
to the smallest scales measured, with possible mild upward curvature 
in the broad vicinity of $k \sim 2 \, h \, \Mpc^{-1}$.
Short of a drastic revision of the current rather successful
cosmological paradigm,
the PSCz nonlinear power spectrum requires scale-dependent galaxy-to-mass bias:
all Dark Matter models without scale-dependent bias are ruled out
with high confidence.

We caution that it is possible that we have underestimated the
PSCz power spectrum systematically at the smallest scales,
$k \ga 100 \, h^{-1} \Mpc$,
because {\it IRAS\/}'s $\sim 1 \farcm 5$ resolution
causes it to miss pairs at the smallest angular separations.
We have attempted to remove most of the systematic
by imposing a lower cutoff of $1 \farcm 5$ in angular separation,
but it is possible that a small residual systematic remains.

The measured nonlinear power spectrum of PSCz clearly contains
valuable information about galaxy-to-mass bias,
and it will be a challenge for $N$-body experiments to reproduce,
and for theories to explain,
the observed power spectra of both {\it IRAS}-selected
and optically-selected galaxies
(White et al. 1987;
Kravtsov \& Klypin 1999;
Col\'{\i}n et al.\ 1999;		% 9809202
Narayanan, Berlind \& Weinberg 2000;	% 9812002
Benson et al.\ 2000;			% 9903343
Ma \& Fry 2000a,b;			% 0001347, 0003343
Seljak 2000, 2001;			% 0001493, 0009016
Peacock and Smith 2000;			% 0005010
Scoccimarro et al.\ 2001).		% 0006319
Because of the wide lever arm in wavenumber,
it is possible that even fairly rudimentary models of nonlinear bias may 
allow interesting constraints to be placed
on certain cosmological parameters, for instance on the primordial
scalar spectral index $n$,
or on deviations from power law behaviour in the primordial spectrum.

%We conclude with a puzzle. 
If the Dark Matter paradigm is correct,
then the fact that the observed power spectrum of PSCz galaxies
is close to a power law over four orders of magnitude in wavenumber
results from a cosmic conspiracy where the funny features in the nonlinear
matter power spectrum are accurately cancelled by scale-dependent bias.
It remains to be seen whether this is merely a cosmic coincidence 
or a hint of interesting underlying physics.

\section*{Acknowledgements}

We thank Chung-Pei Ma and Simon White for helpful comments,
and the referee Will Saunders for many wise suggestions,
notably for emphasizing the important effect of
{\it IRAS\/}'s finite angular resolution on pairs at small separations.
Special thanks to the PSCz team for publishing the data
from this superb survey on a timely basis.
This work was supported by
NASA ATP grant NAG5-7128, NASA LTSA grant NAG5-6034,
NSF grant AST00-71213,
and the University of Pennsylvania Research Foundation.

\appendix

\section{How to prewhiten the power spectrum}
\label{howtoprewhiten}

This part of the Appendix offers some practical hints
on how to prewhiten a power spectrum numerically.
We have had success with two different methods, described below.
The first method uses a logarithmic Fast Fourier Transform technique
to go from Fourier space to real space and back again,
while the second uses a matrix method that works entirely in Fourier space.
The two methods can provide a useful numerical check on each other.

\subsection{How to prewhiten power: Fourier method}
\label{prewhiten1}

The method is:
\begin{itemize}
\item
Fourier transform the power spectrum $P(k)$
to obtain the correlation function $\xi(r)$;
\item
Transform the correlation function $\xi(r)$
to the prewhitened correlation function $X(r)$
in accordance with equation~(\ref{Xr});
\item
Fourier transform the prewhitened correlation function $X(r)$
back to obtain the prewhitened power spectrum $X(k)$.
\end{itemize}
We strongly recommend using the logarithmic FFT
(Talman 1978; H00, Appendix B),
since the normal FFT suffers from serious ringing
when applied to realistic cosmological power spectra
(see Fig.~12 of H00).
Whereas the normal FFT works on linearly spaced points,
the logarithmic FFT,
which we have implemented in a code \FFTLog\ available at
http:/$\!$/\discretionary{}{}{}casa\discretionary{}{}{}.colorado\discretionary{}{}{}.edu/\discretionary{}{}{}$\sim$ajsh/\discretionary{}{}{}FFTLog/,
works on logarithmically spaced points,
easily covering ranges of orders of magnitude
in wavenumber or pair separation with modest numbers of points.

The logarithmic FFT assumes that the function (times some power law)
is periodic in the log.
To reduce artefacts arising from periodicity,
we recommend padding the power spectrum at large and small scales
(for example with a power law $\propto k$ at large scales
and a power law $\propto k^n$ with $n \sim -1.5$ to $-3$ at small scales)
to double the logarithmic range of interest,
and then retaining only the central half of the transformed sequence.

The \FFTLog\ code contains some options.
We recommend the simplest choices, a zero bias exponent $q = 0$,
and a low-ringing value of the relative phasing $kr$
of the $k$ and $r$ logarithmic sequences.

Warning (cf.\ H00, \S4.1):
to avoid artefacts of ringing and aliasing,
the Fourier method should {\em not\/}
be applied over a narrow range of wavenumbers without padding.

\subsection{How to prewhiten power: matrix method}
\label{prewhiten2}

\Xsfig

If for some reason the Fourier method of \S\ref{prewhiten1} is inconvenient,
then the matrix method offers an alternative.
The method is:
\begin{itemize}
\item
Construct the Fourier space version of the matrix
which in real space is diagonal with diagonal entries
$2/\{1 + [1 + \xi(r)]^{1/2}\}$;
\item
Apply this matrix to the power spectrum $P(k)$.
\end{itemize}
Figure~\ref{Xs}
illustrates that the prewhitening matrix in Fourier space
looks essentially like a high-pass filter,
which passes high frequency oscillations in the power,
while reducing any smoothly varying component.

Let $A(r_\alpha,r_\beta)$ denote the matrix which is diagonal in real space
with diagonal entries $\xi(r)$
(H00, eq.\ 58):
\begin{equation}
\label{Ar}
  A(r_\alpha,r_\beta)
    = \deltaD(r_\alpha{-}r_\beta) \, \xi(r_\alpha)
  \ .
\end{equation}
Here $\deltaD(r_\alpha{-}r_\beta)$ is the unit matrix in real space,
a 3-dim\-en\-sional delta function in pair separation $r$,
satisfying $\int \deltaD(r) \, 4\upi r^2 \dd r = 1$.
In the Fourier representation the matrix $A(r_\alpha,r_\beta)$
transforms to
(H00, eq.\ 59):
\begin{equation}
\label{Ak}
  A(k_\alpha,k_\beta)
    = {1 \over 2 \, k_\alpha k_\beta}
      \int_{|k_\alpha-k_\beta|}^{k_\alpha+k_\beta} \!\!\!
      P(k) \, k \dd k
  \ .
\end{equation}
To allow it to be manipulated numerically,
the continuous matrix $A(k_\alpha,k_\beta)$ must be discretized.
To ensure that matrix operations
(matrix multiplication, inversion, diagonalization, etc.)
work in the usual way,
discretization must be done
in such a way that the inner product in continuous Fourier space,
$\int \ddd k/(2\upi)^3$,
translates into ordinary summation in the discrete space
(H00, \S2.3).
This leads to the discretization algorithm:
for each index, $\alpha$, on a vector, matrix, or tensor,
multiply by the square root of the Fourier volume element,
$\Delta V_\alpha^{1/2}$.
Thus $A(k_\alpha,k_\beta)$ should be discretized by
multiplying it by
$(\Delta V_\alpha \Delta V_\beta)^{1/2}$:
\begin{equation}
  \mxA_{\alpha \beta} = A(k_\alpha,k_\beta) \,
  (\Delta V_\alpha \Delta V_\beta)^{1/2}
% (4 \upi k_\alpha^{3/2} k_\beta^{3/2} \Delta\ln k )/(2\upi)^3
\end{equation}
(no implicit summation).
If, for example, points in $k$-space are logarithmically spaced
with spacing $\Delta\ln k$,
then the Fourier volume element is
\begin{equation}
  \Delta V_\alpha = 4\upi k_\alpha^3 \Delta\ln k/(2\upi)^3
  \ .
\end{equation}

 From the discretized matrix $\mxA_{\alpha\beta}$,
construct the prewhitening matrix
\begin{equation}
\label{prewhmx}
  2 \, \left[\mxunit + (\mxunit + \mxA)^{1/2}\right]^{-1}_{\alpha\beta}
  \ .
\end{equation}
This involves the operations:
(1) add the unit matrix $\mxunit_{\alpha\beta}$ to $\mxA_{\alpha\beta}$;
(2) take the square root of the resulting matrix,
$(\mxunit + \mxA)^{1/2}_{\alpha\beta}$,
via an intermediate diagonalization;
(3) add the unit matrix, to form
$\mxunit_{\alpha\beta} + (\mxunit + \mxA)^{1/2}_{\alpha\beta}$;
(4) invert, to get
$[\mxunit + (\mxunit + \mxA)^{1/2}]^{-1}_{\alpha\beta}$;
(5) multiply by $2$.

Note that $1 + \xi(r)$ is necessarily positive,
being an expectation value of products of positive densities in real space.
Thus the matrix $\mxunit + \mxA$ is necessarily positive definite,
with all positive eigenvalues,
and its square root $(\mxunit + \mxA)^{1/2}$ is therefore
always well-defined.

Multiplying the power spectrum by the prewhitening matrix
given by equation~(\ref{prewhmx})
yields the prewhitened power spectrum.
To make this work properly,
the continuous power spectrum $P(k_\alpha)$ must first be discretized
into a vector $P_\alpha$:
\begin{equation}
\label{Pvec}
  P_\alpha = P(k_\alpha) \,
  \Delta V_\alpha^{1/2}
\end{equation}
(no implicit summation).
The discretized prewhitened power $X_\alpha$
is the matrix product of the prewhitening matrix, equation~(\ref{prewhmx}),
with the discretized power $P_\alpha$, equation~(\ref{Pvec}):
\begin{equation}
  X_\alpha =
  2 \, \left[ \mxunit + (\mxunit + \mxA)^{1/2} \right]^{-1}_{\alpha\beta}
    P_\beta
\end{equation}
(implicit summation over $\beta$).
Finally,
undiscretize
\begin{equation}
  X(k_\alpha) = X_\alpha \,
  \Delta V_\alpha^{-1/2}
\end{equation}
(no implicit summation)
to obtain the prewhitened power spectrum $X(k_\alpha)$.

The above prescription describes how to prewhiten the power spectrum
by applying the prewhitening matrix
$2 \, \left[ \mxunit + (\mxunit + \mxA)^{1/2} \right]^{-1}$.
This matrix is not the same as the matrix
$\mxH = (\mxunit + \mxA)^{-1/2}$
that prewhitens the covariance of power, equation~(\ref{DXkDXk}).
Consult equations~(\ref{Xr})--(\ref{DXrDXr})
to see why this distinction arises.
The prewhitening matrix $\mxH$ can be constructed
in a manner similar to the prewhitening matrix
$2 \, \left[ \mxunit + (\mxunit + \mxA)^{1/2} \right]^{-1}$.

Bug alert:
be careful to discretize correctly.

\section{A (failed) attempt at Fisher Compression}
\label{failed}

This Appendix gives an illustrative example of the difficulties
encountered when one tries to compress data (\S\ref{compress})
using a covariance matrix which, being estimated from the data,
contains statistical errors.
The example is that of a single band-power, with a single FKP weighting,
and the aim is to compress the measured even harmonics
of the band-power down to a smaller number of harmonics.

There are $37$ measured even harmonics, up to $\el = 72$.
Assume, according to the prior, equation~(\ref{ladopt}),
that only even harmonics up to $\el \le \el_{\max}$ are nonzero.
The aim is then to compress the 37 harmonics down to $(\el_{\max}/2) + 1$
even harmonics, in optimal fashion.

Let $\hat P_\el$ (with hats) denote the measured amplitudes of the harmonics
of the band-power,
and let $\mxC_{\el m} = \langle \Delta\hat P_\el \Delta\hat P_m \rangle$
denote their covariance matrix,
in the present case also measured from the data
(\S\ref{covariance}).
Let $P_\el$ (without hats) represent the `parameters' of the likelihood,
the true amplitudes of the harmonics.
If the harmonics were uncorrelated with each other,
then the measured amplitudes $\hat P_\el$ of the even harmonics up to
$\el \le \el_{\max}$ would provide the best estimates of $P_\el$.
But in reality the harmonics are correlated,
so measurements of higher harmonics can, in principle,
inform values of lower harmonics.

If the usual simplifying assumption is made that
the measurements $\hat P_\el$ are Gaussianly distributed
with fixed covariance matrix $\mxC_{\el m}$,
then maximizing the likelihood ${\cal L} \propto \e^{-\chi^2/2}$
is equivalent to minimizing $\chi^2$
\begin{equation}
\label{chi2}
  \chi^2 = \sum_{\el m}
    ( \hat P_\el - D_{\el} P_\el )
    \mxC^{-1}_{\el m}
    ( \hat P_m - D_m P_m )
\end{equation}
where $D_{\el} = 1$ or $0$ as $\el \le \el_{\max}$ or $\el > \el_{\max}$.
The minimum $\chi^2$ solution of equation~(\ref{chi2}) is
\begin{equation}
\label{al}
  P_\el = \sum_{mn} F^{-1}_{\el m} D_m \mxC^{-1}_{mn} \hat P_n
\end{equation}
where $F_{\el m}$ is the Fisher matrix of the parameters $P_\el$
\begin{equation}
\label{Flm}
  F_{\el m} = \sum_{lm} D_\el \mxC^{-1}_{\el m} D_m
  \ .
\end{equation}
Equations~(\ref{al}) and (\ref{Flm}) constitute a simple example of
Fisher compression, which in effect reduces here to inverse-variance weighting.
Examination of equation~(\ref{al}) shows
(since the first $(\el_{\max}/2) + 1$ columns of $D_m \mxC^{-1}_{mn}$
(no implicit summation)
are just equal to the Fisher matrix $F_{mn}$)
that the `improved' estimate $P_\el$
is equal to the measured amplitude $\hat P_\el$
plus some linear combination of high order harmonics $\hat P_m$
with $m > \el_{\max}$.
This makes physical sense:
if, according to the prior, the higher order harmonics $\hat P_m$
with $m > \el_{\max}$ are all zero,
then adding judicious combinations of them to the lower order harmonics
can in principle yield more accurate estimates of the latter.

Equations~(\ref{al}) and (\ref{Flm}) are the theory.
Reality is different.

Consider what happens as one adds harmonics into the mix, one at a time,
starting with just the harmonics with $\el \le \el_{\max}$.
The initial situation poses no problem:
one is estimating harmonics up to $\el \le \el_{\max}$
using estimates of harmonics up to $\el \le \el_{\max}$,
and not surprisingly the best estimates are the measured values,
$P_\el = \hat P_\el$.
Now add a harmonic, the one with $\el = \el_{\max}{+}2$.
In most cases this works fine:
the best estimate $P_\el$ of each harmonic acquires a small admixture
of the new harmonic $\hat P_{\el_{\max} + 2}$,
in accordance with equation~(\ref{al}),
and the variance of the best estimate $P_\el$ decreases by a small amount.
As more and and more harmonics are folded into the mix,
the variance creeps down.
So far so good.
Sooner or later, however, the Fisher matrix hits a negative eigenvalue.
Although the negative eigenvalue does not necessarily cause immediate havoc,
it is a sign of doom impending.
Within a few more harmonics, the variance of the `best estimate' has plummeted,
even reaching negative values.
Naturally one is skeptical that a negative eigenvalue could improve
the estimate so.

So how about the idea of stopping one step before the first negative
eigenvalue appears?
At first sight this seems to work fine,
and one is encouraged to take the next step
of computing the estimated real power
$\hat P(\mu{=}0){=}\sum_{\el=0}^{\el_{\max}} P_\el {\cal P}_\el(\mu{=}0)$
from the appropriate linear combination of best-fit harmonics $P_\el$
with Legendre polynomials ${\cal P}_\el(\mu)$.
Typically,
the variance of the best estimate of real power is about half the variance
of the initial, pre-compression estimate.
In a few cases the variance is reduced by as much as a factor of 4,
apparently a serious improvement.

Unfortunately the resulting `best-fit' real power spectrum
does not live up to the advertising,
scattering about unbelievably.

Closer examination reveals the problem.
The powers with the greatest claimed reduction in variance
are the ones with the greatest admixtures of higher order harmonics.
Peering yet closer,
one finds that not only for these powers, but for all the others as well,
the greatest reduction in variance
occurs when some higher order harmonic is mixed in with unusually high weight.
The behaviour is clearly spurious,
an artefact of the compression ferreting out
harmonic combinations that random errors in the covariance matrix
have made appear artificially good.

The problem appears generic:
wherever the reduction in variance is greatest,
it is least believable.
So ends our tale of failed ambition.

\xiktable

\xikdtable

\xirtable

\xikstable

\end{document}